\pdfoutput=1
\documentclass[aps,prb,superscriptaddress,twocolumn,showpacs,a4paper,unsortedaddress,
amsmath,amssymb,byrevtex,floatfix]{revtex4}

\usepackage{graphicx}
\usepackage{times}
\usepackage{color}
\begin{document}

\title{GOLLUM: a next-generation simulation tool for electron, thermal and spin transport}

\author{J. Ferrer}
\affiliation{Departamento de F\'{\i}sica, Universidad de Oviedo, 33007 Oviedo, Spain}
\affiliation{Nanomaterials and Nanotechnology Research Center (CINN), Spain}
\affiliation{Department of Physics, Lancaster University, Lancaster LA1 4YB, UK}


\author{C. J. Lambert}
\affiliation{Department of Physics, Lancaster University, Lancaster LA1 4YB, UK}

\author{V. M. Garc\'{\i}a-Su\'arez}
\affiliation{Departamento de F\'{\i}sica, Universidad de Oviedo, 33007 Oviedo, Spain}
\affiliation{Nanomaterials and Nanotechnology Research Center (CINN), Spain}
\affiliation{Department of Physics, Lancaster University, Lancaster LA1 4YB, UK}

\author{D. Zs. Manrique}
\affiliation{Department of Physics, Lancaster University, Lancaster LA1 4YB, UK}

\author{D. Visontai}
\affiliation{Department of Physics, Lancaster University, Lancaster LA1 4YB, UK}

\author{L. Oroszlani}
\affiliation{Department of Physics, E\"otvos University, Budapest, Hungary}

\author{R. Rodr\'{\i}guez-Ferrad\'as}
\affiliation{Departamento de F\'{\i}sica, Universidad de Oviedo, 33007 Oviedo, Spain}
\affiliation{Nanomaterials and Nanotechnology Research Center (CINN), Spain}

\author{I. Grace}
\affiliation{Department of Physics, Lancaster University, Lancaster LA1 4YB, UK}

\author{S. W. D. Bailey}
\affiliation{Department of Physics, Lancaster University, Lancaster LA1 4YB, UK}

\author{K. Gillemot}
\affiliation{Department of Physics, Lancaster University, Lancaster LA1 4YB, UK}

\author{Hatef Sadeghi}
\affiliation{Department of Physics, Lancaster University, Lancaster LA1 4YB, UK}

\author{L. A. Algharagholy}
\affiliation{AL-Qadisiyah University, Iraq}

\date{\today}

\begin{abstract}
We have developed an efficient simulation tool 'GOLLUM'  for the computation of electrical, spin and thermal
transport characteristics of complex nanostructures. The new multi-scale, multi-terminal tool
 addresses a number of new challenges and functionalities that have emerged
in nanoscale-scale transport over the past few years. To illustrate the flexibility and
functionality of GOLLUM, we present a range of demonstrator calculations
encompassing charge, spin and thermal transport, corrections to density functional theory such as LDA+U and spectral adjustments, transport in
the presence of non-collinear magnetism, the quantum-Hall effect,  Kondo and Coulomb blockade effects, finite-voltage transport, multi-terminal transport, quantum pumps, superconducting nanostructures, environmental effects and
pulling curves and conductance histograms for mechanically-controlled-break-junction experiments.
\end{abstract}


\maketitle
\section{Introduction}
The development of multi-functional codes capable of predicting quantum transport properties of
complex systems is an increasingly active field of research\cite
{Heath09,Bergren09,Poulson09,Cuevasbook,Dattabook}. This is driven in part by the top-down
scaling of the active elements within CMOS devices, for which quantum effects are becoming
 important. It is also driven by the bottom-up demands of communities working on single-molecule electronics and low-dimensional systems, where structures and molecules of increasing size and complexity are of interest. In particular, multi-functional codes are needed to describe the fundamental properties of
quasi-two-dimensional materials such as graphene, silicene, germanene and their integration into
workable devices.  The need to understand the interplay between all of the above structures and
their surrounding environments creates further demands for such codes. At a more fundamental level, 
over the past forty years, a 'standard model' of electron transport has been developed, based on computing 
the scattering matrix of quantum systems connected to external sources  and there is a need for a universal 
code which describes the many realizations of such systems under a common umbrella.

A key task of any quantum transport code is to start from the Hamiltonian describing a system and
calculate the quantum-mechanical scattering matrix $\bf {S}$, from which a wide range of
measurable quantities can be predicted. Unfortunately for most nanoscale systems of interest,
the full many-body atomistic Hamiltonian is too complex to allow this task to be completed and therefore
one usually resorts to a description based on a mean-field Hamiltonian. The system of interest is
then composed of a scattering region, connected to external crystalline leads, which are in turn
connected to external reservoirs. The problem of computing the scattering matrix for such a
system described by an arbitrary mean-field Hamiltonian is solved in reference \cite{jhj}. The
main question therefore is how to obtain the correct mean-field Hamiltonian.
The simplest mean-field approach to describing quantum transport through nanostructures is to
build a tight-binding Hamiltonian, which reproduces key electronic properties near the Fermi
energy. This approach has been available for more than half a century and is still popular today
when describing generic properties of materials such as graphene\cite{kwant}. 

Tight binding parameters
can be obtained by fitting to known band structures and then varied spatially to describe
external fields and other perturbations.
However such an approach does not easily capture the effects of interfaces between different
materials or edge terminations of finite-size systems, whose properties are distinct from those
of bulk materials. Nor does it easily describe finite-voltage effects. To capture these additional features of inhomogeneous 
systems, a more material-specific approach is needed. This problem was solved in part by the non-equilibrium Green's function
technique\cite {natmat,smeagol,NEGF1,NEGF2,NEGF3,NEGF4,NEGF5,spinreview,ferrer88,NEGFA1,NEGFA2,NEGFA3,NEGFA4,NEGFA5,blugel,enkovaara,kelly}, 
that combines with density functional theory (DFT)\cite{DFT1,DFT2} to
obtain the self-consistent mean-field Hamiltonian of the system subject to a finite bias voltage and from it, the
lesser Green's functions providing the non-equilibrium electronic density and current.
This approach is utilized within the SMEAGOL code\cite{natmat,smeagol}, which was the first to describe spin-dependent and 
finite-voltage transport properties of systems with inhomogeneous magnetic moments and in the presence
of spin-orbit scattering. 

It is almost 10 years since the release of SMEAGOL and during this period, we have
developed a new code with increased speed, versatility and functionality, which is particularly
suited to the modeling of larger-scale nanostructures, interacting with their environments. This
new code is called GOLLUM and will be freely available from {\em http://www.physics.lancs.ac.uk/gollum} within the coming weeks.
Our previous experience in the development of SMEAGOL\cite{smeagol,spinreview} allowed us to understand that
non-equilibrium transport codes are quite difficult to handle, in part because of their complex input data structures, 
which can create a steep learning curve, and also because they carry very heavy computational demands. As a consequence, we 
have devised the new code GOLLUM to be more user friendly, with
simple and easy to understand input and output structures, and having no accuracy parameters to tune. We present now a short
summary of the features and functionalities of the two programs to better appreciate its differences. 

SMEAGOL is a NEGF program that computes the charge and spin transport properties of two-terminal junctions subject
to a finite voltage bias. SMEAGOL cannot read a user-defined tight-binding Hamiltonian. Instead, it reads the mean-field 
Hamiltonian from the program SIESTA\cite{siesta} and is tightly bound to the old versions of it. SMEAGOL can read 
from SIESTA Hamiltonians carrying non-collinear spin arrangements as well as the spin-orbit interaction. SIESTA and 
SMEAGOL have indeed been used successfully to simulate the magnetic anisotropies of atomic clusters\cite{noncol,soc,seivane} 
and the spin transport functionalities of several atomic chains and molecular junctions subjected to strong spin-orbit interaction\cite{victor,salva}. 
However, SMEAGOL does not profit from other recent density functionals.  Examples are the van der Waals family of functionals 
or those  based on the LDA+U approach.

GOLLUM is a program that computes the charge and spin, and the electronic contribution to the thermal transport properties of 
multi-terminal junctions. In contrast to NEGF codes, GOLLUM is based on equilibrium transport theory, which means that 
it has a simpler structure, it is faster and consumes less memory. The program has  
been designed for user-friendliness and takes a considerable leap towards the realization of  ab initio multi-scale 
simulations of conventional and more sophisticated transport functionalities. 

The simpler interface of GOLLUM allows it to read 
model tight-binding Hamiltonians. Furthermore, GOLLUM has been designed to interface easily with 
any DFT code that uses a localized basis set. It currently reads information from all the latest 
public flavors of the codes SIESTA\cite{siesta} and FIREBALL\cite{fireball}. These include
functionals that handle the spin-orbit or the van der Waals interactions, or that include strong 
correlations in the spirit of the LDA+U approach. Plans to generate interfaces to other codes are underway. 
Two- and  three-dimensional topological materials display fascinating spin transport properties. GOLLUM can
simulate junctions made of these materials either using parametrized tight-binding Hamiltonians\cite{zhang,kuramoto},
or DFT\cite{pablo1}.

DFT does not handle correctly strong electronic correlation effects, that are inherent many nano-scale electrical
junctions. As a consequence, a number of NEGF programs like SMEAGOL underestimate such effects. GOLLUM includes
several tools to handle strong correlations. These include the above-mentioned interface to the versions of
SIESTA containing the LDA+U functional. A second tool uses a phenomenological but effective approach called the scissors
correction scheme. A third tool  maps the DFT Hamiltonian into an Anderson-like Hamiltonian
that is handled with an impurity solver in the spirit of dynamical mean field theory.
 
The lighter computational demands required by GOLLUM  make it possible to
construct conductance statistics relevant to break-junction and STM measurements of single-molecule 
conductances, therefore making closer contact with experiments. GOLLUM also incorporates 
an interface with some classical molecular dynamics programs, which enables it to handle interactions with the environment.

GOLLUM makes use of the concept of {\em virtual leads}, that allows it to integrate easily a wide range of phenomena by the
use of tight-binding Hamiltonians. These include spintronics, superconductivity, Kondo physics and 
topological phases. In contrast with SMEAGOL, which only computes the magnitudes of transmission coefficients of two-terminal junctions, GOLLUM has access to the full scattering matrix of a multi-terminal junction,
enabling it to compute scattering amplitudes,  phases and Wigner delay times and thereby describe the properties of quantum pumps.

Even though GOLLUM is based on equilibrium transport theory, our experience with the use of the NEGF code SMEAGOL has 
enabled us to incorporate non-equilibrium physics into the  mean-field Hamiltonian. GOLLUM has
therefore the ability to compute non-equilibrium current-voltage curves.

In this
article, our aim is to describe the structure of the code and then present a set of demonstrator
calculations. The latter will illustrate the additional functionality and versatility of GOLLUM and at the
same time constitute a set of new results for the transport properties of selected structures.
All of the functionalities that will be discussed below are available either in the current public version of the code,
or in the current development version, that will be made public in the autumn of 2014.

The layout of this article is as follows. In Section II, we describe the theoretical approach behind the program and
outline the theoretical and practical details of the current implementation. The section
starts with a detailed description of the generic two-probe and multi-probe junction setups available within GOLLUM and introduces the terminology that will
be used throughout the article. This is followed by subsections describing the determination of the surface Green's function of each current-carrying
lead and the full scattering matrix. We then introduce a convenient method that
allows us to describe finite-voltage non-equilibrium effects.
A subsection explaining the concept of virtual leads enables us to describe
hybrid structures containing non-collinear magnetism or superconductivity within the scattering region. Two additional subsections
explain two facilities included in GOLLUM that enable us to describe electronic correlation effects beyond DFT,
including Coulomb blockade and Kondo physics. We then show how to include a gauge field in the GOLLUM Hamiltonian. A final subsection explains the multi-scale methodology used to describe
large-scale junctions and environmental effects, using a combination of classical molecular dynamics for the environment and quantum transport for the central scattering region.

In section III, we present the details and results of the simulations
of a series of sixteen different demonstrator systems. The purpose of each demonstrator is to present one or more
of the functionalities of GOLLUM. We start with a few model junctions, described by tight-binding Hamiltonians, which show basic capabilities and demonstrate how easily and flexibly the program can analyze
non-trivial physical effects. These include two and four-terminal normal-metal junctions, a two-dimensional system
showing the  quantum Hall effect, and two hybrid structures, the first containing two superconducting islands
sandwiched by two normal metal electrodes, and the second containing a non-collinear spin structure. There follow a series of DFT-based calculations, that describe spin-active junctions;
graphene junctions, where the use of van der Waals functionals is crucial and junctions enclosing metallo-organic
molecules, where a treatment of strong-correlations beyond DFT is mandatory. These are handled using three
different approaches. In the first, the  properties of  metalloporphyrin junctions are described using the LDA+U
methodology; in the second, the LDA spectrum of an OPE derivative is adjusted to improve the agreement with experimental data;
in the third, we describe Coulomb blockade and Kondo features of model or simple gold junctions. The next 
demonstrator shows how GOLLUM can compute the thermoelectric transport properties of a junction.This is followed by a discussion of
the transport properties of a carbon-nanotube-based four-probe junction. The  next three
examples require the use of multi-scale techniques, where we use a three-step methodology described later in the
article. The first demonstrator describes how liquid environmental effects modify the transport properties of a
single-molecule junction. The second example demonstrates that single strands of DNA can be trans-located though
graphene nanopores, where the strands effectively gate the nanopore structure yielding a highly sensitive 
DNA sensor based on field-effect-transistor concepts. A final demonstrator shows how GOLLUM can compute full 
sequences of pulling and pushing cycles in single-molecule junctions resembling the opening and closing cycles of
Mechanically Controllable Break Junction (MCBJ) experiments, enabling the construction of theoretical conductance histograms.
The final demonstrator shows how GOLLUM has access to the phase of the full scattering matrix,
and describes non-trivial quantum pumping effects related to the phase evolution of the scattered wave function. A concluding section summarizes  the features delivered by the program and an appendix
illustrates our method to decimate Hamiltonians and overlap matrices.

\section{Theoretical approach}
\subsection{Description of the transport methodology}
\subsubsection{The generic setup and construction of the Hamiltonian}
GOLLUM describes open systems comprising an extended scattering region (colored dark blue in Figs. 1 and 2) connected to external crystalline leads (colored light blue in Figs. 1 and 2). Depending on the 
problem of interest and the language used to describe the system, the material (M) of interest forming the central part of the scattering region could comprise  a
single molecule, a quantum dot, a mesoscopic cavity, a carbon nanotube, a two-dimensional mono- or multi-layered material,
a magneto-resistive element or a region containing one or more superconductors.

Figure 1 shows an example of a 4-lead system whose central scattering region (generically labelled M throughout the paper) is a molecule.
It is important to note that in an accurate ab initio description of such a structure, the properties of the leads 
closest to the molecule (or more generally the central scattering material) will be modified by the presence of the central scattering region (M) and by the fact 
that the leads terminate. In what follows, we refer to those affected portions of the leads closest to the central scatterer as 'branches' and include them as part of the 'extended scatterer' (denoted EM throughout this article). Consequently within GOLLUM, a typical structure 
consists of an extended scatterer (EM), formed from both the central scatterer (M) and the  branches. 
The extended scattering region is connected to crystalline current-carrying leads of constant cross-section, shown in light blue
in the Figs. 1 and 2. For an accurate description of a given system, the branches are chosen to be long enough such that they 
join smoothly with the (light blue) crystalline leads. Crucially, the properties of this interface region between the central scatterer M 
and the leads are determined by their mutual  interaction and are not properties of either M or the electrodes alone. 

\begin{figure}
\includegraphics[trim=3cm 2cm 3.5cm 0cm, clip=true, width=1.0\columnwidth]{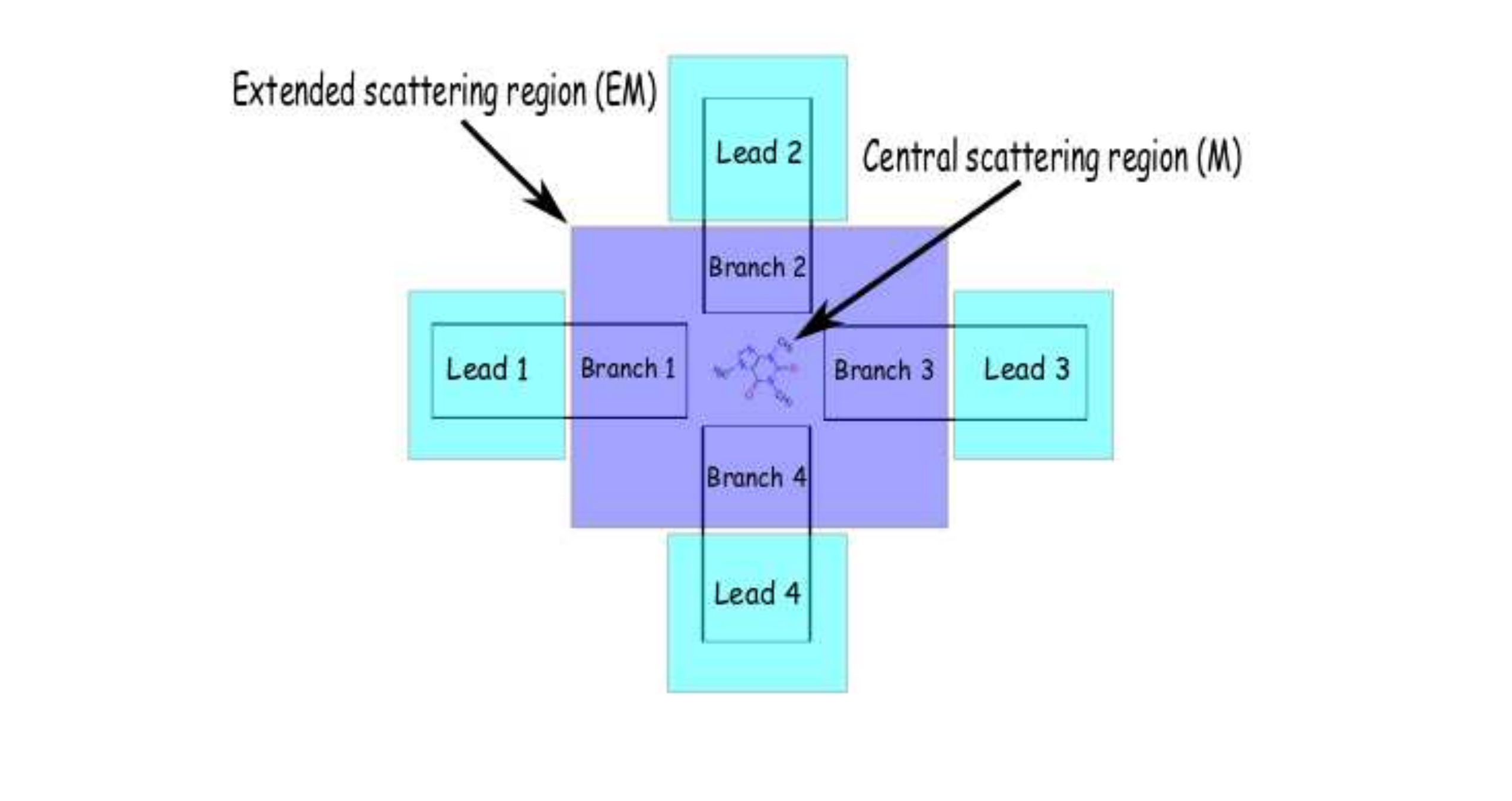}
\caption{(Color online) (a) Schematic plot of a four-terminal device, which includes an extended 
scattering region and four leads. }
\end{figure}

\begin{figure}
\includegraphics[width=\columnwidth]{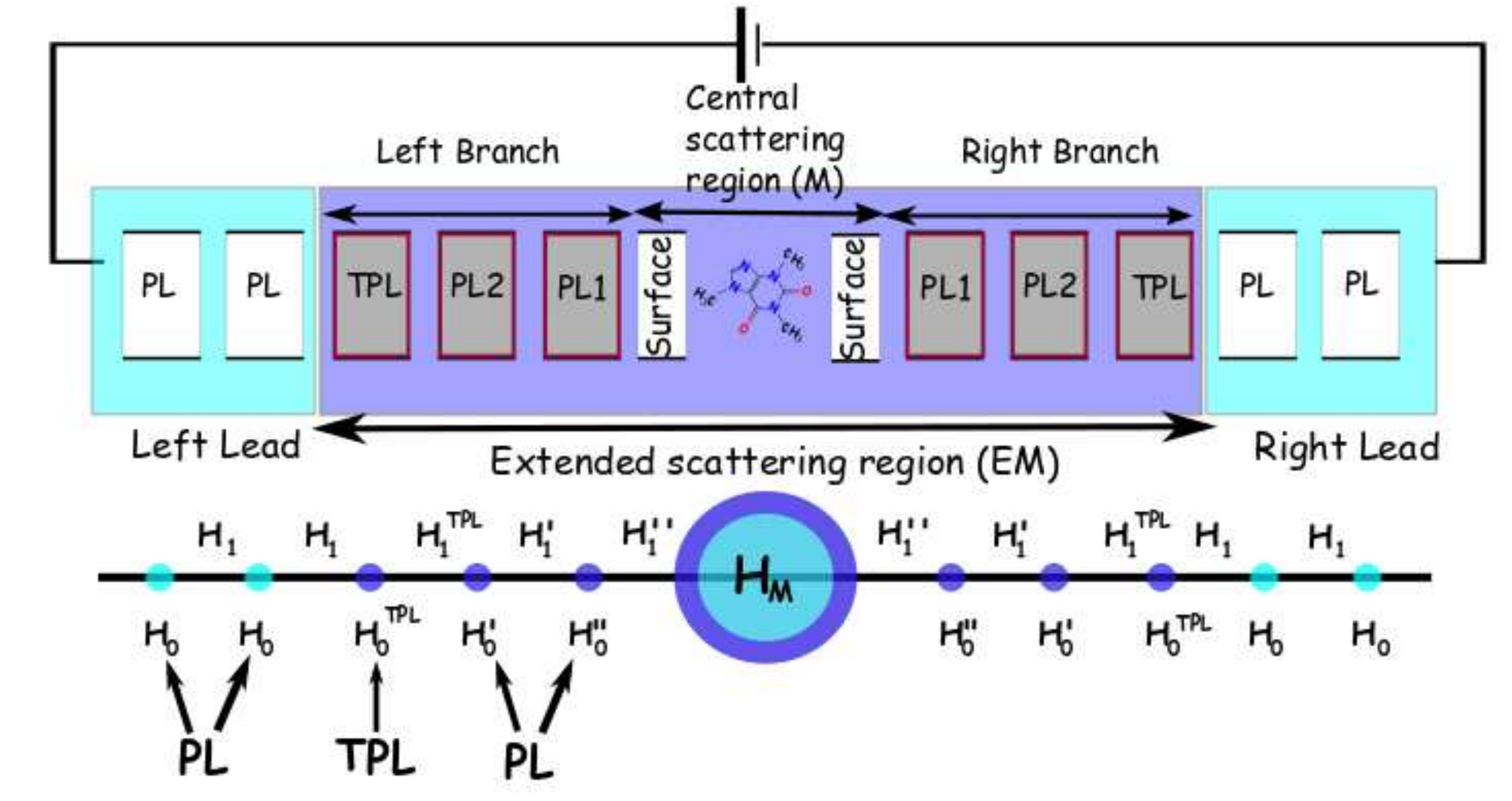}
\caption{(Color online) (Top) Schematic two-terminal device, where electrons are driven from the left to the right 
lead through the Extended scattering region. The leads are possibly kept at chemical potentials $\mu_{L,R}=\pm e\, V/2$, 
where $V$ is an applied bias. (Bottom) Each lead is composed of an infinite chain of identical PL with Hamiltonian $H_0$ 
coupled with each other via coupling Hamiltonians $H_1$. The extended scattering region comprises the
actual central scattering region and several PL in each branch up to the TPL. The TPL connect the EM region to 
the leads. The central scattering region consists in this example of the electrodes surfaces and a molecule.
}
\end{figure}

Fig. 2 shows a two-terminal device in more detail and introduces further terminology to be used
throughout the paper. The regions in 
light blue are called electrodes or leads and are described by perfect periodic Hamiltonians subject to chosen chemical 
potentials. Each lead $i$ is formed by a semi-infinite series of identical layers of constant cross section, which we refer to 
as principal layers (PLs).  Fig. 2 shows only two PLs per lead (colored white), although an infinite number is implied. Furthermore in the figure, the leads are identical and therefore the lead index $i$ has been dropped. These PLs are described mathematically by intra-layer 
Hamiltonians $H^i_0$.  PLs must be chosen so that they are coupled only to their nearest neighbors 
by the Hamiltonians $H^i_1$, which means that in the presence of long-range couplings, a PL may contain more than one longitudinal unit cell of the lead. Then, if each PL contains $N^i$ orbitals, then $H^i_0$ and $H^i_1$ are square $N^i\times N^i$ 
matrices. 
The extended scatterer (EM) in dark blue is composed of a central scattering region (M)
and branches. Each branch contain several PLs. These PLs have an identical atomic arrangement as the PLs in the 
leads. However, their Hamiltonians  differ from $H^i_0$ and $H^i_1$ due to the presence of the central scattering region. 
PL numbering at each branch starts at the PL beside the central scattering region.   
The outermost PL at each branch of the EM region is called the terminating principal layer (TPL) and must be described by 
Hamiltonians  
$H^{i,TPL}_0$ and $H^{i,TPL}_1$  which are close enough to $H^i_0$ and $H^i_1$,   to match
smoothly with the corresponding lead Hamiltonian. For this reason, GOLLUM requires that the EM contain at the very least one PL. 
The central scatterer (M) itself
is described by an intra-scatterer Hamiltonian $H_\mathrm{M}^0$ and coupling matrices to the closest PLs of the branches.
In the example in Fig. 2, the central scattering region M comprises a molecule and the atoms forming the electrode surfaces. The surfaces
in GOLLUM include all atoms belonging to the electrodes whose atomic arrangements cannot be cast exactly as a PL, due to 
surface reconstructions, etc. For simplicity, Fig. 2 shows the case of a symmetric system, although no such symmetries 
are imposed by GOLLUM. All Hamiltonians are spin-
dependent, but again for notational simplicity, the spin index $\sigma$ will not be written explicitly here. 

This means that the Hamiltonian ${\cal H}^{i}$ for a given lead $i$ can be written as the semi-infinite matrix:
\begin{equation}
{\cal H}^{i}=\left(\begin{array}{cccccccc}
.&.&.&.&.&.&.&.\\
.&0&H_{-1}^i&H_0^i&{H}_1^i&0&.&.\\
.&.&0&H_{-1}^i&H_0^i&{H}_{1}^i&0&.\\
.&.&.&0& H_{-1}^i& H_{0}^i &H_{1}^i&0\\
.&.&.&.&0&H_{-1}^i&H_0^i&H_1^i\\
.&.&.&.&.&0&H_{-1}^i&H_0^i\\
\end{array}\right)\;
\end{equation}
When using a non-orthogonal basis set, overlap matrices must be defined with the same structure as
the Hamiltonian matrices: $S_{0,\pm1}^i$ and ${\cal S}^i$.
It is convenient to introduce the notation
\begin{eqnarray}
K^i_{0,\pm1}&=&H^i_{0,\pm1}-E\,S^i_{0,\pm1}\nonumber\\
{\cal K}^i&=&{\cal H}^i-E\,{\cal S}^i
\end{eqnarray}
For notational simplicity, from now we will consider the case where all leads are equal so that the super-index $i$ can be omitted, although GOLLUM imposes no such restrictions.
The program can assume either open or periodic boundary conditions in the plane perpendicular to the transport direction.
In this last case, unit cells are chosen in the plane perpendicular to the transport direction and the Hamiltonians and overlap 
matrices acquire a specific dependence on the transverse
k-vector, $k_\perp$,
\begin{eqnarray}
 K_{0,\pm1}^{\mu,\nu}(k_\perp)=\sum_{R_\perp} K_{0,\pm1}^{\mu,\nu'}( R_\perp) \,e^{i\,k_\perp\cdot R_\perp}
\end{eqnarray}
where $\mu$ and $\nu$ label the $N$ orbitals in the unit cell (ie PL) at the origin of $R_\perp$, while $\nu'$ denote
orbitals equivalent to $\nu$, but placed in adjacent unit cells located at transverse positions $R_\perp$. For $K_0$, $\mu$ and $\nu$ must belong to the
same PL $n$, while for $K_1$, $\nu'$ must belong to the PL $n'=n+1$. Finally $R_\perp$ are vectors in the two-dimensional
Bravais lattice, joining the unit cell taken as origin with its neighboring unit cells.

To illustrate how an EM is connected to leads, we now consider the 4-lead example of Fig.1, where we assume that the TPL is
the third PL in each branch. To describe such a multi-terminal 
setup, the Hamiltonian matrix ${\cal K}^\mathrm{EM}$ of the EM 
is arranged in a non-conventional way. The first matrix block corresponds to the central scattering region  $K^0_\mathrm{M}$, the second matrix block corresponds to the PLs in the EM branch connecting to lead 1, 
the third matrix block to the PLs in the EM branch connecting to lead 2, and so on. 
\begin{widetext}
\begin{equation}
{\cal K}^\mathrm{EM}=
\left(\begin{array}{cccc|ccc|ccc|ccc|ccc}
.        &   .           &.        &.        &  0      &0       &0    &0       &0      &0   &K''_1   &0      &0   &0       &0      &0\\
.        &K_\mathrm{M}^0 &.        &.        &  0      &0       &0    &K''_1   &0      &0   &0       &0      &0   &0       &0      &0  \\
.        &   .           &.        &.        &  0      &0       &0    &0       &0      &0   &0       &0      &0   &K''_1   &0      &0\\
.        &   .           &.        &.        & K''_1   &0       &0    &0       &0      &0   &0       &0      &0   &0       &0      &0\\
\hline
0        &   0           &0        &K''_{-1} & K''_0   &K'_1    &0    &0       &0      &0   &0       &0      &0   &0       &0      &0\\
0        &   0           &0        &0        & K'_{-1} &K'_0    &K^{TPL}_1  &0       &0      &0   &0       &0      &0   &0       &0      &0\\
0        &   0           &0        &0        & 0       &K^{TPL}_{-1}  &K^{TPL}_0  &0       &0      &0   &0       &0      &0   &0       &0      &0\\
\hline
0        &   K''_{-1}    &0        &0        & 0       &0       &0    &K''_0   &K'_1   &0   &0       &0      &0   &0       &0      &0\\
0        &   0           &0        &0        & 0       &0       &0    &K'_{-1} &K'_0   &K^{TPL}_1 &0       &0      &0   &0       &0      &0\\
0        &   0           &0        &0        & 0       &0       &0    &0       &K^{TPL}_{-1} &K^{TPL}_0 &0       &0      &0   &0       &0      &0\\
\hline
K''_{-1} &   0           &0        &0        & 0       &0       &0    &0      &0       &0    &K''_0   &K'_1   &0   &0       &0      &0\\
0        &   0           &0        &0        & 0       &0       &0    &0      &0       &0    &K'_{-1} &K'_0   &K^{TPL}_1 &0       &0      &0\\
0        &   0           &0        &0        & 0       &0       &0    &0      &0       &0    &0       &K^{TPL}_{-1} &K^{TPL}_0 &0       &0      &0\\
\hline
0        &   0           &K''_{-1} &0        & 0       &0       &0    &0      &0       &0    &0       &0      &0   &K''_0   &K'_1   &0   \\
0        &   0           &0        &0        & 0       &0       &0    &0      &0       &0    &0       &0      &0   &K'_{-1} &K'_0   &K^{TPL}_1 \\
0        &   0           &0        &0        & 0       &0       &0    &0      &0       &0    &0       &0      &0   &0       &K^{TPL}_{-1} &K^{TPL}_0 
\end{array}\right)\;
\end{equation}
\end{widetext}

Finally, the EM described by ${\cal K}^\mathrm{EM}$ and the leads described by ${\cal K}^i$ are coupled via matrices ${\cal K}^\mathrm{iM}$ to yield a
  four-terminal junction described by the infinite matrix:
\begin{equation}
{\cal K}
=
\left(\begin{array}{ccccc}
{\cal K}^\mathrm{1}   & 0                    & 0                    & 0                    & {\cal K}^\mathrm{1M}\\
0                     & {\cal K}^\mathrm{2}  & 0                    & 0                    & {\cal K}^\mathrm{2M} \\
0                     & 0                    & {\cal K}^\mathrm{3}  & 0                    & {\cal K}^\mathrm{3M}\\
0                     & 0                    & 0                    & {\cal K}^\mathrm{4}  & {\cal K}^\mathrm{4M}\\
{\cal K}^\mathrm{M1}  & {\cal K}^\mathrm{M2} & {\cal K}^\mathrm{M3} & {\cal K}^\mathrm{M4} & {\cal K}^\mathrm{EM}\\
\end{array}\right)\;
\end{equation}

where, ${\cal K}^\mathrm{iM}=({\cal K}^\mathrm{Mi})^\dagger$ couple ${\cal K}^\mathrm{EM}$ and for example,
\begin{equation}
{\cal K}_\mathrm{1M}=\left(\begin{array}{cccccc|ccc|ccc|ccc|ccc}
.&.&.&.&.&.&.&.&.&.&.&.&.&.&.&.&.&.\\
0&0&0&0&0&0&0&0&0&0&0&0&0&0&0&0&0&0\\
0&0&0&0&0&0&0&0&K_1&0&0&0&0&0&0&0&0&0
\end{array}\right)
\end{equation}
This arrangement of the Hamiltonian  enables the straightforward generalization of the approach to an arbitrary number of leads.

The transport properties of the junction are encapsulated in its scattering matrix $\mathbb{S}$, which
can be obtained by computing the Green's function of the whole junction
\begin{equation}
{\cal G}=\left(\begin{array}{ccccc}
{\cal G}^\mathrm{1}   & {\cal G}^\mathrm{12} & {\cal G}^\mathrm{13} & {\cal G}^\mathrm{14} & {\cal G}^\mathrm{1M}\\
{\cal G}^\mathrm{21}  & {\cal G}^\mathrm{2}  & {\cal G}^\mathrm{23} & {\cal G}^\mathrm{24} & {\cal G}^\mathrm{2M} \\
{\cal G}^\mathrm{31}  & {\cal G}^\mathrm{32} & {\cal G}^\mathrm{3}  & {\cal G}^\mathrm{34} & {\cal G}^\mathrm{3M}\\
{\cal G}^\mathrm{41}  & {\cal G}^\mathrm{42} & {\cal G}^\mathrm{43} & {\cal G}^\mathrm{4}  & {\cal G}^\mathrm{4M}\\
{\cal G}^\mathrm{M1}  & {\cal G}^\mathrm{M2} & {\cal G}^\mathrm{M3} & {\cal G}^\mathrm{M4} & {\cal G}^\mathrm{EM}\\
\end{array}\right)
\end{equation}
by solving the infinite system of equations
\begin{equation}
 -{\cal K}\,{\cal G}={\cal I}
\end{equation}
This equation can be simplified by replacing the semi-infinite lead Greens functions
${\cal G}^\mathrm{i}$
 by their surface Green's functions $G_{S,0}^\mathrm{i}$, whose dimensions are $N\times N$.
The remaining system of equations takes the form
\begin{widetext}
\begin{equation}
\left(\begin{array}{cccc|c}
(G_{S,0}^\mathrm{1})^{-1} & 0                         & 0                         & 0                         & -K^\mathrm{1M}\\
0                         & (G_{S,0}^\mathrm{2})^{-1} & 0                         & 0                         & -K^\mathrm{2M} \\
0                         & 0                         & (G_{S,0}^\mathrm{3})^{-1} & 0                         & -K^\mathrm{3M}\\
0                         & 0                         & 0                         & (G_{S,0}^\mathrm{4})^{-1} & -K^\mathrm{4M}\\
\hline
-K^\mathrm{M1}            & -K^\mathrm{M2}            & -K^\mathrm{M3}            & -K^\mathrm{M4}            & -{\cal K}^\mathrm{EM}\\
\end{array}\right)\;\left(\begin{array}{cccc|c}
G_S^\mathrm{1} & G^\mathrm{12}  & G^\mathrm{13}  & G^\mathrm{14}  & G^\mathrm{1M}\\
G^\mathrm{21}  & G_S^\mathrm{3} & G^\mathrm{23}  & G^\mathrm{24}  & G^\mathrm{2M} \\
G^\mathrm{31}  & G^\mathrm{32}  & G_S^\mathrm{3} & G^\mathrm{34}  & G^\mathrm{3M}\\
G^\mathrm{41}  & G^\mathrm{42}  & G^\mathrm{43}  & G_S^\mathrm{4} & G^\mathrm{4M}\\
\hline
G^\mathrm{M1}  & G^\mathrm{M2}  & G^\mathrm{M3}  & G^\mathrm{M4}  & {\cal G}^\mathrm{EM}\\
\end{array}\right) ={\cal I}
\label{GF}
\end{equation}
\end{widetext}
where the coupling Hamiltonians are now
\begin{eqnarray}
K^\mathrm{1M}&=&\left(\begin{array}{cccccc|ccc|ccc|ccc|ccc}
0&0&0&0&0&0&0&0&K_1&0&0&0&0&0&0&0&0&0
\end{array}\right)
\end{eqnarray}
and the surface Green's functions of the isolated leads $G_{S,0}^i$ can be obtained
as described in Section II.A.3 below. The equation for the full Green's function can be written in a more compact form as

 \begin{eqnarray}
\left(\begin{array}{cc}
        G_{S,0}^{-1}& -K^\mathrm{coup}\\-(K^\mathrm{coup})^\dagger&-{\cal K}^\mathrm{EM}
       \end{array}\right)\;
       \left(\begin{array}{cc}
        G_S& G^\mathrm{SM}\\G^\mathrm{MS}&{\cal G}^\mathrm{EM}
        \end{array}\right)=&I.
\end{eqnarray}

The scattering matrix can be computed using the Green's functions matrix elements $G_S^{ij}$ connecting the different
leads, which can be obtained by inverting only the upper matrix box in Eq. (\ref{GF})
\begin{equation}
 G_S=\left(G_{S,0}^{-1}+K^\mathrm{coup}\,{\cal K}^\mathrm{EM}\,(K^\mathrm{coup})^\dagger\right)^{-1}
\end{equation}
In contrast, access to the local electronic and current densities at the EM region is obtained from
\begin{equation}
{\cal G}^\mathrm{EM}=-\left({\cal K}^\mathrm{EM}+ (K^\mathrm{coup})^\dagger\,G_{S,0}^{-1}\,K^\mathrm{coup}\right)^{-1}
\end{equation}

The above expressions for the Hamiltonians are very general. Any appropriate tight-binding Hamiltonian could be introduced by hand to
allow computation of the transport properties of a parametrized model. Alternatively, any DFT code using localized basis
sets can provide them. In this case the DFT program  produces
the Hamiltonians and Fermi energy of the EM region ${\cal H}^\mathrm{EM}$, ${\cal S}^\mathrm{EM}$ and $E_F^\mathrm{EM}$,
and of each lead $H_{0,\pm1}^i$, $S_{0,\pm1}^i$ and $E_F^i$ in separate runs.
GOLLUM has an interface to the latest versions of the DFT program SIESTA (SIESTA 3.1, SIESTA VDW and SIESTA LDA+U) and of FIREBALL and
more interfaces will be developed in the future.
Spin degrees of freedom in spin-active systems are handled as follows: if the spins are all collinear, then we compute separate
Hamiltonians and perform separate transport calculations for the spin-up and -down degrees of freedom. However, if the junction
has non-collinear spins or is subject to spin-orbit interactions\cite{noncol,soc}, then the spin-up and down-degrees of freedom
are regarded as two distinct sets of orbitals in the Hamiltonian, whose distinct labels allow the 
computation of spin-currents or magneto-resistive behaviors.

\begin{figure}
\includegraphics[width=0.9\columnwidth]{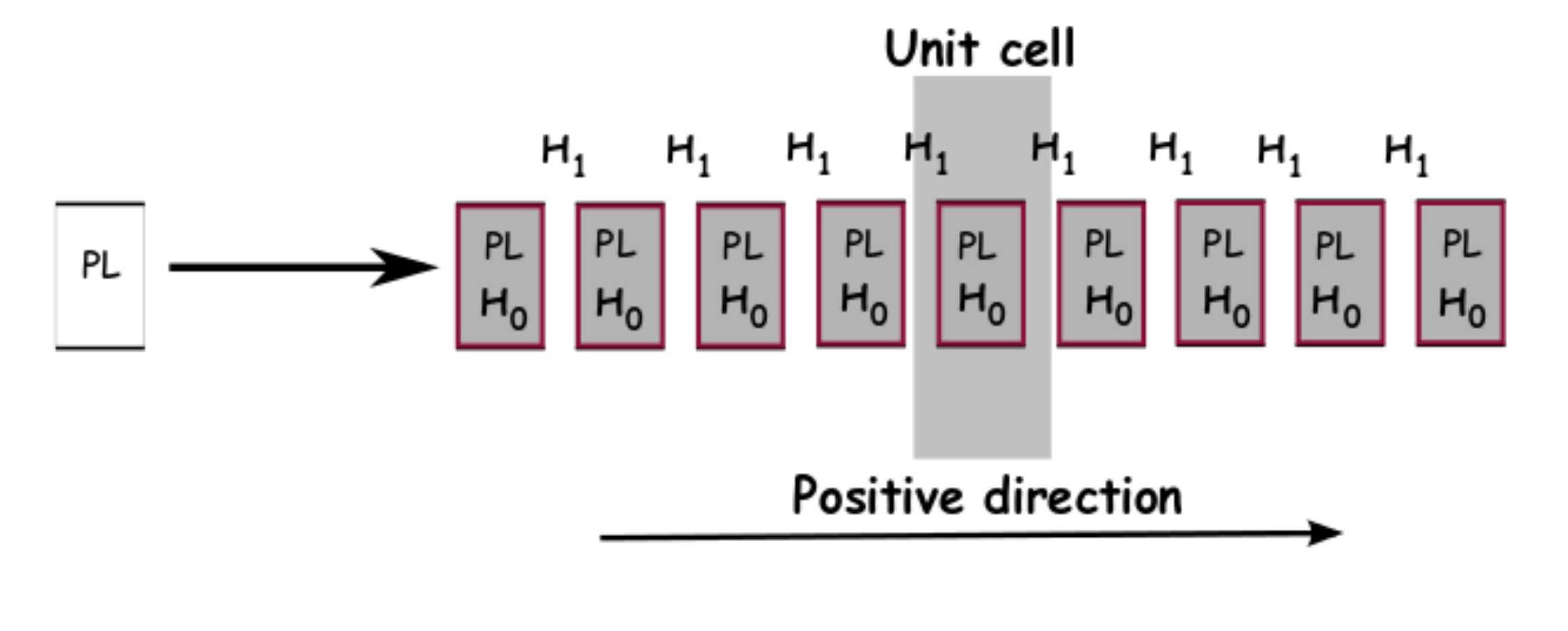}
\caption{\label{fig:lead-generation}(Color online) Infinite system used to generate the leads Hamiltonians $K_0^i$ and $K_1^i$. A positive direction is defined to be towards the scatterer. }
\end{figure}

\subsubsection{Generating the lead surface Green functions $G_{S,0}^i$}
Each of the lead Green's functions $G_{S,0}^i$ is determined following the procedures described in Refs. 
[\onlinecite{jhj,smeagol,Rungger08}] with some
minor modifications. To do so, we start by associating to each semi-infinite lead $i$ a periodic infinite system, whose unit cell contains a single PL,
as sketched in Fig. (\ref{fig:lead-generation}). $K_0^i$ and $K_1^i$ can then be created for this infinite system 
by hand as model Hamiltonians, or can be generated by a DFT program in a dedicated simulation.
Notice that we will drop the $i$ super-index until the end of the section for simplicity.

By expanding the Bloch eigenstates of the infinite system in a localized basis set
\begin{equation}
| \Psi(k)\rangle=\sum_{n,\mu}\,e^{i k n } \,c_\mu(k)\,\,|\,\varphi(n,\mu)\,\rangle
\end{equation}
where $n$, $\mu$ are indices for its unit cells and the orbitals within them, and $k$ is a dimensionless, longitudinal 
Bloch wave-vector, the following secular $N\times N$ equation can be deduced
\begin{equation}
 \left(K_0+K_1\,e^{i\,k}+K_{-1}\,e^{-i\,k}\right)\,C(k)=0
 \label{GEP1}
\end{equation}
where the column vector $C(k)$ contains the wave-function coefficients $c_\mu(k)$. The above equation is
usually solved by choosing a wave vector $k$ and solving for the eigen-energies and the corresponding eigenvectors.
However, in the present transport problem, we do the opposite: we choose the energy $E$ and solve  for the allowed
wave vectors and corresponding eigenstates. For a given energy $E$, the above equation has $2N$ solutions with
either real or complex wave vectors $k_p$, $p=1,...,2N$. To obtain the latter, the equation can be recast as
\begin{equation}
 \left(\begin{array}{cc}
 -K_0&-K_{-1}\\I_N&0_N
 \end{array}\right)\,{\bf C}(k_p)\,=\,e^{i k_p}\,
 \left(\begin{array}{cc}
 K_1&0_N\\0_N&I_N
 \end{array}\right)\,{\bf C}(k_p)
 \label{GEP}
\end{equation}
where $I_N$ and $0_N$ are the $N\times N$ identity and zero matrices,
\begin{equation}
 {\bf C}(k_p)=\left(\begin{array}{c} e^{i k_p /2}\\e^{-i k_p /2}\end{array}\right)\,C(k_p)
\end{equation}
and $C(k_p)$ is a N-component column vector.
GOLLUM solves equation (\ref{GEP}) as it is superior in numerical terms to equation (\ref{GEP1}).
We compute the group velocities of the states corresponding to real wave vectors as
\begin{eqnarray}
 v(k_p)&=&\frac{\langle\Psi(k_p)|\hat{v}|\Psi(k_p)\rangle}{\langle\Psi(k_p)|\Psi(k_p)\rangle}\\
 &=&i\,\frac{C(k_p)^\dagger\,\left(K_1\,e^{i k_p}-K_{-1}\,e^{-i k_p}\right)\,C(k_p)}
 {C(k_p)^\dagger\,\left(S_0+S_1\,e^{i k_p}+S_{-1}\,e^{-i k_p}\right)\,C(k_p)}\nonumber
 \label{group-velocity}
\end{eqnarray}
Note that $v(k_p)$ has units of energy and therefore the real, fully-dimensioned group velocity is $v(k_p)(a/\hbar)$, 
where $a$ is the spacing between neighboring PLs in a given lead.

We now divide the $2N$ wave vectors obtained from Eq. (\ref{GEP}) into two sets, each containing N values.The first set denoted \{$k_p$\} are real (complex) wave vectors that have $v_p>0$ ($\mathrm{Im}(k_p)>0$)  and therefore propagate 
propagate (decay) to the right of the figure. They are consequently called positive open (closed) channels.
The second set denoted \{$\bar k_p$\} are real (complex) wave-vectors that have $v_p<0$  ($\mathrm{Im}(k_p)<0$) propagate (decay) to the left of the
figure. They are called negative open (closed) channels. 

We introduce the dual vectors $D(k_p)$, $D(\bar k_p)$,
which satisfy $D(k_p)^\dagger\cdot C(k_q)=\delta_{p,q}$ and  $D(\bar k_p)^\dagger\cdot C(\bar k_q)=\delta_{p,q}$.
These can be found by inverting the $N\times N$ matrices
\begin{eqnarray}
 Q&=&\left(C(k_1),...,C(k_N)\right)=\left(C_1,...,C_N\right)\nonumber\\
 \bar Q&=&\left(C(\bar k_1),...,C(\bar k_N)\right)=\left(\bar C_1,...,\bar C_N\right)\nonumber\\
  \left(D_1,...,D_N\right)&=&(Q^{-1})^\dagger\nonumber\\
\left({\bar D}_1,...,{\bar D}_N\right)&=&({\bar Q}^{-1})^\dagger
\label{duals}
\end{eqnarray}
The  above eigenvectors can be used to construct the following transfer matrices
\begin{eqnarray}
 {\cal T}&=&\sum_1^N\,C_n\, e^{i k_n}\, D_n^\dagger\nonumber\\
 {\cal {\bar T}}&=&\sum_1^N\,{\bar C}_n\, e^{-i {\bar k}_n}\, {\bar D}_n^\dagger\nonumber
\end{eqnarray}
These transfer matrices allow us to build the coupling matrix $V$, the self-energies $\Sigma$, and the
surface Green's functions $G_{S,0}^i$:
\begin{eqnarray}
 V&=&K_{-1}\,({\cal T}^{-1}-{\cal {\bar T}})\nonumber\\
 \Sigma&=&K_1\,T\nonumber\\
 G_{S,0}^i&=&-(K_0+\Sigma)^{-1}
\end{eqnarray}

The procedure described above to compute the lead Green's functions can fail, because of the singular
behavior of the Hamiltonians matrices $K_1$, which lead to numerical inaccuracies in the solution of Eq. (\ref{GEP}), and is usually
manifested in the program producing a number of positive and negative channels different from $N$. Notice that if 
the number of positive and negative channels is different from $N$, then the dual vectors cannot be found by inverting 
$Q$ and $\bar Q$. There exist several schemes to regularize $K_1$, based on decimating out the offending degrees of freedom.
These procedures are explained in detail in Refs. [\onlinecite{smeagol,Rungger08}]. GOLLUM uses a suitable adaptation of
these methods, which is described in the appendix.

\begin{figure}
\includegraphics[width=0.9\columnwidth]{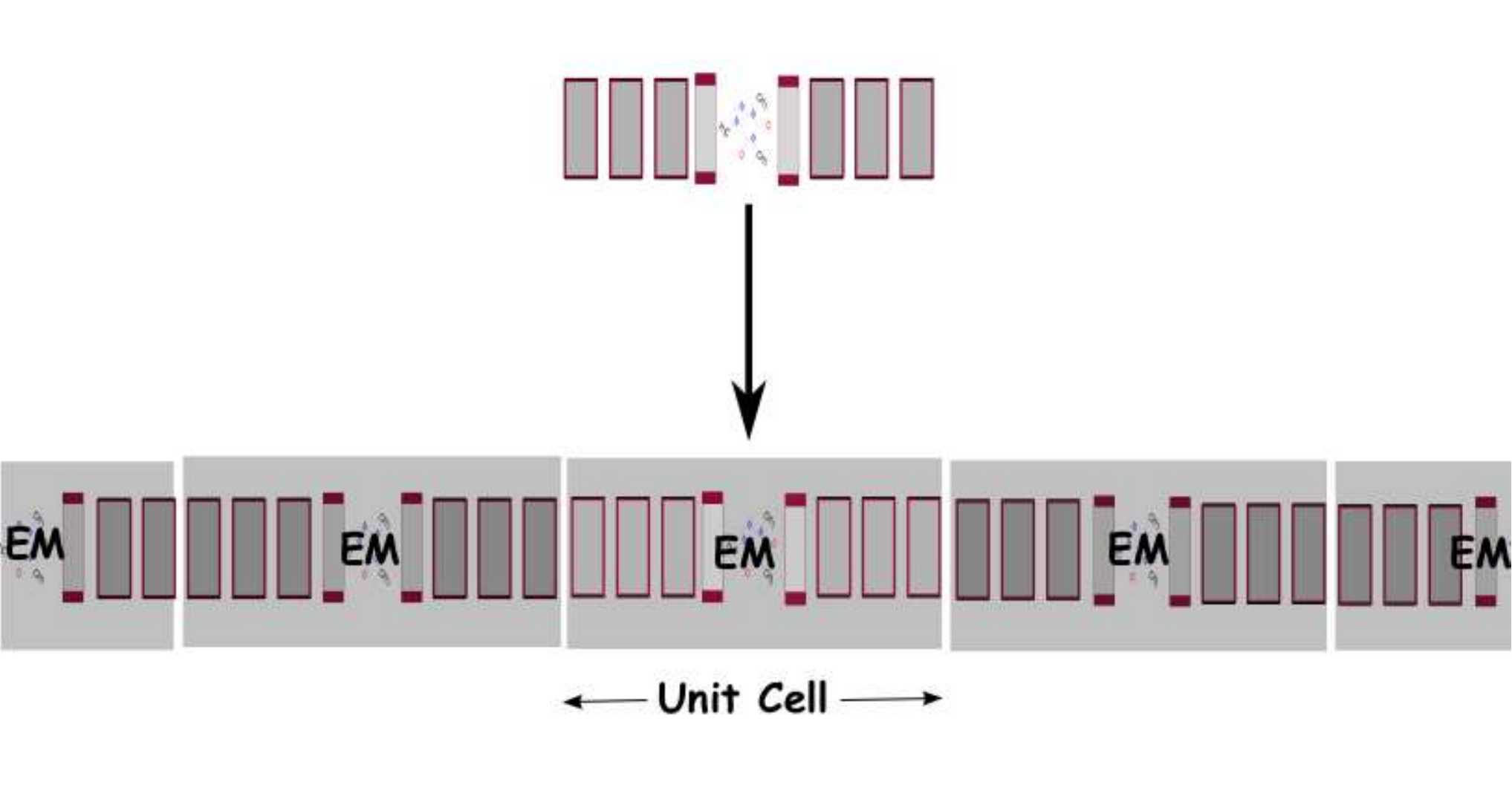}
\caption{\label{fig:em1}(Color online) Infinite system
whose unit cell is the EM region. This is linked to neighboring EM cells by
periodic boundary conditions.}
\end{figure}

\begin{figure}
\includegraphics[width=0.9\columnwidth]{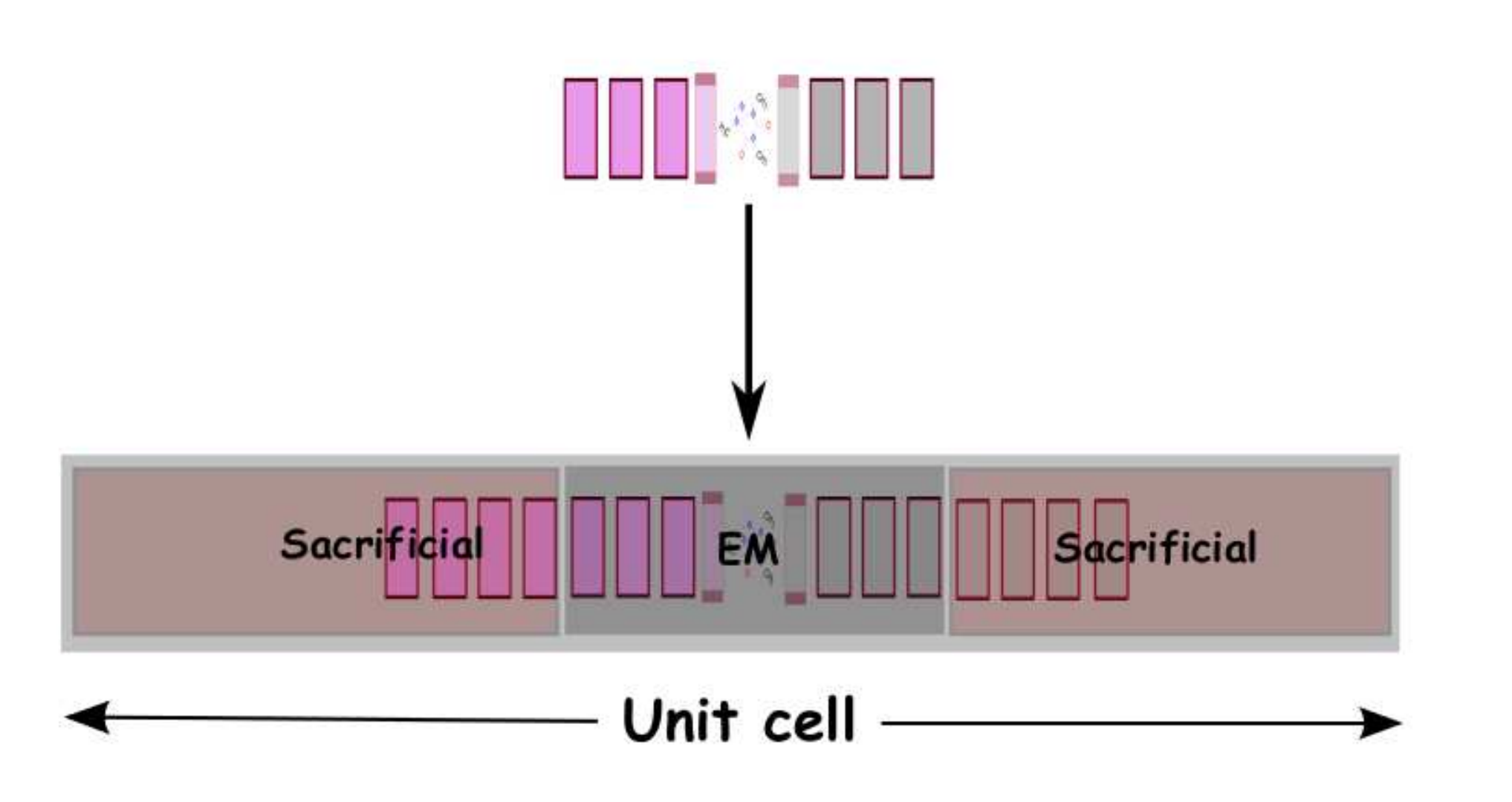}
\caption{\label{fig:em2}(Color online) Super cell containing the extended scattering region EM.
The EM region is surrounded by buffer vacuum regions to its left and right.}
\end{figure}

\subsubsection{Generating the Hamiltonian of the extended scattering region ${\cal K}^{EM}$}
As noted above, ${\cal K}^{EM}$ can be provided as a model Hamiltonian, or generated
by a DFT or other material-specific program. One of the strengths of GOLLUM is an ability to treat interfaces with high accuracy. In a tight-binding 
description, tight-binding parameters of a particular material are often chosen by fitting to a band structure. However 
this does not solve the problem of choosing parameters to describe the interface between two materials. Often this 
problem is finessed by choosing interface parameters to be a combination of pure-material parameters such as an arithmetic 
or geometric mean, but there is no fundamental justification for such approximations. 

Therefore we describe here methods to generate ${\cal K}^{EM}$ using a DFT program, where the inclusion of  branches as 
part of the extended scatterer occurs naturally. Fig. (\ref{fig:em1}) shows an example of a junction where the electrodes are identical. The system is composed of super cells formed from a central scatterer and PLs. There are periodic 
boundary conditions in the longitudinal direction, such that the TPL of one branch of a super-cell is linked smoothly to the TPL of a neighboring super-cell.
Running a DFT program for such a super-cell then automatically generates  ${\cal K}^{EM}$. Provided the super cells contain sufficient PLs, the Hamiltonians $K_0^{TPL}$ and $K_1^{TPL}$
associated with the TPLs will be almost identical to those generated from a calculation involving an infinite periodic 
lead. ie. if the Hamiltonians $K_0$ and $K_1$  associated with the PLs are generated from a calculation involving an infinite periodic lead,  then  provided the super cells contain sufficient PLs,
these will be almost identical to  $K_0^{TPL}$ and $K_1^{TPL}$ respectively.
In this case  then there will be minimal scattering caused by the junction between the TPL and the lead.
Clearly there is a trade-off between accuracy and CPU time, because inserting more PLs increases the size and cost of the calculation. In practice, the number of PLs retained in such a super-cell is increased in stages until the results do not change significantly as the number of PLs is increased further.

There exist situations where the electrodes are dissimilar, either chemically, or because of their different crystalline 
structure, or because their magnetic moments are not aligned. In these cases, there cannot be a 
smooth matching between TPLs of neighboring super cells in Fig. 4.
To address this situation, we use a setup similar to that displayed in Fig. (\ref{fig:em2}), where additional PLs
are appended to the branches in the EM region. These additional PLs are terminated by artificial 
surfaces and surrounded by vacuum. The TPLs are then chosen to be one of the PLs near the middle of each branch 
and should be surrounded by enough PLs both towards its artificial surface and towards the central scattering 
region. Then the PLs placed between the TPL and the artificial surface are discarded. These sacrificial PLs 
ensure that the chosen TPL is unaffected by the presence of the artificial vacuum boundary. 
Clearly, calculations of this sort are more expensive in numerical terms than those performed with 
super cells generated as in Fig. (\ref{fig:em1}), because they contain many more atoms.

\subsubsection{Hamiltonian assembly}
In an ab initio calculation of the transport properties of a junction, the DFT program  produces
the Hamiltonians and Fermi energy of the EM region ${\cal H}^\mathrm{EM}$, ${\cal S}^\mathrm{EM}$ and $E_F^\mathrm
{EM}$, and of each lead $H_{0,\pm1}^i$, $S_{0,\pm1}^i$ and $E_F^i$ in separate runs. 
Notice that the Hartree potential is defined up to a constant, which is usually different for the EM and for each lead. 
This usually means that the energy origin of the
 EM and of the corresponding lead PLs, as well as their Fermi energies do not agree with each other, 
so Eq. (2) must be rewritten as follows:
\begin{eqnarray}
K_{0,\pm1}^i=H_{0,\pm1}^i-(E_F^i+E)\,S_{0,\pm1}
\end{eqnarray}
where we have referred the energy of each lead to its own Fermi energy. To fix the Hamiltonian mismatch we define a 
realignment variable for each lead as follows:
\begin{eqnarray}
 \Delta^i=H_0^i(\mu,\mu)-H_0^\mathrm{EM}(\mu,\mu)
\end{eqnarray}
where $\mu$ indicates a relevant orbital or group of orbitals. Then, the Hamiltonian of each lead is realigned with
that of the EM
\begin{eqnarray}
 \bar K_{0,\pm1}^i&=&H_{0,\pm1}^i-(E_F^\mathrm{EM}+E+\Delta^i)\,S_{0,\pm1}\nonumber\\&=&H_{0,\pm1}^i-(\bar E_F^i+E)\,S_{0,\pm1}
\end{eqnarray}
It turns out that the renormalized $\bar E_F^i$ and bare $E_F^i$ Fermi energies of each lead 
do not match perfectly with each other if the number of PLs in the EM region is not sufficiently large. This is the case
when for efficiency reasons, it is desirable to artificially minimize the size of the EM. Sometimes it is advisable to choose
the Fermi energy of one of the leads $E_F^I$ as the reference energy. In this case, a second overall shift can be performed using 
either
$\Delta^I$ or  the quantity $\delta=E_F^I-E_F^\mathrm{EM}$.

\subsubsection{Scattering matrix and transmission in multi-terminal devices}
We note that the most general scattering state in a given lead $i$ at a given energy $E$ can be written as a linear combination
of open and closed channels as follows
\begin{equation}
 |\Phi^i(E)\rangle=\sum_{k_i}\,o_{k_i}\,\frac{|\Psi^i(k_i)\rangle}{\sqrt{v(k_i)}}+
 \sum_{\bar{q}_i}\,o_{\bar q_i}\,\frac{|\Psi^i({\bar q_i})\rangle}{\sqrt{v({\bar q_i})}} +|\chi^i\rangle
\end{equation}
where $k_i$ and ${\bar q_i}$ denote here open positive and negative channels and $|\Psi^i(k_i)\rangle$ 
and $|\Psi^i({\bar q_i})$  are their normalized kets. Here, the contribution of all the
closed channels in lead $i$ is described by the ket $|\chi^i\rangle$. 
Consequently, the number of electrons per unit time flowing between two adjacent PLs within the lead is
\begin{equation}
 j^i(E)=\,\sum_{k_i\in i} |o_{k_i}|^2 - \,\sum_{\bar q_i\in i} |o_{\bar q_i}|^2\label{ok}
\end{equation}
We pick in this section the convention that positive direction in the lead means flow towards the EM region 
and vice versa. So positive (negative) open channels are also called incoming (outgoing) channels of lead $i$. 
With this notation. the wave-function coefficients of the incoming open channels of a given lead are 
determined by the properties of the reservoir connected to the lead.

The wave function coefficients of the open outgoing channels of lead $i$ are obtained from the amplitudes of all 
incoming channels by
\begin{equation}
 o^i_{\bar q_i}=\sum_{j,k}s^{ij}_{{\bar q_i}k_j}\,\,o_{k_j}^j
\end{equation}
where ${\bar q_i}$ ($k_j$) is an outgoing (incoming) dimensionless wave-vector of lead $i$£ ($j$).
It is therefore convenient to assemble the wave-functions of the $M^i$ outgoing and $M^i$ incoming open channels of
a given lead $i$ in the column vectors $\bar {\cal O}^i$, ${\cal O}^i$, and
all the scattering matrix elements connecting leads $i$ and $j$ into the matrix block ${\cal S}^{ij}$. Notice that
the dimensions of ${\cal S}^{ij}$ are $M^i\times M^j$. Then the above equation can be written more compactly as
\begin{equation}
\left(\begin{array}{c}\bar {\cal O}^1\\\bar {\cal O}^2\\...\\\bar {\cal O}^P\end{array}\right)\,\,=\,\,
\left(\begin{array}{cccc}
          S^{11}&S^{12}&...&S^{1P}\\
          S^{21}&S^{22}&...&S^{2P}\\
          .     &.     &.  &      \\
          S^{P1}&S^{P2}&...&S^{PP}
         \end{array}\right)\,\,
\left(\begin{array}{c}{\cal O}^1\\{\cal O}^2\\...\\{\cal O}^P\end{array}\right)
\end{equation}

By normalizing the Bloch eigenvectors $C(k)$, $C({\bar q})$ and their duals to unit flux,
\begin{eqnarray}
 {\cal C}^i(k_i)&=&C^i(k_i)/\sqrt{v_{k_i}},\,\,\,\,\,{\cal D}^i(k_i)=\sqrt{v_{k_i}}\,\,D^i(k_i)\nonumber\\
 {\cal C}^i(\bar q_i)&=&C^i(\bar q_i)/\sqrt{v_{\bar q_i}},\,\,\,\,\,{\cal D}^i(\bar q_i)=\sqrt{v_{\bar q_i}}\,\,D^i(\bar q_i)
 \end{eqnarray}
the matrix elements of the scattering matrix block connecting leads $i$ and $j$ can be written as.
\begin{equation}
s_{{\bar q_i}k_j}^{ij}=\bar {\cal D}^i (\bar q_i) \,\,\left(\,G_S^{ij}\,V^i-I\,\delta_{ij}\,\right)\,\,{\cal C}^j(k_j)\,
\end{equation}
Here $G_S^{ji}$ is the off-diagonal block of the surface Green's function defined in Eqs. (11) and (12), that connects
leads $i$ and $j$ and $V^i$ is the matrix defined in Eq. (20).

With the above notation, if the incoming channel $k_i$ of lead $i$ is occupied with probability $f_{k_i}(E)$ 
(ie if in Eq. (\ref{ok}), $o_{k_i} = 1$ with probability $f_{k_i}(E)$) then the number of electrons per unit time, 
entering the scattering region from reservoir $i$ along channel $k_i$ with energy between $E$ and $E+dE$ is
\begin{equation}
dI^{\rm in}_{k_i}(E)=(dE/h)f_{k_i}(E)
\end{equation}
 and the number per unit time, per unit energy leaving the scatterer and entering reservoir $i$ along 
 channel $\bar q_i$ with energy between $E$ and $E+dE$ is
\begin{equation}
dI^{\rm out}_{\bar q_i}(E)=(dE/h)\sum_{\bar q_i, j, k_j}\vert s^{ij}_{{\bar q_i}k_j}\vert^2 f_{k_j}(E)
\end{equation}

In many cases, the incoming and outgoing channels of each lead $i$ can be grouped into channels possessing particular 
attributes (ie quantum numbers)  labeled $\alpha_i, \,\, \beta_i \,.........$ etc. This occurs when all incoming 
channels of a particular type $\alpha_i$ in lead $i$ possess the same occupation probability $f^i_\alpha(E)$. For 
example, all quasi-particles of type $\alpha_i$ in reservoir $i$ may possess a common chemical potential 
$\mu_{\alpha_i}$ and $f^i_{\alpha_i}(E)$ may take the form $f^i_{\alpha_i}(E)=f(E-\mu_{\alpha_i})$, where 
$f(E)$ is the Fermi function.
In this case, if the incoming and outgoing channels of type $\alpha_i$ belonging to lead $i$ possess wave-vectors 
$k_{{\alpha_i}}$, $\bar q _{{\alpha_i}}$, then the number of quasi-particles per unit time of type $\alpha_i$
leaving reservoir $i$  with energy between $E$ and $E+dE$ is

\begin{equation}
dI^{i}_{\alpha_i}(E)=(dE/h)\sum_{ j,\beta_j}P^{i,j}_{{\alpha_i},{\beta_j}} f^j_{\beta_j}(E)  \label{xxx}
\end{equation}
where
\begin{equation}
P^{i,j}_{\alpha_i,\beta_j} =  M^i_{\alpha_i}(E) \delta_{i,j}\delta_{\alpha_i,\beta_j} - 
\sum_{\bar q_{\alpha_i}, k_{\beta_j}}\vert s^{ij}_{  \bar q_{\alpha_i}k_{\beta_j}  }\vert^2
\end{equation}
and $M^i_{\alpha}(E) $  is the number of open incoming channels of type $\alpha$, energy $E$ in lead $i$. 
Note that in the above summation, $ \bar q_{\alpha_i}$ runs over all outgoing wave-vectors of energy $E$ 
and type $\alpha_i$ of lead $i$ and $k_{\beta_j}$ runs over all incoming wave-vectors of energy $E$ and 
type $\beta_j$ in lead $j$.

If $i$ and $j$ are different leads, then $s_{{\bar q_i}k_j}$ is often called the transmission amplitude and 
denoted $t_{{\bar q_i}k_j}$, while
if they are the same lead, then $s_{{\bar q_i}k_i}$ is called the reflection amplitude $r_{{\bar q_i}k_i}$. 
Similarly, for $i\ne j$, it is common to define the transmission coefficient $ T^{i,j}_{\alpha_i,\beta_j}$
as
\begin{equation}
 T^{i,j}_{\alpha_i,\beta_j}= \sum_{\bar q_{\alpha_i}, k_{\beta_j}}\vert s^{ij}_{  \bar q_{\alpha_i}k_{\beta_j}  }\vert^2\label{T}
\end{equation}
and for $i=j$, we define the reflection coefficient as
\begin{equation}
 R^{i,i}_{\alpha_i,\beta_i}= \sum_{\bar q_{\alpha_i}, k_{\beta_i}}\vert s^{ij}_{  \bar q_{\alpha_i}k_{\beta_j}  }\vert^2
\end{equation}
so that
\begin{eqnarray}
dI^{i}_{\alpha_i}(E)&=&\frac{dE}{h}\{ \sum_{\beta_i}[ M^i_{\alpha_i}(E)\delta_{\alpha_i\beta_i}-  
R^{i,i}_{\alpha_i,\beta_i}]f^i_{\beta_i}(E) \nonumber \\
&-& \sum_{ {j\ne i},{\beta_j}}T^{i,j}_{\alpha_i,\beta_j} f^j_{\beta_j}(E)\}\label{x1}
\end{eqnarray}

Note that unitarity of the scattering matrix requires
\begin{equation}
\sum_{i,\alpha_i,\bar q_{\alpha_i}}\vert s^{ij}_{  \bar q_{\alpha_i}k_{\beta_j}  }\vert^2 =
\sum_{j,\beta_j k_{\beta_j}}\vert s^{ij}_{  \bar q_{\alpha_i}k_{\beta_j}  }\vert^2 =1
\end{equation}

Hence the sum of the elements of each row and column of the matrix $P$ is zero:
\begin{equation}
\sum_{ j,\beta_j}P^{i,j}_{\alpha_i,\beta_j}=\sum_{ i,\alpha_i}P^{i,j}_{\alpha_i,\beta_j}=0\label{x2}
\end{equation}
or equivalently,
\begin{equation}
\sum_{\beta_i} R^{i,i}_{\alpha_i,\beta_i} + \sum_{{j\ne i},\beta_j} T^{i,j}_{\alpha_i,\beta_j} = M^i_{\alpha_i}\label{x3}
\end{equation}
and
\begin{equation}
\sum_{\alpha_i} R^{j,j}_{\alpha_i,\beta_j} + \sum_{{i\ne j},\alpha_i} T^{i,j}_{\alpha_i,\beta_j} = M^j_{\beta_j}
\end{equation}

From Eqs. (\ref{x1}) and (\ref{x3}), if $f^j_{\beta_j}(E)$ is independent of $j$ and $\beta_j$ 
then $dI^i_{\alpha_i}(E)=0$ for all $i$ and $\alpha_i$, as expected. For this reason, in the above 
equations, $f^j_{\beta_j}(E)$ can be replaced by $\bar f^j_{\beta_j}(E)=f^j_{\beta_j}(E)-f(E)$, 
where $f(E)$ is an arbitrary function of energy, which in practice is usually chosen to be a Fermi 
function, evaluated at a convenient reference temperature and chemical potential.

When comparing theory with experiment, we are usually interested in computing the flux of some quantity $Q$ 
from a particular reservoir. From Eq. (\ref{xxx}), if the amount of $Q$ carried by quasi-particles of 
type $\alpha_i$ is $Q_{\alpha_i}(E)$, then the flux of $Q$ from reservoir $i$ is
\begin{equation}
I^{i}_{Q}=\int(dE/h) \sum_{\alpha_i, j,\beta_j}Q_{\alpha_i}(E)P^{i,j}_{\alpha_i,\beta_j} \bar f^j_{\beta_j}(E)  \label{xxxx}
\end{equation}
In the simplest case of a normal conductor, choosing $Q_{\alpha_i}=-e$, independent of $\alpha_i$, the above equation yields 
the electrical current from lead $i$. Within GOLLUM $\alpha_i$ may represent spin and in the presence of superconductivity 
may represent hole ($\alpha_i= h$) or particle ($\alpha_i= p$) degrees of freedom. In the latter case, the charge $Q_p$ 
carried by particles is -e, whereas the charge $Q_h$ carried by holes is +e.

\subsection{Incorporation of non-equilibrium effects in the transmission coefficients}
GOLLUM starts from a mean-field Hamiltonian provided either by the user or by an outside material-specific DFT code. It then 
computes the scattering matrix and its related transport properties. When finite voltages are applied to the electrodes, they 
change
the distribution of incoming and outgoing electrons and therefore the underlying Hamiltonian. For example, a finite voltage in a two-terminal 
device may introduce an electrostatic potential, which should be included in the Hamiltonian. A key feature of many NEGF 
codes including SMEAGOL is that such effects can be treated self-consistently, albeit at the cost of a greatly increased 
computing overhead. 
To avoid this overhead, GOLLUM assumes that the user is able to provide a modified Hamiltonian at finite voltages.

Based on our experience on the development and usage of NEGF programs and as demonstrated in in section III.J below, we have found that in many cases, the following intuitive 
modification of the initial 
zero-voltage Hamiltonian yields reasonable-accurate voltage-dependent transmission 
coefficients $T_{ij}(E,V)$ connecting leads $i$ and $j$. The scheme enables the simulation of non-trivial $I-V$ curves which 
compare favorably to those obtained 
using NEGF techniques and enables the modeling of generic  non-equilibrium transport phenomena such as negative differential
resistance (NDR) and current rectification in close agreement with NEGF codes\cite{Garcia12}. 

Consider the case where each lead has a different voltage $V^i$. Then the finite-voltage Hamiltonian takes the form
\begin{widetext}
\begin{equation}
{\cal K}
=
\left(\begin{array}{ccccc}
{\cal K}^\mathrm{4}-e V^4 \,{\cal S}^4   & 0                    & 0                    & 0                    & {\cal K}^\mathrm{4M}-e V^4 \,{\cal S}^\mathrm{4M}\\
0                     & {\cal K}^\mathrm{3}-e V^3\, {\cal S}^3  & 0                    & 0                    & {\cal K}^\mathrm{3M}-e V^3 \,{\cal S}^\mathrm{3M} \\
0                     & 0                    & {\cal K}^\mathrm{2}-e V^2\, {\cal S}^2  & 0                    & {\cal K}^\mathrm{2M}-e V^2 \,{\cal S}^\mathrm{2M}\\
0                     & 0                    & 0                    & {\cal K}^\mathrm{1} -e V^1\, {\cal S}^1  & {\cal K}^\mathrm{1M}-e V^1 \,{\cal S}^\mathrm{1M}\\
{\cal K}^\mathrm{M4}-e V^4 \,{\cal S}^\mathrm{M4}  & {\cal K}^\mathrm{M3}-e V^3 \,{\cal S}^\mathrm{M3} & {\cal K}^\mathrm{M2}-e V^2\, {\cal S}^\mathrm{M2} &
{\cal K}^\mathrm{M1}-e V^1\, {\cal S}^\mathrm{M1} & {\cal K}^\mathrm{EM}\\
\end{array}\right)\;
\end{equation}
\end{widetext}

We find that ${\cal K}^\mathrm{EM}$ needs only be computed at zero voltage in most cases; the effect of a finite bias can be 
accounted for by a suitable re-alignment of the energies of the orbitals in the EM region with the shifted energy levels of the electrodes.
Mathematically, we apply a simple shift to the Hamiltonian matrix elements at each orbital $n$ in the EM region 
\begin{equation}
{\cal K}^\mathrm{EM} \longrightarrow {\cal K}^\mathrm{EM}(V)={\cal K}^\mathrm{EM}-eV_n\,{\cal S}^\mathrm{EM}\label{voltage}
\end{equation}
where these local shifts $V_n$ depend on  the 
junction electrostatics, which in many cases are known. 
For example, in the case of a highly-transparent  junction, the shifts can be modeled by a linear voltage ramp connecting the 
matrix elements of the orbitals at the  TPLs of the EM region. In contrast, when the central scattering region Hamiltonian 
$K_\mathrm{M}^0$ is connected to the PL  Hamiltonians $K''_0$ in each branch of the EM region by weaker links $K''_1$, 
the voltage drop and  therefore the resistance is  concentrated at these spots. In this case, we take $V_n= V^i$ for all 
orbitals 
in branch $i$, starting at the TPL and up to the linker atoms,  and $V_i=0$ for all the orbitals inside the M region itself.
This scheme performs specially well for systems where the states around the Fermi level (HOMO or LUMO) are localized at or 
close to the contact atoms. It enables us to mimic accurately junctions displaying non-trivial negative differential resistance, 
as well
as rectification effects for asymmetric molecules\cite{Garcia12}. 

\subsection{Virtual leads versus physical leads.}
What is the difference between a lead and a channel? From a mathematical viewpoint, channels connect an extended scattering region to a 
reservoir and the role of lead $i$ is simply to label those channels $k_i, \bar q_i$, which connect to a particular 
reservoir $i$. Conceptually, this means that from the point of view of solving a scattering problem at energy $E$, a single 
lead with $N(E)$ incoming channels can be regarded as $N(E)$ virtual leads, each with a single channel. GOLLUM takes advantage 
of this equivalence by regarding the above groups of channels with wave-vectors $k_{\alpha_i}, \bar q_{\alpha_i}$ as virtual 
leads and treating them on the same footing as physical leads. From this viewpoint, Eq. (\ref{xxx}) and (\ref{x1}) yield the  
number of quasi-particles per unit time "from virtual lead $\alpha_i$ " entering the scattering region  with energy between 
$E$ and $E+dE$.

This viewpoint is particularly useful when the Hamiltonians $H_0^i$, $H_1^i$ describing the PLs of the physical lead $i$ are 
block diagonal with respect to the quantum numbers associated with $k_{\alpha_i}, \bar q_{\alpha_i}$. For example, this occurs 
when the leads possess a uniform magnetization, in which case the lead Hamiltonian is block diagonal with respect to the local 
magnetization axis of the lead and $\alpha$ represents the spin degree of freedom $\sigma$.  This occurs also when the leads 
are normal metals, but the scattering region contains one or more superconductors, in which case the lead Hamiltonian is block 
diagonal with respect to particle and hole degrees of freedom and $\alpha$ represents either particles $p$ or holes $h$. 
More generally, in the presence of both magnetism and superconductivity, or combinations of singlet and triplet 
superconductivity, $\alpha$ would represent combinations of spin and particles and holes degrees of freedom.

In all of these cases,  $H_0^i$, $H_1^i$ are block diagonal and it is convenient to identify virtual leads $\alpha_i$ with 
each block, because GOLLUM will compute the channels $k_{\alpha_i}, \bar q_{\alpha_i}$ belonging to each block in separate calculations 
and therefore guarantees that all such channels can be separately identified. This is advantageous, because if all channels of $H_0^i$, $H_1^i$ 
were calculated simultaneously, then in the case of degeneracies, arbitrary superpositions of channels with different quantum 
numbers could result and therefore it would be necessary to implement a separate unitary transformation to sort channels into the 
chosen quantum numbers. By treating each block as a virtual lead, this problem is avoided.
Examples of this approach are presented below, when describing the scattering properties of magnetic or 
normal-superconducting-normal systems.

\subsection{Charge, spin and and thermal currents}
In the presence of non-collinear magnetic moments, provided the lead Hamiltonians are block
diagonal in spin indices (in general relative to lead-dependent magnetization axes) choosing $\alpha_i =\sigma_i$ and $Q_{\alpha_i}=-e$ in 
Eq. (\ref{xxxx}) yields for the total electrical current
\begin{equation}
I^{i}_{e}=-e\int(dE/h) \sum_{\sigma_i, j,\sigma_j}P^{i,j}_{\sigma_i,\sigma_j} \bar f^j_{\beta_j}(E)  \label{xxxx1}
\end{equation}
Note that in general it is necessary to retain the subscripts $i,j$ associated with $\sigma_i$ or $\sigma_j$, because the 
leads may possess different magnetic axes.

Similarly the thermal energy from reservoir $i$  per unit time is
\begin{equation}
I^{i}_{q}=\int(dE/h) \sum_{\sigma_i, j,\sigma_j}(E-\mu_i)P^{i,j}_{\sigma_i,\sigma_j} \bar f^j_{\beta_j}(E)  \label{xxxx2}
\end{equation}

For the special case of a normal  multi-terminal junction having collinear magnetic moments, $\alpha_i = \sigma$ for 
all $i$ and since there is no spin-flip scattering, $P^{i,j}_{\sigma,\sigma'} = P^{i,j}_{\sigma,\sigma}\delta_{\sigma,\sigma'}$. 
In this case, the total Hamiltonian of the whole system is block diagonal in spin indices and the scattering matrix can be 
obtained from separate calculations for each spin.
We assume that initially the junction is in thermodynamic equilibrium, so that all reservoirs possess the same chemical 
potential $\mu_0$. Subsequently. we apply to each reservoir $i$ a different voltage $V_i$, so that its chemical potential is 
$\mu_i=\mu_0-e\,V_i$. Then from equation (\ref{xxx}),  the charge per unit time 
per spin entering the scatterer from each lead can be written as
\begin{equation}
I^{i}_e=(-e)\int(dE/h) \sum_{\sigma, j}P^{i,j}_{\sigma,\sigma} \bar f^j_{\sigma}(E)  \label{xx3}
\end{equation}
and the thermal energy per spin per unit time is
\begin{equation}
I^{i}_q=\int(dE/h) \sum_{\sigma, j}(E-\mu_i)P^{i,j}_{\sigma,\sigma} \bar f^j_{\sigma}(E)  \label{x4}
\end{equation}
where  $e=\vert e \vert$ and $\bar f^i_\sigma(E)=f(E-\mu_i)-f(E-\mu)$ is the deviation in Fermi distribution 
of lead $i$ from the reference distribution $f(E-\mu)$.

In the limit of small potential differences or small differences in reservoir temperatures, the deviations in 
the distributions from the reference distribution $ \bar f^j_{\sigma}(E)$ can be approximated by differentials and therefore to 
evaluate currents, in the presence of collinear magnetism, GOLLUM provides the following spin-dependent integrals
\begin{equation}
 L^n_{ij,\sigma}(T)=\int_{-\infty}^{\infty}\,dE\,(E-\mu_0)^n\,T^{ij}_{\sigma,\sigma}(E,V=0)\,\left(-\frac{\partial f}{\partial E}\right)
\end{equation}

In the presence of two leads labeled $i=1,2$, the spin-dependent low-voltage electrical conductance $G(T)$, the thermopower 
(Seebeck coefficient) $S^e(T)$,
the Peltier coefficient $\Pi(T)$ and the thermal conductance
$\kappa(T)$ can be obtained as
\begin{eqnarray}
 G_{\sigma}(T)&=&(e^2/h)\,L_{12,\sigma}^0\nonumber\\
 S^e_{\sigma}(T)&=&-\frac{1}{e\,T}\,\frac{L^1_{12,\sigma}}{L^0_{12,\sigma}}\nonumber\\
 \Pi_{\sigma}(T)&=&T\,S^e_{\sigma}(T)\nonumber\\
 \kappa_{\sigma}(T)&=&\frac{1}{h\,T}\left(L_{12,\sigma}^2-\frac{(L_{12,\sigma}^1)^2}{L_{12,\sigma}^0}\right)
\end{eqnarray}
so that the equivalent spin-summed magnitudes are 
\begin{eqnarray}
 G(T)&=&\sum_{\sigma}G_{\sigma}(T)\nonumber\\
 S^e(T)&=&\sum_{\sigma} S^e_{\sigma}(T)\nonumber\\
 \Pi(T)&=&\sum_{\sigma} \Pi_{\sigma}(T)\nonumber\\
 \kappa(T)&=&\sum_{\sigma} \kappa_{\sigma}(T)
\end{eqnarray}

Note that the thermal conductance is guaranteed to be positive, because the expectation value of the square of 
a variable is greater than or equal to the square of the expectation value. For a two-terminal system, the above expressions 
allow us to obtain the electronic contribution to the thermoelectric  figure of merit\cite{zt}:
\begin{equation}
ZT=\frac{1}{\frac{L^0_{12}\,L^2_{12}}{(L^1_{12})^2}
-1}
\end{equation}

\subsection{Additional functionalities}
\subsubsection{Spectral adjustment}
A phenomenological scheme that improves the agreement between theoretical simulations and
experiments in, for example, single-molecule electronics consists of shifting the occupied and unoccupied levels of the M region downwards and upwards respectively
to increase the energy gap\cite{Neaton06,Quek09,Mowbray08,Cehovin08,Gar11} of the M region. 
The procedure is conveniently called spectral adjustment in nanoscale transport (SAINT).
At the request of a user, GOLLUM modifies the Hamiltonian operator of the 
M region as follows:
\begin{equation}
 \hat{K}_\mathrm{M}=\hat{K}_\mathrm{M}^0+\Delta_\mathrm{o}\,\sum_{no}\,|
 \Psi_{no}\rangle\langle\Psi_{no}|+\Delta_\mathrm{u}\,\sum_{nu}\,
 |\Psi_{nu}\rangle\langle\Psi_{nu}|
\end{equation}
where $\Delta_\mathrm{o,u}$ are energy shifts and ($no$, $nu$) denote the occupied and unoccupied states, respectively. 
By using the definition of the density matrix operator of the M region
\begin{eqnarray}
 \hat{\rho}_\mathrm{M}=\sum_{no}\,|\Psi_{no}\rangle\langle\Psi_{no}|\nonumber\\
  \hat{I}=\sum_{n}\,|\Psi_{n}\rangle\langle\Psi_{n}|
 \end{eqnarray}
The above Hamiltonian can be rewritten as
\begin{equation}
 \hat{K}_\mathrm{M}=\hat{K}_\mathrm{M}^0+(\Delta_\mathrm{o}-\Delta_\mathrm{u})\,
 \hat{\rho}_\mathrm{M}+\Delta_\mathrm{u}\,\hat{I}
\end{equation}
The equation can also be written in matrix form as
\begin{equation}
K_\mathrm{M}=K_\mathrm{M}^0+(\Delta_\mathrm{o}-\Delta_\mathrm{u})\,S_\mathrm{M}\,
\rho_\mathrm{M}\,S_\mathrm{M}+\Delta_\mathrm{u}\,S_\mathrm{M}
\end{equation}
To find the density matrix, we first solve the generalized eigenvalue problem:
\begin{eqnarray}
 H_\mathrm{M}^0\,\vec{c}_n&=&\epsilon_n^0\,S_\mathrm{M}\,\vec{c}_n\\
 \vec{c'}_n\,&=&\frac{\vec{c}_n}{\sqrt{\vec{c}_n\,S_\mathrm{M}\,\vec{c}_n}}\\
 R&=&\left(\vec{c'}_1\,,...,\vec{c'}_n\,\right)\\
 R^\dagger\,K_\mathrm{M}^0\,R&=& \varepsilon^0_\mathrm{M}-E\,I_P
\end{eqnarray}
where $I_P$ is the $P \times P$ identity matrix, and we have arranged the eigen-energies $\epsilon_n^0$ into
a diagonal matrix $\varepsilon_M^0$. Then
\begin{equation}
(\rho_\mathrm{M})_{\mu,\mu'}=\sum_{no}\,c_{n,\mu}'\,c_{n,\mu'}'^*
\end{equation}
In the simplest case, for a single-molecule 
junction, the  shifts $\Delta_\mathrm{o,u}^0$ are chosen to align the
highest occupied and lowest unoccupied molecular orbitals (ie the HOMO and LUMO) with (minus) the ionization potential (IP) and 
electron affinity (EA)
of the isolated molecule
\begin{eqnarray}
 \Delta_\mathrm{o}^0&=&\epsilon_\mathrm{HOMO}+IP\nonumber\\
 \Delta_\mathrm{u}^0&=&-(\epsilon_\mathrm{LUMO}+EA)
\end{eqnarray}
However the Coulomb interactions in the isolated molecule are screened if the molecule is placed in close proximity to the 
metallic electrodes. Currently, GOLLUM takes this effect by using a simple image charge model\cite{Neaton06}, where the molecule 
is replaced by a point charge located at the middle point of the molecule and where the image planes are placed {1 \AA} 
above the electrodes' surfaces. Then the shifts are corrected by screening effects as follows:
\begin{eqnarray}
 \Delta_\mathrm{o}&=&\Delta_\mathrm{o}^0+\frac{e^2}{8\,\pi\epsilon_0}\,\frac{\ln 2}{a}\nonumber\\
  \Delta_\mathrm{u}&=&\Delta_\mathrm{u}^0-\frac{e^2}{8\,\pi\,\epsilon_0}\,\frac{\ln 2}{a}
\end{eqnarray}
where $a$ is the distance between the image plane and the point image charge.

\subsubsection{Coulomb blockade and Kondo physics}
Many nanoscale-scale junctions are expected to show Coulomb blockade behavior, and in specific situations also Kondo features \cite{quantum_dots}.
These features can be demonstrated by gating the junction, and should appear as Coulomb and Kondo diamond lines
in contour density plots of the low-voltage conductance as a function of bias and gate voltages. These strong correlation 
effects are completely missing in conventional DFT. 
Accurate parametrizations of the ground-state energy density functional of the single-channel Anderson model
exist that allow a correct description of those phenomena\cite{balda,kieron}. However, most nanojunctions
are better modeled in terms of a multi-channel Anderson model as we have chosen to do in GOLLUM. 
This model is described by the Hamiltonian
\begin{equation}
\label{anderson-Hamiltonian}
\hat{\cal H}_\mathrm{And}=\hat{\cal H}_\mathrm{Leads}+\hat{\cal H}_\mathrm{M}+\hat{\cal H}_\mathrm{Coupling}
\end{equation}
where the Hamiltonian at the central scattering region is given by
\begin{equation}
\hat{\cal H}_\mathrm{M}=\sum_m\,\epsilon_{m}\,\hat d_m^\dagger\hat d_m\,+\,U\,\sum_{m>l}\hat{n}_{m}\,\hat{n}_{l}
\label{eq:anderson-hamiltonian}
\end{equation}
Here the $m$-sum runs over the $M$ correlated degrees of freedom and includes the spin index, $\epsilon_m$ denote the 
on-site energies and $U$ is the electronic Coulomb repulsion, that is assumed to be the same for all degrees of freedom.

We map the central scattering region Hamiltonian $K^0_\mathrm{M}$ in Eq. (4) into $\hat{\cal H}_\mathrm{Mol}$ to extract the self-energies
$\Sigma_\mathrm{M}$ of the correlated degrees of freedom, in the spirit of Dynamical Mean Field Theory\cite{gabi}. 
Following Ref. (\onlinecite{nicolas}), our correlated degrees of freedom are a subset of the eigenstates of $K^0_\mathrm{M}$,
that we will call here molecular orbitals. This contrast with the approach followed in Ref. (\onlinecite{gabi}) where the 
correlated degrees of freedom where taken to be atomic orbitals of transition metal atoms. 
For a four-lead device, the details of our implementation are as follows. We take the EM Hamiltonian 
\begin{equation}
{\cal K}^\mathrm{EM,0}=
\left(\begin{array}{c|cccc}
K_\mathrm{M}^0 & K_\mathrm{M1}        & K_\mathrm{M2}       & K_\mathrm{M3}        & K_\mathrm{M4}       \\
\hline
K_\mathrm{1M}  & K_\mathrm{branch 1}  & 0                   & 0                    & 0                   \\    
K_\mathrm{2M}  & 0                    & K_\mathrm{branch 2} & 0                    & 0                   \\
K_\mathrm{3M}  & 0                    & 0                   & K_\mathrm{branch 3}  & 0                   \\
K_\mathrm{4M}  & 0                    & 0                   & 0                    & K_\mathrm{branch 4} \\
\end{array}\right)\;
\end{equation}
where $K_\mathrm{M}^0$ has dimensions $P\times P$ and therefore describes $P$ molecular orbitals. We first solve the 
generalized eigenvalue problem at the M region as in Eqs. (54-57) in the previous section to find the rotation matrix $R$
that diagonalizes $K_\mathrm{M}^0$:
\begin{eqnarray}
 R^\dagger\,K_\mathrm{M}^0\,R&=& \varepsilon^0_\mathrm{M}-E\,I_p
\end{eqnarray}
where $I_p$ is the $P \times P$ identity matrix. We then perform the direct product $\,U=R\otimes I$ to enlarge the
size of the matrix $R$ to the dimensions of the EM Hamiltonian matrix. We then rotate
the EM Hamiltonian and compute the Green's function of the EM 
region
\begin{eqnarray}
{\cal K}^\mathrm{EM\,'}&=&U\,{\cal K}^\mathrm{EM,0}\,U^\dagger\\ 
{\cal G}^\mathrm{EM\,'}&=&-\left({\cal K}^\mathrm{EM\,'}+ (K^\mathrm{coup})^\dagger\,G_{S,0}^{-1}\,K^\mathrm{coup}\right)^{-1}
\end{eqnarray}
We now choose which molecular orbitals $i=1,...M$ with associated on-site energies $\epsilon_i^0$ are the correlated degrees of 
freedom, 
and use the projectors $P_i$ to find their projected Green's functions and occupancies
\begin{eqnarray}
 g_i'&=&P_i\,{\cal G}^\mathrm{EM\,'}\,P_i\\
 N_i^\mathrm{DFT}&=&-\frac{1}{\pi}\,\int\,\mathrm{d}E\,\mathrm{Imag}[g_i']\,f(E_F)
\end{eqnarray}
where $E_F$ is the EM Fermi level. We then shift the on-site energy of the correlated orbitals by the conventional 
double counting term\cite{anisimov93,pethukov03}
\begin{eqnarray}
\label{eq:double_counting}
\epsilon_i^0&\longrightarrow&\epsilon_i=\epsilon_i^0-U\,(N_i^\mathrm{DFT}-1/2)\\
 {\cal K}^\mathrm{EM\,'}&\longrightarrow& {\cal K}^\mathrm{EM}
\end{eqnarray}

As a consequence, we have to recompute again the Green's functions
\begin{eqnarray}
 {\cal G}^\mathrm{EM}&=&-\left({\cal K}^\mathrm{EM}+ (K^\mathrm{coup})^\dagger\,G_{S,0}^{-1}\,K^\mathrm{coup}\right)^{-1}\\
 g_i&=&P_i\,{\cal G}^\mathrm{EM}\,P_i
\end{eqnarray}
This is cast in the form
\begin{equation}
 g_i=\frac{1}{E-\epsilon_i-\Delta_i^0}\,\longrightarrow\Delta_i^0=E-\epsilon_i-\frac{1}{g_i}
\end{equation}
which allows us to extract the hybridization function $\Delta_i^0$.

The initial ingredients in the solution of the multichannel Anderson model are the set $(\epsilon_i,\Delta_i^0, U)$. They 
allow us to extract the self-energies $\Sigma_i(\epsilon_i,\Delta_i^0, U)$ using a impurity solver. 
These are added again to the on-site energies 
\begin{equation}
 \epsilon_i\longrightarrow\epsilon_i+\Sigma_i
\end{equation}
leading to a new EM Hamiltonian ${\cal K}^\mathrm{EM}$ and associated Green's function ${\cal G}^\mathrm{EM}$.
From here we compute a new hybridization function
\begin{eqnarray}
\Delta_i=E-\varepsilon_i-\frac{1}{g_i}
\end{eqnarray}
with which new self-energies $\Sigma_i$ are determined. The cycle is repeated until self-consistency in $\Delta_i$ 
and $\Sigma_i$ is
achieved. The resulting ${\cal K}^\mathrm{EM}$ is inserted back into Eq. (12) and the surface Green-function matrix
$G_S$ is computed to extract the transport properties of the correlated junction.

We have decided to include in GOLLUM a finite-$U$ impurity solver. This way, we can subtract the double-counting terms and 
place the molecular orbitals at their correct bare energy positions by using Eq. (\ref{eq:double_counting}).
There exist a variety of finite-$U$ impurity solvers based on perturbation expansions on the Coulomb interaction $U$, on the 
hybridization function $\Delta_i^0$, on interpolative approaches, on Monte-Carlo algorithms (see Ref. (\onlinecite{gabi}) for a detailed
account of some of these solvers), or on Numerical Renormalization Group techniques\cite{Bulla} (NRG). 
NRG techniques have superior accuracy, but they bring high computational demands. Slave-boson-based expansions on $\Delta_i^0$
like the OCA\cite{Haule01,Haule10} are rather accurate and less expensive numerically.

The impurity solver used in GOLLUM is based on the Interpolative Perturbation Theory approach\cite{yeyati,gabirmp}, where 
the second-order in $U$ expression for the electron self-energy is interpolated to match the atomic self-energy, and adjusted to
satisfy consistency equations for the high-energy moments together with Luttinger's theorem. This approach is computationally very simple, but
has been proven to provide reasonable results for the multi-channel finite-$U$ Anderson model\cite{ferrer87,gabirmp}. Its main shortcoming is that it
overestimates the Kondo Temperature, as we discuss in Section (\ref{section_kondo}). Specifically, the impurity solver 
that we have implemented to handle the multi-channel Hamiltonian 
(\ref{eq:anderson-hamiltonian}) is described in Ref. (\onlinecite{yeyati}), although we have corrected errors in some 
of the equations in
that reference. We note that this impurity solver handles $M\ge 2$ spin-degenerate correlated degrees of freedom, so that 
$M$ must be an even number. In other words, these channels must come as Kramers pairs. We stress that other impurity solvers 
can be implemented straightforwardly, due to the modular
nature of GOLLUM.

\subsubsection{Inclusion of a Gauge field}
To compute transport properties in the presence of a magnetic field GOLLUM allows the user to introduce a Peierls substitution by changing the phase factors 
of the coupling elements\cite{peierls} between atomic orbitals.
For example in the case of a nearest-neighbor tight-binding Hamiltonian, the inter- site matrix element $H_{ij}$ between site i and site j is replaced with the modified element,
\begin{eqnarray}
H_{ij}^{B}=H_{ij}e^{-i\frac{e}{\hbar}\int_{\mathbf{r}_{j}}^{\mathbf{r}_{i}}\mathbf{A}(\mathbf{r})d\mathbf{r}},
\end{eqnarray}
where $\mathbf{r}_{i}$ and $\mathbf{r}_{j}$ are the positions of site i and j and $\mathbf{A}$ is the vector potential.
The gauge is chosen such that the principal layers of the leads remain translational 
invariant after the substitution. As an example, below we demonstrate how GOLLUM describes
the quantum Hall effect in a disordered square lattice, with a perpendicular uniform magnetic field.

\subsubsection{Superconducting systems}
\begin{figure}
\includegraphics[trim=1cm 1cm 1cm 1cm, width=0.8\columnwidth]{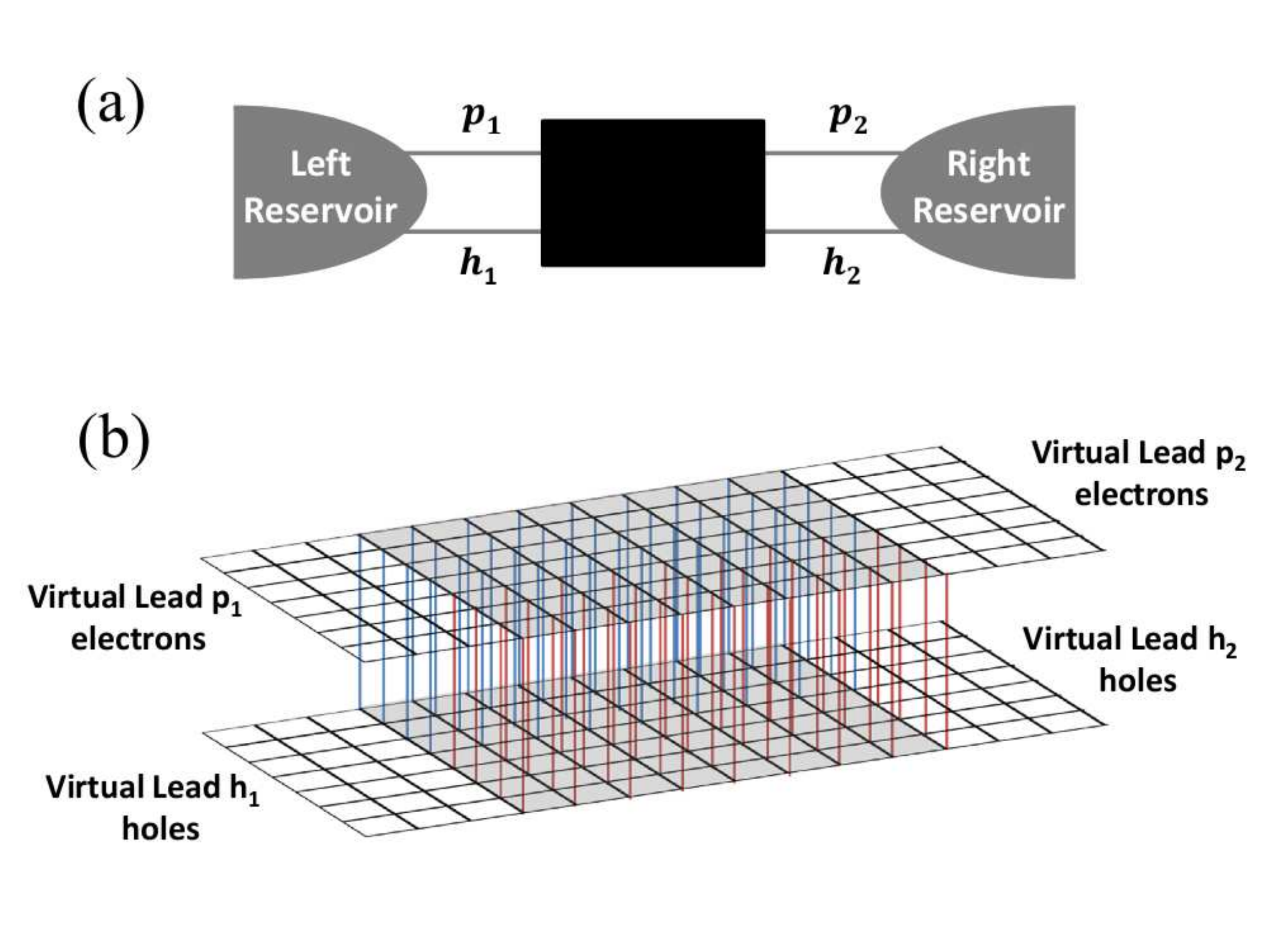}
\caption{Two-probe device consist of reservoirs $\alpha$ and $\beta$ connected to a superconductor}
\label{fig:superconductor}
\end{figure}

Figure \ref{fig:superconductor}(a) shows a two-probe normal-superconductor-normal (N-S-N) device with left and right normal  
reservoirs  connected to a scattering region containing one or more superconductors. If the complete Hamiltonian describing a normal system of the type shown in Fig. 2 is  $H_N$,
then in the presence of superconductivity within the extended scattering region, the new system is described by the 
Bogoliubov-de Gennes Hamiltonian
\begin{equation}
 H=\left(\begin{array}{cc}
          H_N&\Delta\\
          \Delta^*&-H_N^*\\
         \end{array}\right)\,\,\label{super}
\end{equation}
where the elements of the matrix $\Delta$ are non-zero only in the region occupied by a superconductor,  as indicated in Figure 
\ref{fig:superconductor}(b).
Physically, $H_N$ describes particle degrees of freedom, $-H_N^*$ describes hole degrees of freedom and 
$\Delta$ is the superconducting order parameter.

The multi-channel scattering theory for such a normal-superconducting-normal (N-S-N) structure was first derived by Lambert in 
Ref. [\onlinecite{Lambert_1991}], where the following current-voltage relation was presented:
\begin{equation}
 \left(\begin{array}{cc}
          I_{left}\\
          I_{right}\\
         \end{array}\right)
         =\frac{2\,e^2}{h}\,a\,\left(\begin{array}{cc}
          (\mu_{left}-\mu)/e\\
          (\mu_{right}-\mu)/e\\
         \end{array}\right)\,\,\label{x7}
\end{equation}
where $I_{left}$ ($I_{right}$) is the current from the left (right) reservoir, $\mu_{left}-\mu$ ($\mu_{right}-\mu$) is 
the difference between the chemical potential of the left (right) reservoir and the chemical potential $\mu$ of the 
superconducting condensate and the voltage difference between the left and right reservoirs is 
$(\mu_{left}-\mu_{right})/e$. This expression is the low-voltage limit of more general current-voltage 
relations discussed in [\onlinecite{Lambert_1991,hui1}]. The generalization to multi-probe structures is
described in Refs. \onlinecite{Lambert_1998_1,allsopp},  to thermoelectric properties of superconducting nanostructures 
in Refs. [\onlinecite {thermo1,thermo2}] and to ferromagnetic-superconducting structures in Refs. [\onlinecite{f1,f2,f3}].
In this equation,
\begin{equation}
 a=\left(\begin{array}{cc}
          M_{left}-R_o+R_a & -T_o'+T_a'\\
          -T_o+T_a& M_{right}-R_o'+R_a'\\
         \end{array}\right)\,\,
\end{equation}
where $M_{left}$ ($M_{right}$) is the number of open channels in the left (right) lead, $R_o, T_o$ ($R_a, T_a$) 
are normal (Andreev) reflection and transmission coefficients for quasi-particles emitted from the right lead, 
$R_o', T_o'$ ($R_a', T_a'$) are normal (Andreev) reflection and transmission coefficients from the left lead and 
all quantities are evaluated at the Fermi energy $E=\mu$.
As a consequence of unitarity of the scattering matrix, these satisfy $R_o+ T_o+ R_a+ T_a= M_{left}$ 
and $R_o'+ T_o'+ R_a'+ T_a'= M_{right}$.

The current-voltage relation of Equ. (\ref{x7}) is fundamentally different from that encountered for normal systems, because 
unitarity of the s-matrix does not imply that the sum of each row or column of the matrix $a$ is zero. 
Consequently, the currents do not automatically depend solely of the applied voltage difference 
$(\mu_{left}-\mu_{right})/e$ (or more generally on the differences between incoming quasi-article 
distributions). In practice such a dependence arises only after the chemical potential of the 
superconductor adjusts itself self-consistently to ensure that the current from the left reservoir is equal to the 
current entering the right reservoir. Insisting that $I_{left}=-I_{right}=I$, then yields
\begin{equation}
\frac{2\,e^2}{h} \left(\begin{array}{cc}
           (\mu_{left}-\mu)/e\\
          (\mu_{right}-\mu)/e\\
         \end{array}\right)=a^{-1}
        \left(\begin{array}{cc}
          I\\
          -I\\
         \end{array}\right)\,\,
\end{equation}
and therefore the two-probe conductance $G=I/((\mu_{left}-\mu_{right})/e)$ takes the form of
\begin{equation}
G=\frac{2\,e^2}{h}\,\frac{a_{11}a_{22}-a_{12}a_{21}}{a_{11}+a_{22}+a_{12}+a_{21}}
\label{eq:Gsup}
\end{equation}

The above equation demonstrates why a superconductor possesses zero resistivity, because if the 
superconductor is disordered, then as the length $L$ of the superconductor increases, all transmission 
coefficients will vanish. In this limit, the above equation reduces to $(h/2e^2)G=2/R_a + 2/R_a'$. 
In contrast with a normal scatterer, this shows that in the presence of Andreev scattering, as $L$ 
tends to infinity, the resistance ( = 1/conductance) remains finite and therefore the resistivity (ie resistance per unit 
length) vanishes.

In the notation of Eqs. (\ref{xxx}) and (\ref{x1}), the above current-voltage relations and 
their finite-temperature, finite voltage generalizations can be obtained from Eq. 
(\ref{xxxx}) by writing $\alpha_i = p_i$ or $h_i$ to yield
\begin{equation}
I^{i}_{e}=\int_0^\infty(dE/h) \sum_{\alpha_i=p_i,h_i}\sum_{j=1,2}
\sum_{\beta_j=p_j,h_j}Q_{\alpha_i}(E)P^{i,j}_{\alpha_i,\beta_j} \bar f^j_{\beta_j}(E)  \label{x5}
\end{equation}
Since $Q_{p_i}=-e$ and $Q_{h_i}=+e$, this becomes
\begin{equation}
I^{i}_{e}=(-e/h)\int_0^\infty dE \sum_{j=1,2}\sum_{\beta_j=p_j,h_j}[P^{i,j}_{p_i,\beta_j}-P^{i,j}_{h_i,\beta_j}] 
\bar f^j_{\beta_j}(E)  \label{x6}
\end{equation}
Since, in the low-bias limit, $\bar f^j_{p_j}(E)=-\bar f^j_{h_j}(E)=(-df(E)/dE)(\mu_j-\mu)$, where $f(E)$ is 
the Fermi distribution with chemical potential $\mu$, this simplifies to
\begin{equation}
I^{i}_{e}=(e^2/h)\sum_{j=1,2}A_{ij}(\mu_j-\mu)  \label{x66}
\end{equation}
where
\begin{equation}
A_{ij}=\int_0^\infty dE (-df(E)/dE)\sum_{\beta_j=p_j,h_j}[P^{i,j}_{p_i,\beta_j}-P^{i,j}_{h_i,\beta_j}] \label{x77}
\end{equation}
The total current is obtained by multiplying Eq. (\ref {x6}) by a factor of 2 to account for spin. 
On the other hand, in the limit of zero temperature,
$\int_0^\infty dE (-df(E)/dE)=1/2$ Hence in this limit, the current-voltage relation (\ref{x6}) reduces to Eq. (\ref{x7}).

\begin{figure*}
\includegraphics[width=\textwidth]{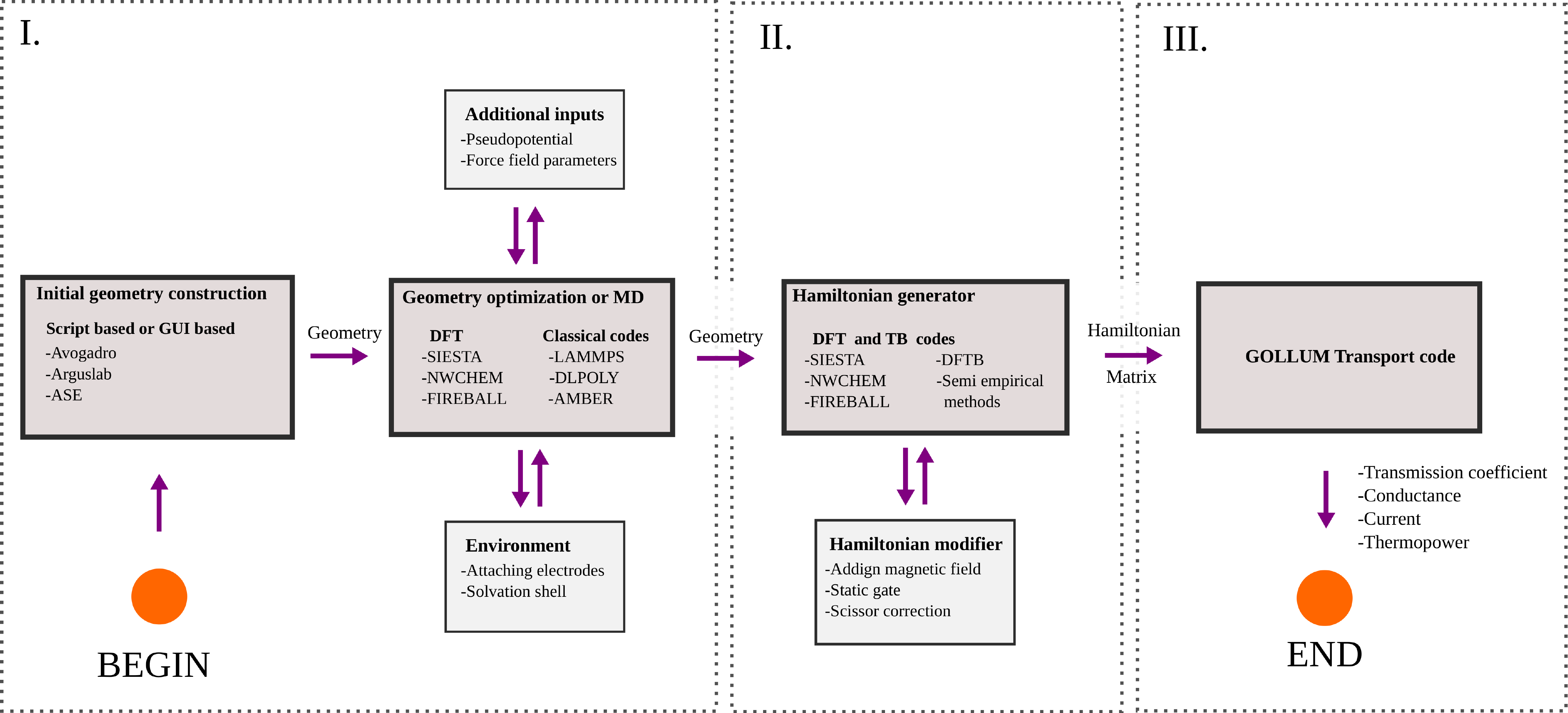}
\caption{\label{multiscale} Typical GOLLUM work-flow with various optional software tools}
\end{figure*}

\subsection{Multiscale tools}

Simulation of the transport properties of a nanoscale-scale junction involves three distinct tasks. First, 
model geometries must be generated. Secondly, the Hamiltonian for each geometry must be constructed. Thirdly, the s-matrix can be calculated and transport properties of the junction calculated. GOLLUM separates these three tasks into
three different processes. An overview of the work-flow of a generic GOLLUM calculation is shown in Figure \ref{multiscale}. 
The three consecutive stages of the work process are denoted by the three dotted rectangular boxes. 
The initial step consists usually of modeling the atomistic arrangement of the junction. An initial structure is 
usually guessed, followed by geometry optimization or molecular dynamics simulations to obtain a more realistic atomic arrangement.
This task can be performed by either ab-initio or classical molecular-dynamics methods. For systems containing a few hundred 
atoms, a quantum-mechanical DFT-based simulation is usually the method of choice. However, experiments are often performed under ambient
conditions or in a liquid environment. In these cases that the microscopic model should include the atomic structure 
of the environment, as we show below in section III.GOLLUM addresses this task by using classical molecular dynamics to model the environment and in the spirit of the Born-Oppenheimer approximation, feeding snapshots of the associated electrostatic field into the the DFT-based mean-field Hamiltonian.

A similar approach is used to model the evolution of mechanically-controlled break junctions upon stretching, where the atomistic arrangement of the junction evolves slowly  in time. In this case, if the same experiment is repeated
a number of times, the junction geometry will be slightly different each time. Therefore, a proper statistical analysis of the 
junction geometries is mandatory and calculations of the associated distribution of transmission coefficients is required. The task of generating junction geometries is also better suited for classical molecular dynamics situations. GOLLUM also facilitates the use of combined DFT and classical molecular dynamics approaches  to gain accurate, yet quicker 
simulation results\cite{indium}. A non-comprehensive set of software tools is listed in Figure \ref{multiscale}.
Once the atomic arrangements are generated, these are fed into the second stage, where the Hamiltonian matrix is generated. 
This stage is in practice independent of the previous geometry construction and can be run separately, taking only the output 
geometries of the first stage. The junction Hamiltonian can be generated using a variety of tools, some of which are listed 
in box II in Figure \ref{multiscale}. A popular approach is the use of DFT codes that are able to write the Hamiltonian in a tight-binding
language. In this way, model tight-binding Hamiltonians can also be easily generated. Other approaches involve the use of Slater-Booster or 
semi-empirical methods. In addition, GOLLUM has the ability to modify suitably these Hamiltonian matrices as  discussed above.
 For example, the Hamiltonian matrix can be modified to include scissor corrections, Coulomb-blockade physics, a gate 
or bias voltage, a magnetic phase factor or a superconducting order parameter. 
Finally, stage III is the actual quantum transport calculation. This takes the Hamiltonian matrix as an input and calculates the s-matrix and
associated physical quantities, such the electrical or spin current, the conductance, or the thermopower.

\section{Demonstrator calculations }
In this section, we present a diversity of  calculations, which demonstrate the broad capabilities of GOLLUM. 
For simplicity, we begin with a set of calculations on model  Hamiltonians, which demonstrate that GOLLUM can easily 
handle tight-binding models for a range of physical systems. We then move on to more material-specific calculations, 
in which the Hamiltonian is obtained from DFT. These include examples exhibiting Kondo physics, Coulomb blockade and 
non-linear, finite-voltage effects.  Next we present more computationally challenging calculations involving 
van der Waals interactions, environmental effects and 
series of geometries associated with break-junction measurements. 
Finally an example of a quantum pump is presented, which requires access to the phase of scattering amplitudes.
We define the conductance quantum $G_0=2\,e^2/h$, that will be used frequently below.

\subsection{Simple one-dimensional tight-binding two and four terminal device}

\begin{figure}
\includegraphics[width=\columnwidth]{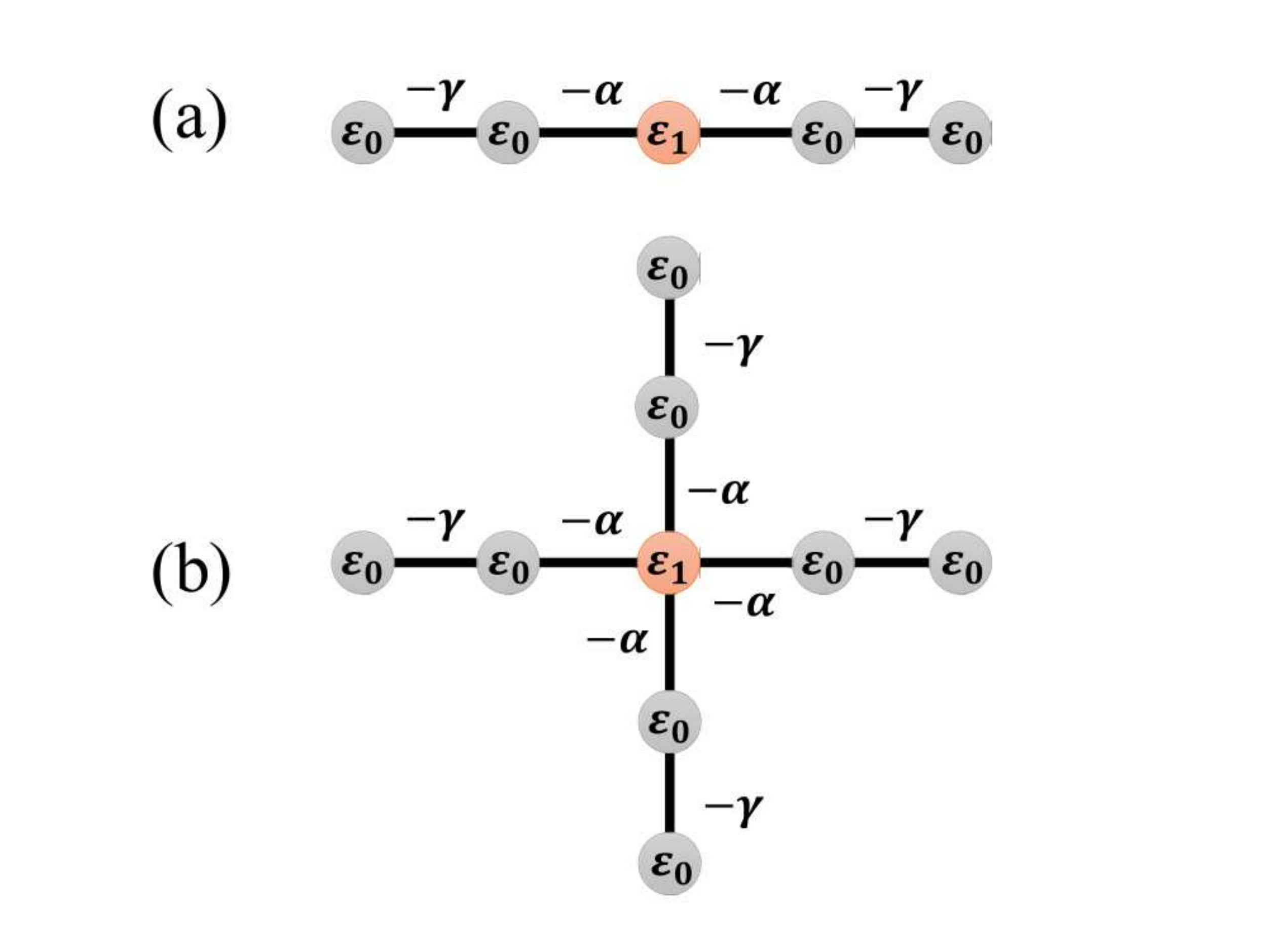}
\caption{Simple tight-binding one-dimensional (a) two probe (b) four probe systems containing a single orbital per PL
and a single impurity orbital at the EM region. 
The parameters of the tight-binding model are $\varepsilon_0=0$, $\varepsilon_1=1$, $\gamma=1$ and $\alpha=1.5$ (taken
in arbitrary units).}
\label{fig:2-4probeTB}
\end{figure}

\begin{figure}
\includegraphics[width=\columnwidth]{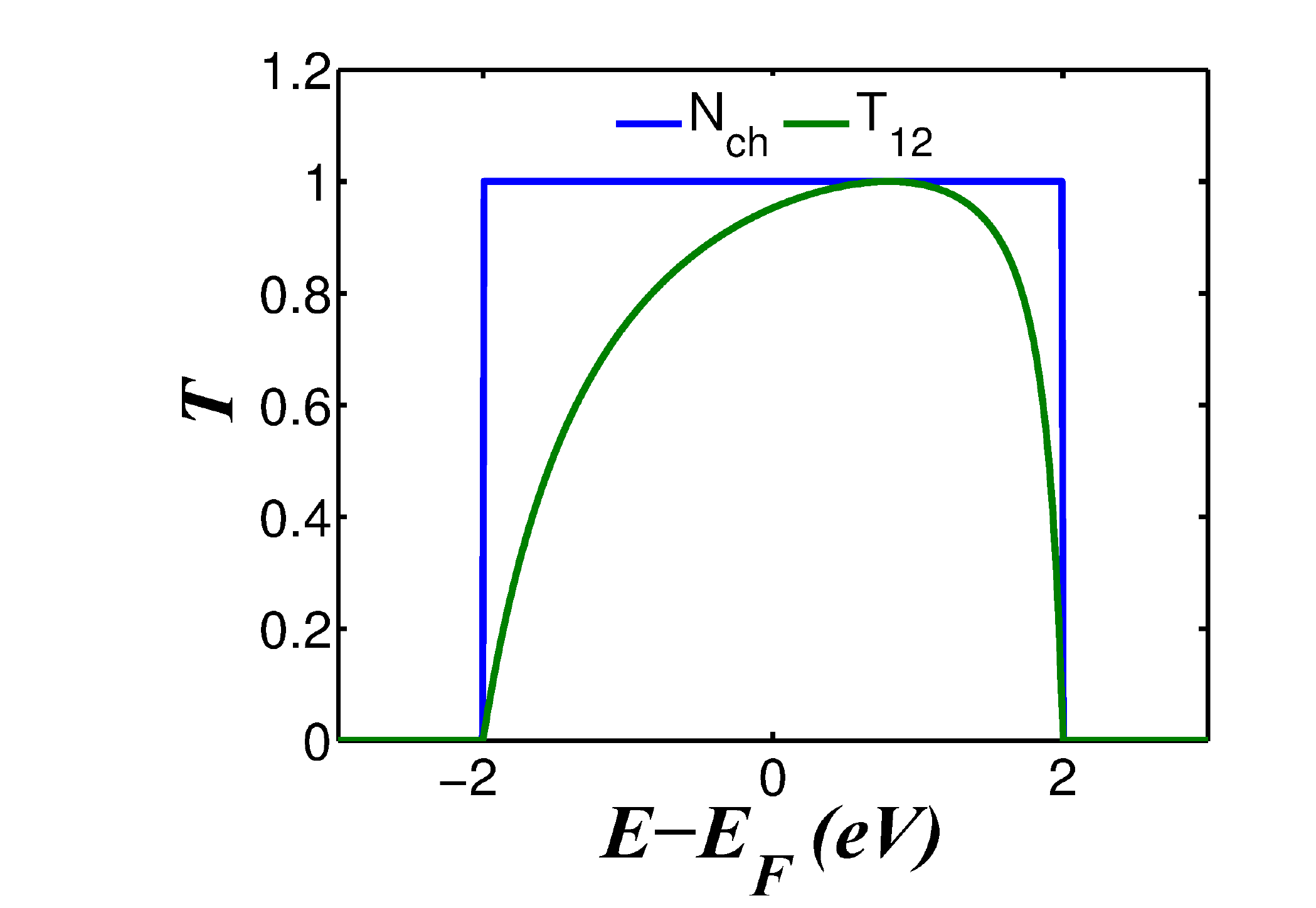}
\caption{Transmission and number of open channels for the simple tight-binding one-dimensional chain shown in Fig. 
\ref{fig:2-4probeTB}(a) as a function of the energy. Energies are referred to the Fermi energy $E_F$ and are 
given in units of $\gamma$.} \label{fig:2probeTB}
\end{figure}

\begin{figure}
\includegraphics[trim=21cm 0cm 0cm 0cm, clip=true, width=\columnwidth]{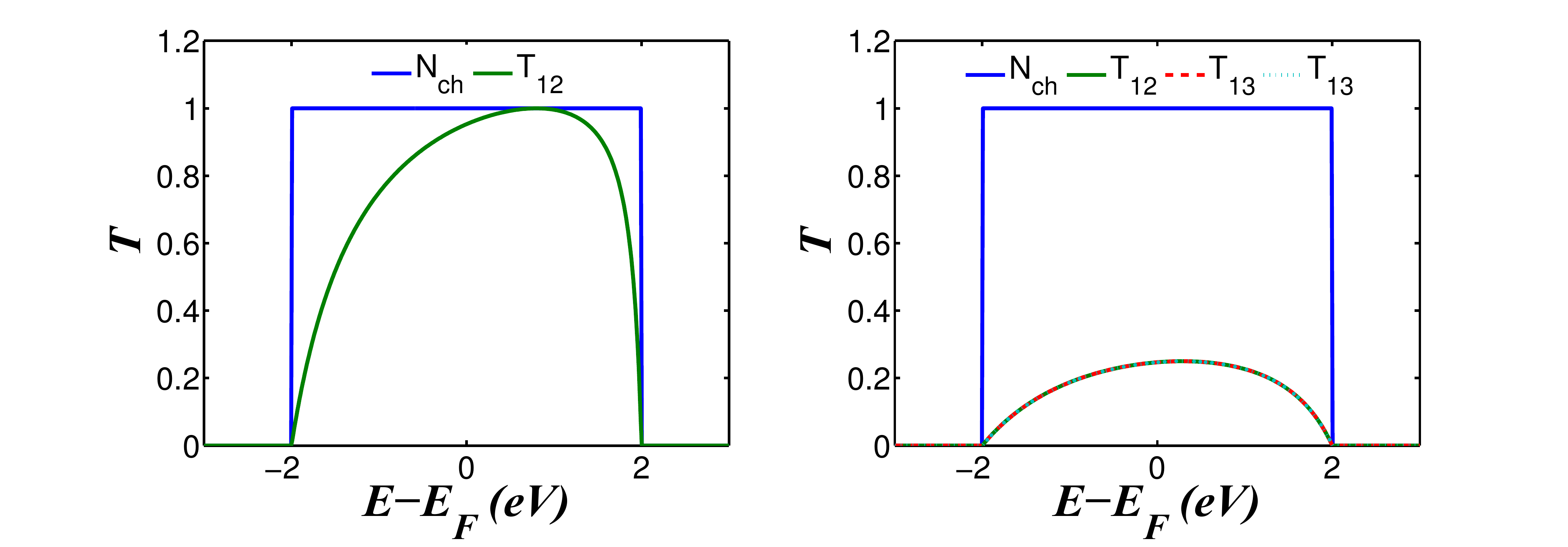}
\caption{Transmission and number of open channels for the four-probe device shown in Fig. \ref{fig:2-4probeTB}(b)
that has four one-dimensional chain leads. as a function of the energy. Energies are referred to the Fermi energy $E_F$ and are 
given in units of $\gamma$.}
\label{fig:2probeTB4}
\end{figure}

As a first example, we consider a simple one-dimensional tight-binding chain containing a single orbital per PL and a single impurity
orbital in the EM region, as shown in fig \ref{fig:2-4probeTB}(a). We take the following parameters, that are given in arbitrary units. 
Within the leads, the site energies are $\varepsilon_0=0$, 
and the nearest neighbor couplings are $-\gamma$. The impurity has a site energy  $\varepsilon_1=1$ and is coupled to the leads 
by a hopping element $-\alpha$. Results are shown for $\gamma=1$ and $\alpha=1.5$. The transmission coefficient for this chain is shown in figure \ref{fig:2probeTB}.

As a second example, we consider the four-probe structure of  Fig. \ref{fig:2-4probeTB}(b), that shares the same set of parameters as
the two-probe model above.
The various transmission coefficients for this structure are shown in figure \ref{fig:2probeTB4}. By symmetry, these are all identical.

\subsection{The quantum Hall effect}
\begin{figure}
\includegraphics[width=\columnwidth]{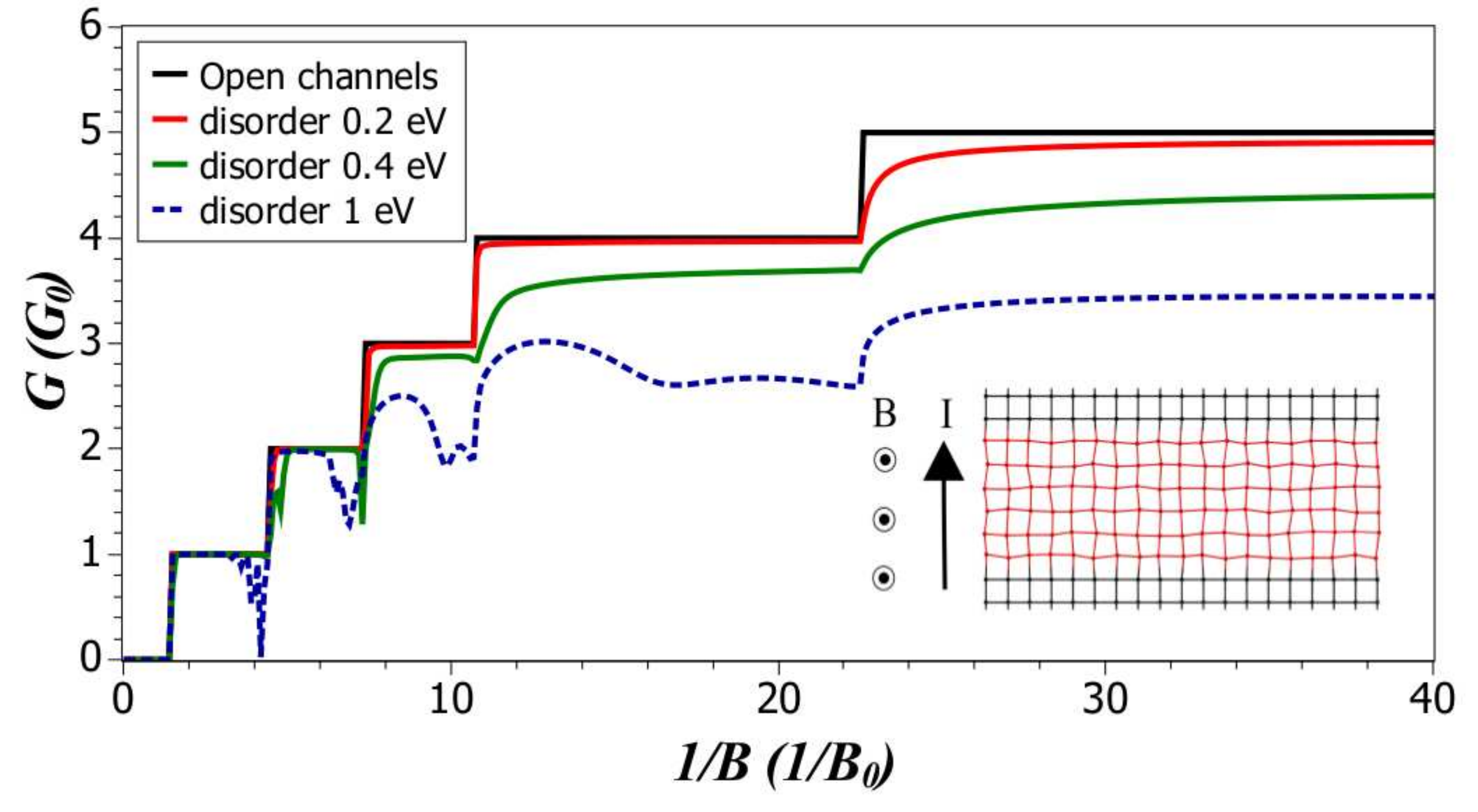}
\caption{\label{quantumhall} Conductance $G$ in units of $G_0$ as a function of inverse magnetic field with various level of disorder. The 
magnetic field unit is set to $B_0=6.58\times10^{4}$ Tesla. The inset shows the square lattice used for the calculation.
The black area denotes a perfect square lattice. The red area denotes a disordered portion of the lattice, where the inter-site 
distances are slight modified from $1$ \AA to perturb the phase contribution.
The onsite energy for the regular lattice is $\epsilon=3.35$ eV, the coupling with zero magnetic field is $\gamma=1$ eV and the Fermi energy is chosen as zero.
In the disordered area the onsite energy is randomly varied as $\epsilon' = \epsilon + \xi$, where $\xi$ is a random number distributed with uniform probability in
the range $(-0.2,0.2)$, 
$(-0.4,0.4)$ and $(-1,1)$ eV (red, green and blue dashed curves, respectively).} 
\end{figure}

As an example of a quantum transport calculation with a magnetic field, we demonstrate the quantum Hall effect within the simple 
tight-binding square lattice  shown in the inset of Fig. \ref{quantumhall}. The lattice constant is set to 
$a=1$ \AA. The onsite energies of the perfect lattice are $\epsilon=3.35$ eV, the hopping integrals 
at zero magnetic field are $\gamma=1$ eV and the Fermi energy is set at zero.
The red area in the figure denotes a disordered portion of the lattice. In this disordered area, the onsite energies are 
randomly varied as $\epsilon' = \epsilon + \xi$, where $\xi$ is a random number distributed with uniform probability in
the range $(-0.2,0.2)$, 
$(-0.4,0.4)$ and $(-1,1)$ eV (red, green and blue dashed curves, respectively).

The transport direction is chosen to be the $y$ axis (e.g.: from bottom to top) while the $x$ axis goes along the horizontal direction.
To demonstrate the quantum Hall effect we introduce a homogeneous magnetic field perpendicular to the square lattice, pointing out 
of the paper, which is expressed in units of $B_0 = 6.58\times10^{4}$ Tesla. With this setup the vector potential is chosen so that the lead remains 
translationally invariant along the $y$ direction. This means that we implement a Peierls substitution of the form
\begin{equation}
\gamma_{ij}^B=\gamma e^{-i\frac{B}{B_0} \frac{(y_i-y_j)(x_i+x_j)}{2 a^2 }},
\end{equation}
where $x_i$ and $y_i$ are the coordinates of the site i.   With this modified Hamiltonian the conductance calculated by GOLLUM is shown 
in Fig. (\ref{quantumhall}). This clearly shows the presence of quantum Hall plateaus, which are resilient to the presence of disorder.

\subsection{Superconductivity}
\begin{figure}
 \includegraphics[trim=1cm 10cm 1cm 0cm, clip=true, width=\columnwidth]{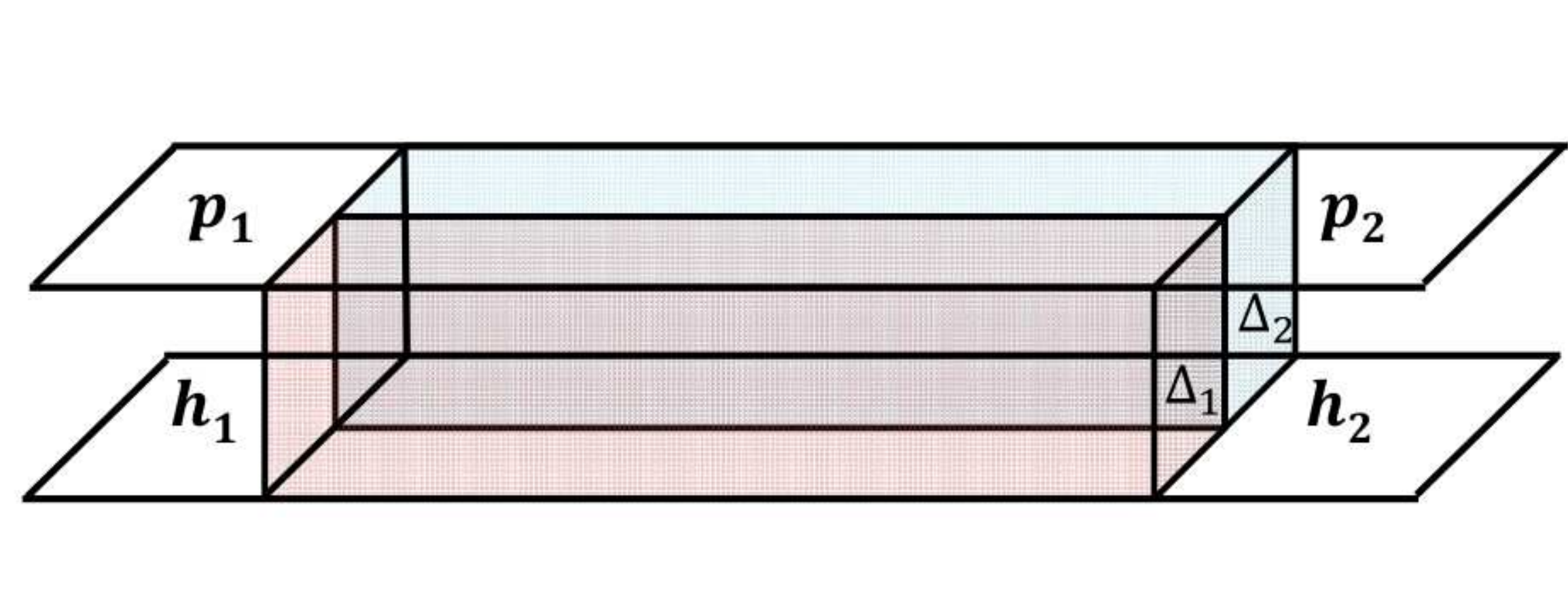}
 \caption{Two-terminal device consisting of two physical leads connected to a scattering region containing two 
 superconductors with order parameters $\Delta_1$ and $\Delta_2$. The left (right) physical lead consists of two 
 virtual leads $p_1$ and $h_1$ ( $p_2$ and $h_2$) carrying particle and hole channels respectively.}
  \label{fig:sup-1}
\end{figure}

As an example of scattering in the presence of superconductivity, we now compute the electrical conductance of the N-S-N structure 
shown in Fig. (\ref{fig:sup-1}), which contains two superconducting regions with order parameters $\Delta_1$ and 
$\Delta_2=\Delta_1 e^{i\theta}$. Such a structure is known as an Andreev interferometer and was first analyzed in 
Refs. [\onlinecite{lam2,hui2}], where it was predicted that the electrical conductance is a periodic function of the 
order-parameter phase difference $\theta$, with period $2\pi$. At that time, this effect was completely missing 
from the more traditional quasi-classical description of superconductivity. When the missing terms were restored, 
good agreement between quasi-classical theory and scattering theory was obtained \cite{volkov}.

\begin{figure}
 \includegraphics[trim=5cm 0cm 0cm 0cm, clip=true, width=\columnwidth]{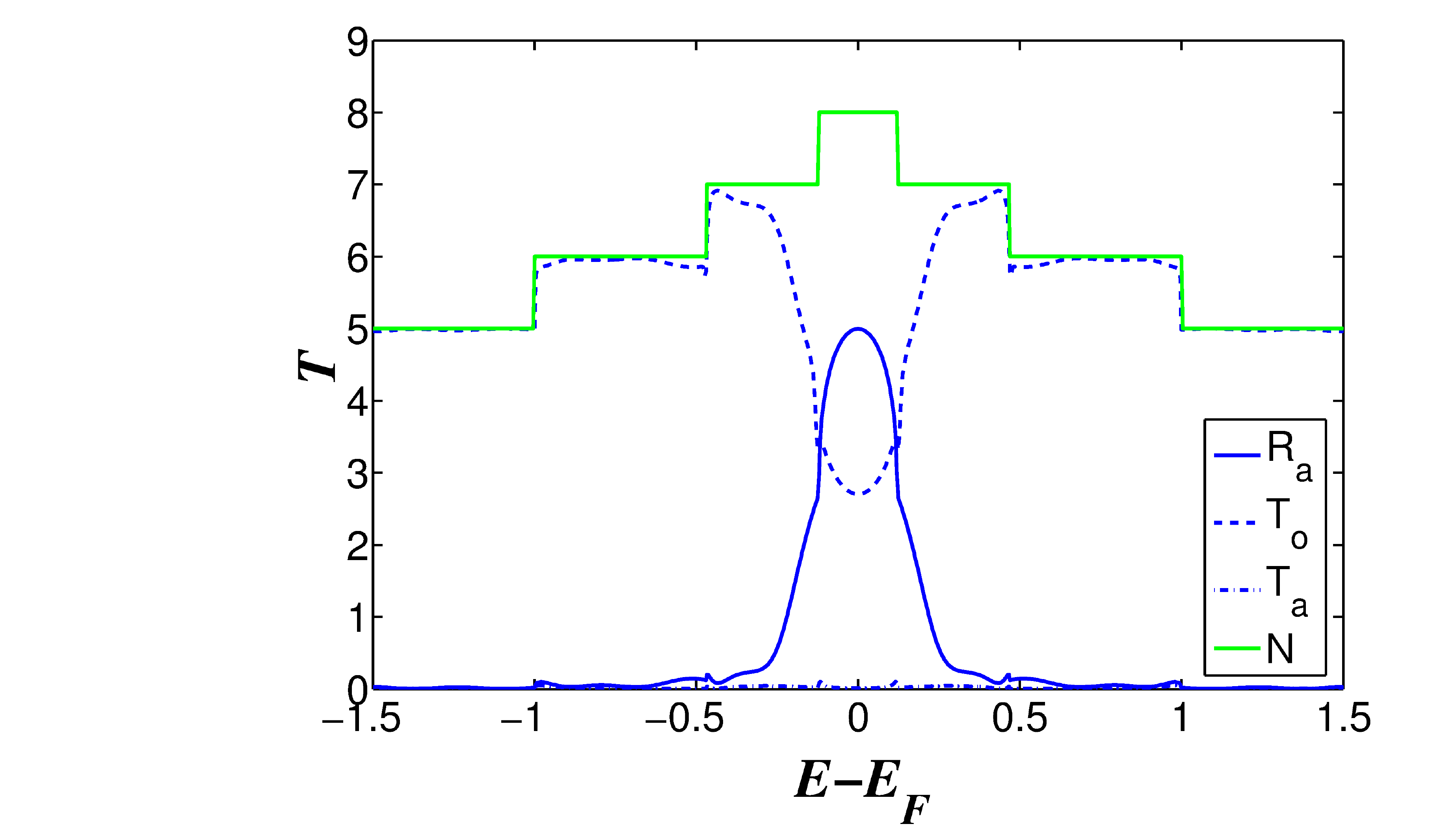}
 \caption{Transmission coefficients ($R_{o}$ (dot dashed line line), $R_{a}$ (solid blue line), $T_{o}$ (dashed line)  
 and the number of open channels in the left lead for the device shown in Fig. (\ref{fig:sup-1}), as a function of the energy.
 as a function of the energy. Energies are referred to the Fermi energy $E_F$ and are given in units of $\gamma$. }
  \label{fig:sup-2}
\end{figure}

\begin{figure}
  \includegraphics[width=\columnwidth]{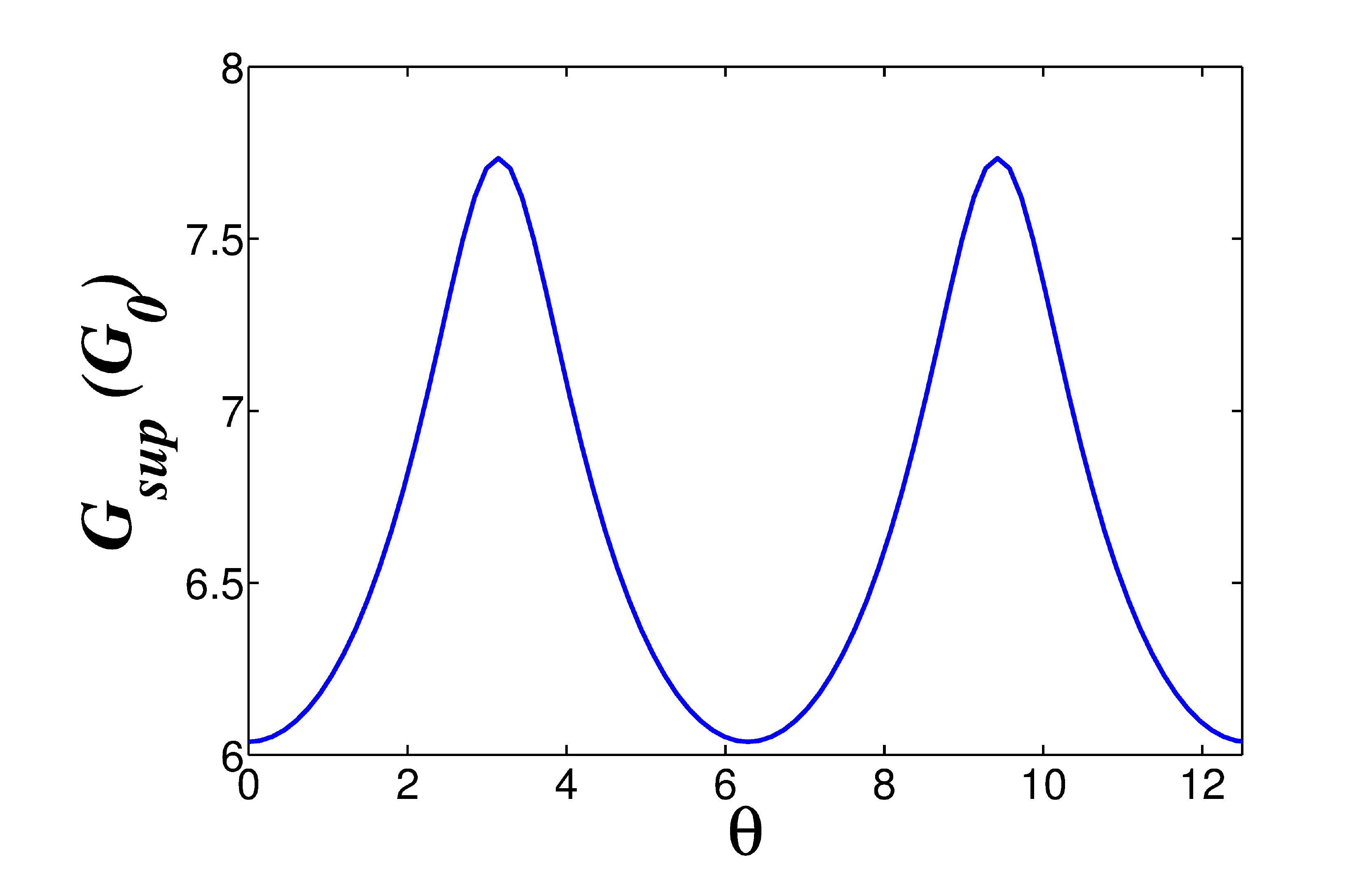}
   \caption{Two-probe conductance $G$ in units of $G_0$, for a N-S-N structure shown in Fig. (\ref{fig:sup-1}) as a function of the 
   phase difference $\theta$ between the two order parameters. }
  \label{fig:sup-3}
\end{figure}

In the following calculation, the Hamiltonian $H_N$ of Eq. (\ref{super}) is simply a nearest neighbor tight-binding 
Hamiltonian on a square lattice, with diagonal elements $\varepsilon_0=0$ and nearest-neighbor couplings with $\gamma=1$
(in arbitrary units). 
Within the regions occupied by superconductor $j$, (where $j=1$ or $2$) the top (particle) sites are coupled to the bottom (hole) sites by 
$\Delta_j$, with $\vert \Delta_j\vert=0.1$ given in units of $\gamma$. For $\theta=0$, Figure (\ref{fig:sup-2}) shows the energy dependence of the 
Andreev refection coefficient $R_a$ and the normal and Andreev transmission coefficients $T_o$ and $T_a$ respectively.  
The green line in Figure (\ref{fig:sup-2}) represents the number of open channels in electron (hole) conducting leads. 
As expected, the Andreev reflection coefficient is large for small energies and decreases for energies above $\vert \Delta_1\vert$.
Substituting the values of these coefficients at $E=0$ into Eq. (\ref{eq:Gsup}) and evaluating them for all $\theta$ yields the 
conductance versus $\theta$ plot shown in Figure \ref{fig:sup-3}. As expected, the conductance is an oscillatory function of the order-parameter phase difference
$\theta$ with period $2\pi$.

\subsection{Non-collinear magnetism}
\begin{figure}
  \includegraphics[width=\columnwidth]{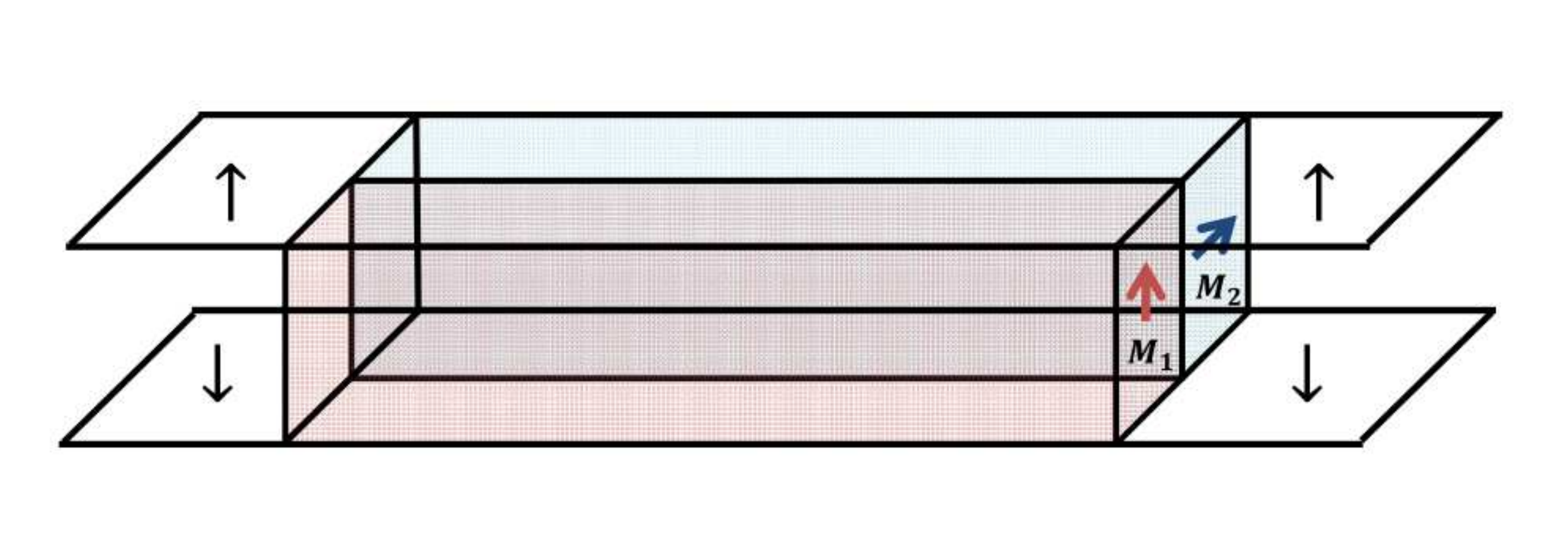}
   \caption{Two terminal device consisting of two physical leads connected to a scattering region containing two 
   ferromagnetic islands with  magnetic moments $(M^1_x, M^1_y,0)$ and $(M^2_x, M^2_y,0)$ The left (right) physical 
   lead consists of two virtual leads $\uparrow_1$ and $\downarrow_1$ ( $\uparrow_2$ and $\downarrow_2$) carrying 
   up-spin and down-spin channels respectively.}
  \label{fig:mag1}
\end{figure}

\begin{figure}
  \includegraphics[width=\columnwidth]{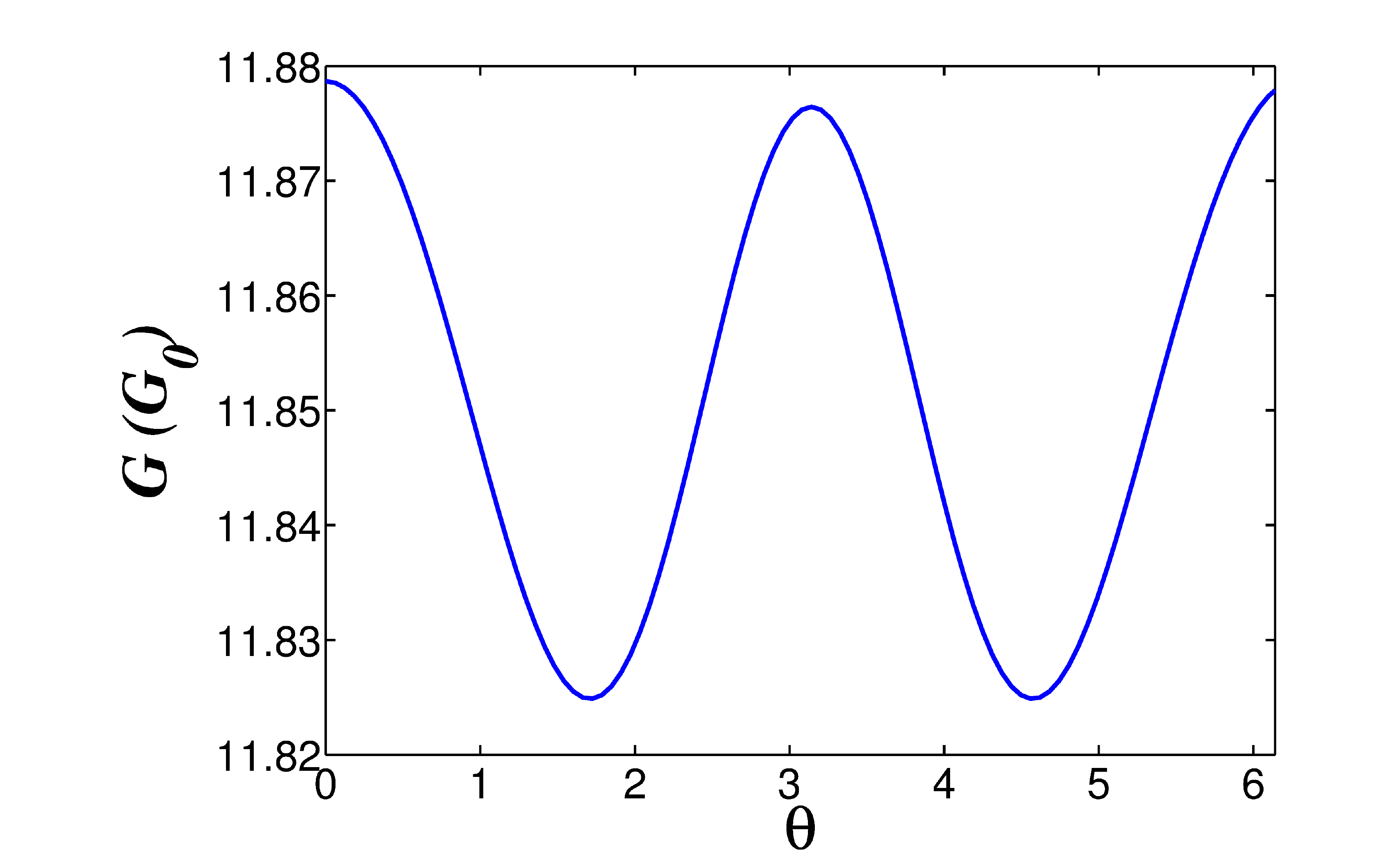}
   \caption{Two-probe conductance $G$ in units of $G_0$, for the structure shown in shown in Fig. (\ref{fig:mag1}) as a function of the angle $\theta$ 
   between the two magnetic moments.}
  \label{fig:mag2}
\end{figure}

In this section, we compute the electrical conductance of the structure shown in Figure (\ref{fig:mag1}), which we again 
describe using a simple tight-binding model of the form
\begin{equation}
 H=\left(\begin{array}{cc}
          H_N+M_z&M_x-iM_y\\
          M_x+iM_y&H_N-M_z\\
         \end{array}\right)\,\,\label{magn}
\end{equation}
The Hamiltonian $H_N$ is simply a nearest neighbor Hamiltonian on a square lattice,
with diagonal elements $\varepsilon_0=0$ and nearest-neighbor couplings with $\gamma=1$ (in arbitrary units) and for simplicity we choose $M_z=0$ everywhere.
The systems consists of two magnetic islands with magnetic moments $(M^1_x, M^1_y,0)$ and $(M^2_x, M^2_y,0)$, connected to non-magnetic leads. Choosing the
Fermi energy to be $E_F=0$, and evaluating Eq. (\ref{xxxx1}) at zero temperature, Figure  (\ref{fig:mag2}) shows the resulting electrical conductance as a 
function of the angle $\theta$ between the two magnetic moments. As expected, the conductance is an oscillatory function of the magnetic angle $\theta$.

Having discussed model systems described by simple tight-binding Hamiltonians, we now turn to more material-specific
descriptions based on DFT. We will use the program SIESTA in most of the calculations below, and will
provide many of the simulation parameters to help people to reproduce our calculations.

\begin{figure}
\includegraphics[trim=1cm 1cm 1cm 22.5cm, clip=true, width=0.9\columnwidth]{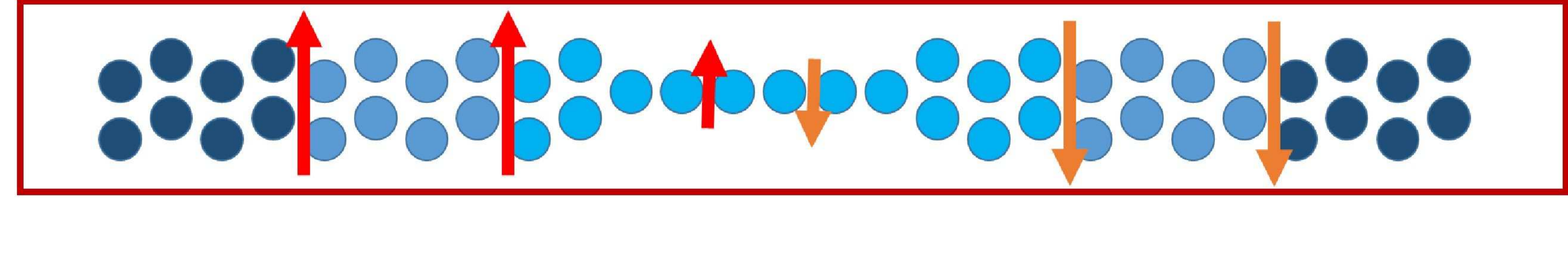}
\caption{Sketch of the EM setups used in the calculation of spin-resolved transport through
nickel electrodes that corresponds to the EM unit cell shown in Fig. (5). The scattering region is hown in light blue,
the first PL is shown in greyish blue, and the second PL is shown in dark blue. The second PL is followed by vacuum.}
\label{em_victor1}
\end{figure}

\begin{figure}
\includegraphics[trim=0cm 1cm 0cm 2cm, width=\columnwidth]{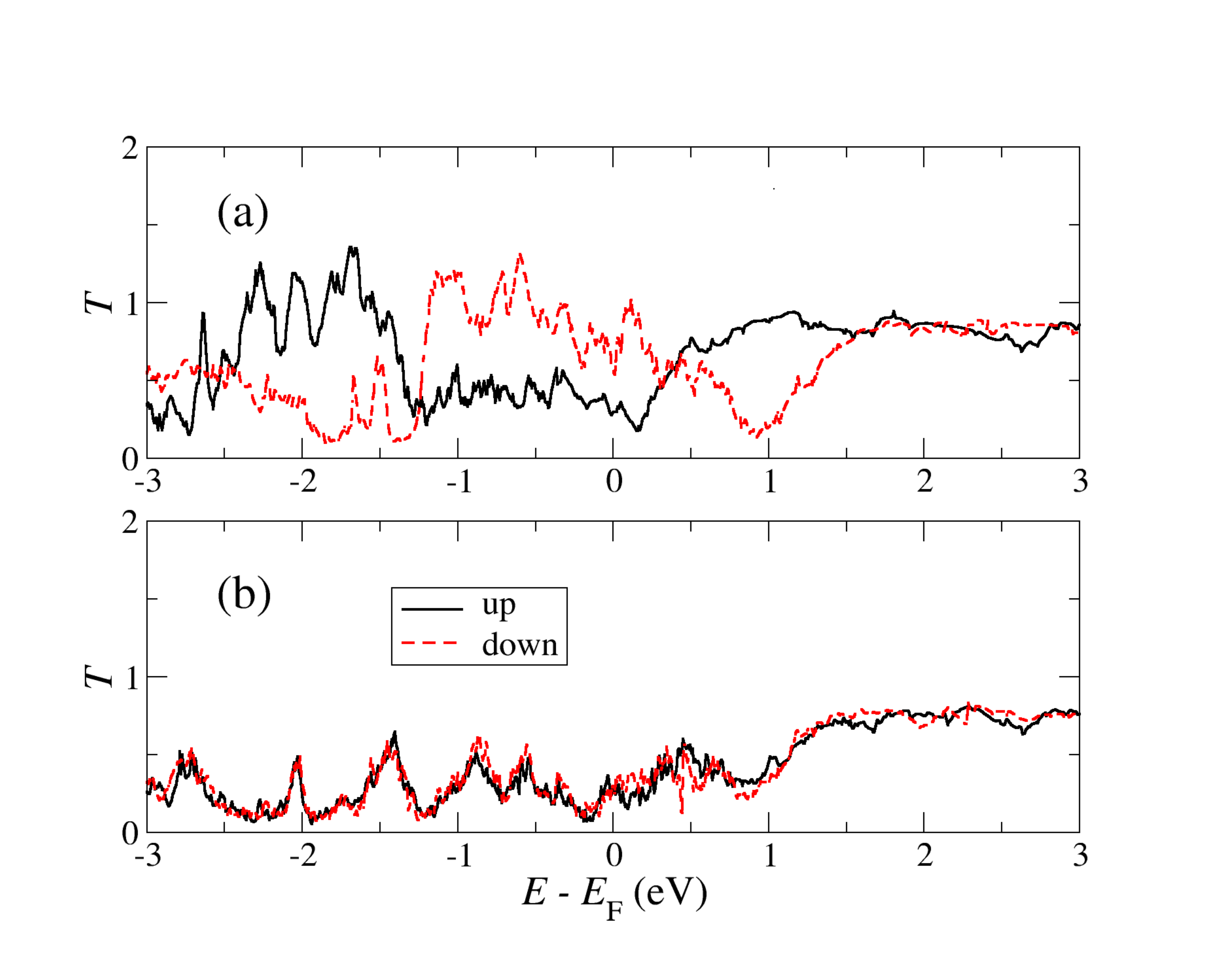}
\caption{\label{k-points}(Color online) Spin-resolved transmission
coefficients as a function of energy for nickel-chain junctions. The 
different curves correspond to simulation with different levels of accuracy
in the $k_perp$ summations: (a), (b), (c) and (d)
correspond to 1, 4, 16 and 64 $k$-points, respectively. (1) and (2)
correspond to parallel and anti-parallel configurations, respectively.}
\label{fig:victor2}
\end{figure}

\subsection{Spin polarized transport and magnetoresistance in nickel chains}
GOLLUM can describe voltage-dependent spin-polarized transport in spin-active junctions, made from a variety of 
metals, including iron \cite{noncol,soc}, platinum or palladium \cite{single}.
To demonstrate this, we describe here the voltage-dependent spin-filtering and magneto-resistive
behavior of a two-terminal junction where (001) fcc nickel electrodes with parallel (P)
or anti-parallel (AP) spin
orientations are connected by a nickel atomic chain\cite{Jac04,smeagol,Jac08,Haf09}. 
Notice
that because we may have electrodes with AP spin orientations, we are forced to use
EM setups such as those shown in Fig. (5). We sketch in Fig. (\ref{em_victor1}) the Scattering Region
used in the present calculation. It comprises a 6-atom-long nickel chain, 
the electrodes surfaces and the two branches. The electrodes surfaces contain 2/3 atomic layers with 4 atoms 
each. The left and right branches contain two PLs that have 4 atoms each. The second PL is followed by vacuum.
We have checked that the transport results in this example 
are  reasonably converged if we choose PL1 as the TPL, which means that PL2 is sacrificial. 

To find the junction Hamiltonian, we use the program SIESTA. We
use the Generalized Gradient approximation (GGA) functional\cite{gga} and take the theoretical GGA 
lattice constant of 3.45 \AA ~ for the PLs as well as inter-atomic distances
of 2.27 \AA ~ along the chain. We have employed a single-$\zeta$ (SZ) 
basis set to span the valence states and a mesh cutoff of 400 Ry to define 
the real-space grid where the density, potential and matrix elements are calculated.

\begin{figure}
\includegraphics[width=\columnwidth]{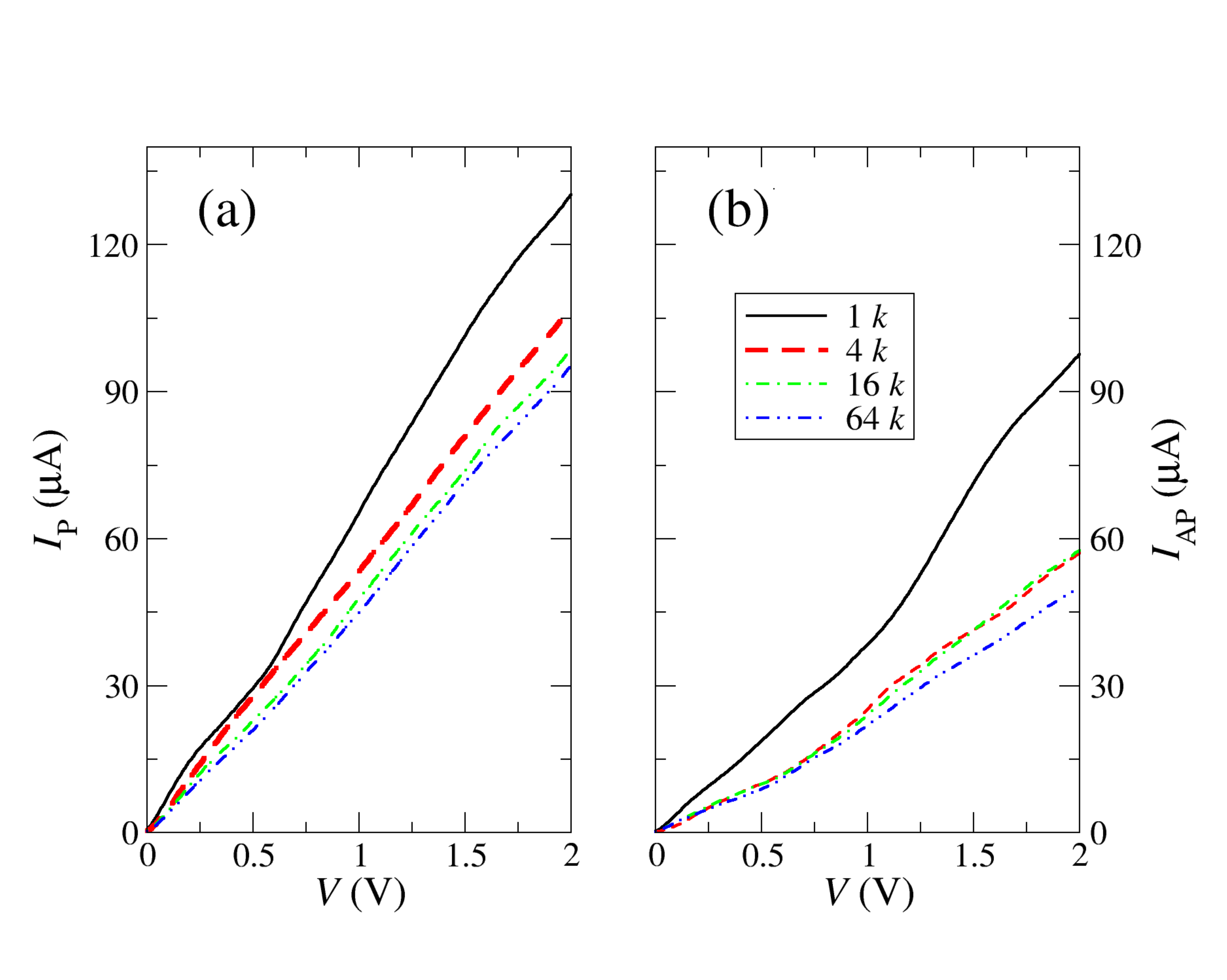}
\caption{(Color online) Charge current of the junction shown in Fig. (\ref{em_victor1})
in the (a) P and (b) AP spin configurations as a function of the bias voltage applied to the junction. The 
different curves correspond to different numbers of $k_\perp$-points.}
\label{fig:victor3}
\end{figure}

To understand the spin-polarized transport properties of the junction, we will
analise below  the spin-dependent transmission coefficients  $T_\sigma(E)$,
together with the spin-dependent charge currents. These are computed using the 
approximate expression
\begin{equation}
 I_\sigma\approx\,\frac{e}{h}\,\int_{-eV/2}^{eV/2}\,dE\,\,T_\sigma(E,V=0)
\end{equation}
\noindent The above approximation is quantitatively accurate for small enough bias voltages 
($\le 0.5$ V) and also shows the expected qualitative behavior at larger voltages. 
We have found that the transmission coefficients and currents depend sensitively on the number of 
transverse $k_\perp$-points taken along the plane perpendicular to 
the transport direction. As we will show below, we need to use at least 16
$k_\perp$-points to achieve convergence. In other words, 
a $\Gamma$-point calculation  provides a poor estimate of the transport properties of
these junctions.

\begin{figure}
\includegraphics[width=\columnwidth]{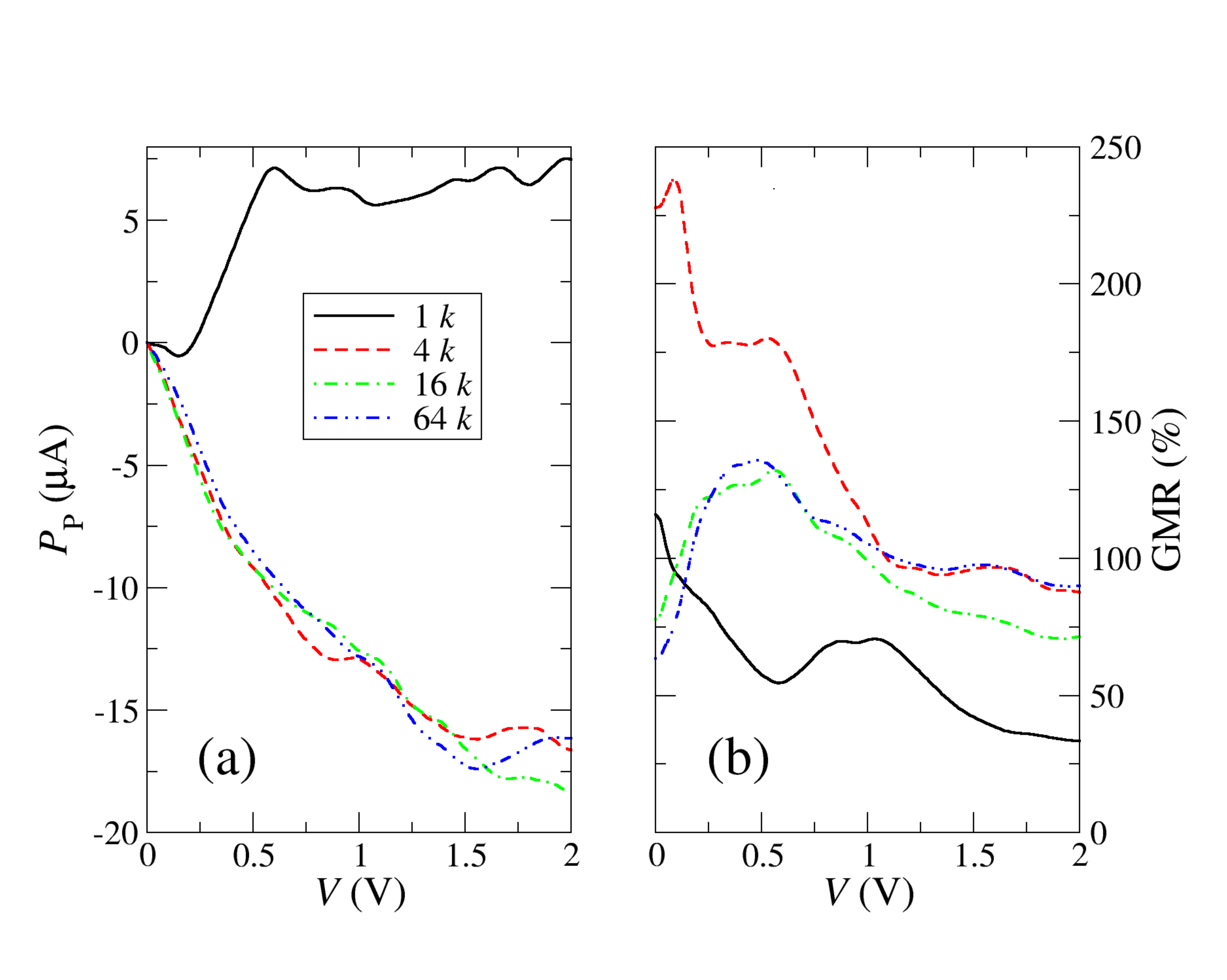}
\caption{(Color online) (a) Spin polarization of the current in the P configuration, measured
in $\mu A$ and (b) GMR ratio plotted as a function of the bias voltage applied to the junction shown in 
Fig. (\ref{em_victor1}). The different
curves correspond to different numbers of $k_\perp$-points.}
\label{fig:victor4}
\end{figure}

We plot $T_\sigma(E)$ as a function of the energy referred to
the Fermi energy of the Scattering Region for P and AP spin orientations  in Fig. (\ref{fig:victor2}). 
The upper 
panel of the figure shows that the transmission coefficients for the P configuration are strongly spin-polarized.
This polarization remains at the Fermi level, which suggests that these junctions could act 
as spin filters. The bottom panel of the figure shows the transmission coefficients for the AP configuration.
The fact that these are different from those of the P spin orientation hints that these junctions could show
significant GMR ratios. To quantify these statements, we compute the charge current of the junction in the P and AP 
configurations $I_P$ and $I_{AP}$ and plot them in Fig. (\ref{fig:victor3}). The figure shows that indeed these
junctions show magnetorresistive properties. The figure also demonstrates that the currents depend on the
number of $k_\perp$-points.  To further give quantitative estimates of the spin activity of the junctions, we
define the spin polarization in the P arrangement and the GMR ratio as
\begin{eqnarray}
P_\mathrm{P}&=&I_\mathrm{P,\uparrow}-I_\mathrm{P,\downarrow}\\
\mathrm{GMR}(\%)&=&\frac{I_\mathrm{P}-I_\mathrm{AP}}{I_\mathrm{AP}}\times
100.
\end{eqnarray}
Where $I_{P,\sigma}$ are the spin-dependent currents for the P orientation. 
We show in Fig. (\ref{fig:victor4}) these two magnitudes as a functions of the bias voltage applied to
the junction. We indeed find large spin signals for these devices. Furthermore, the figure demonstrates that 
at least 16 $k_\perp$-points are needed to achieve converged results.

\begin{figure}
\includegraphics[width=\columnwidth]{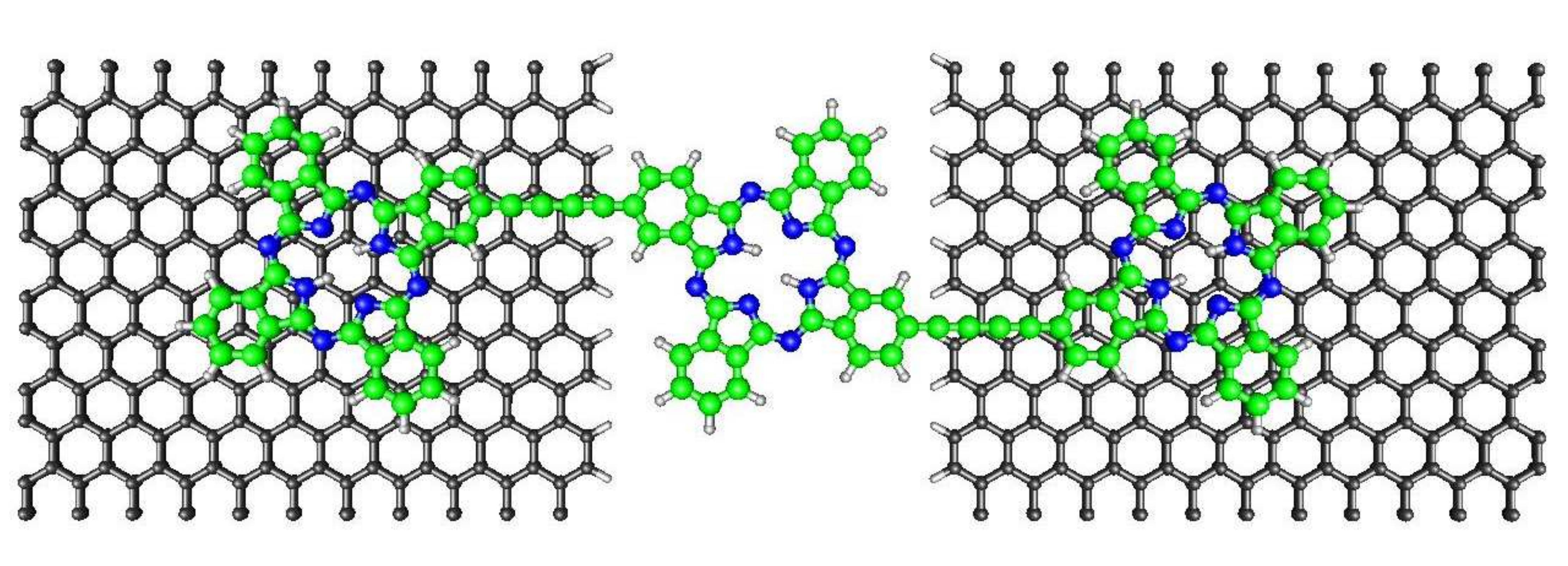}
\caption{(Color online) A junction showing a molecule made of three phthalocyanine units connected via butadiyne linkers
placed on top of two graphene electrodes separated by a physical gap of length 17.265 \AA. The molecule is placed
3.4 \AA above the graphene sheets. }
\label{fig:graphene-junction}
\end{figure}

\subsection{Simulation of a graphene-based junction using a van der Waals Density Functional}
GOLLUM  can profit from the improved chemical accuracy
delivered by the most advanced density functionals. As an example, we discuss here the transport properties of the junction shown
in Fig. (\ref{fig:graphene-junction}), where a single phthalocyanine trimer molecule bridges two graphene electrodes
separated by a physical gap of length 17.265 \,\AA.
These graphene sheets are armchair-terminated and passivated by hydrogen atoms. Periodic boundary conditions are
applied in the two directions across the graphene plane. The phthalocyanine units are linked
by butadiyne chains. The planar anchors couple to the graphene via interaction of the $\pi$-clouds and therefore
an accurate description of the chemical bonding and transport properties can only be achieved by the use of a
van der Waals density functional. We use here the implementation of Dion et al. in the SIESTA program \cite{Dio04,Rom09}.
We have computed the Hamiltonian
and overlap matrix elements using a double-zeta basis set for all the elements in the simulation, together with
a grid fineness of 200 Rydberg. By minimizing the energy, we find that the molecule sits at a height of 3.4 \,\AA   \, above
the sheets.

\begin{figure}
\includegraphics[width=\columnwidth]{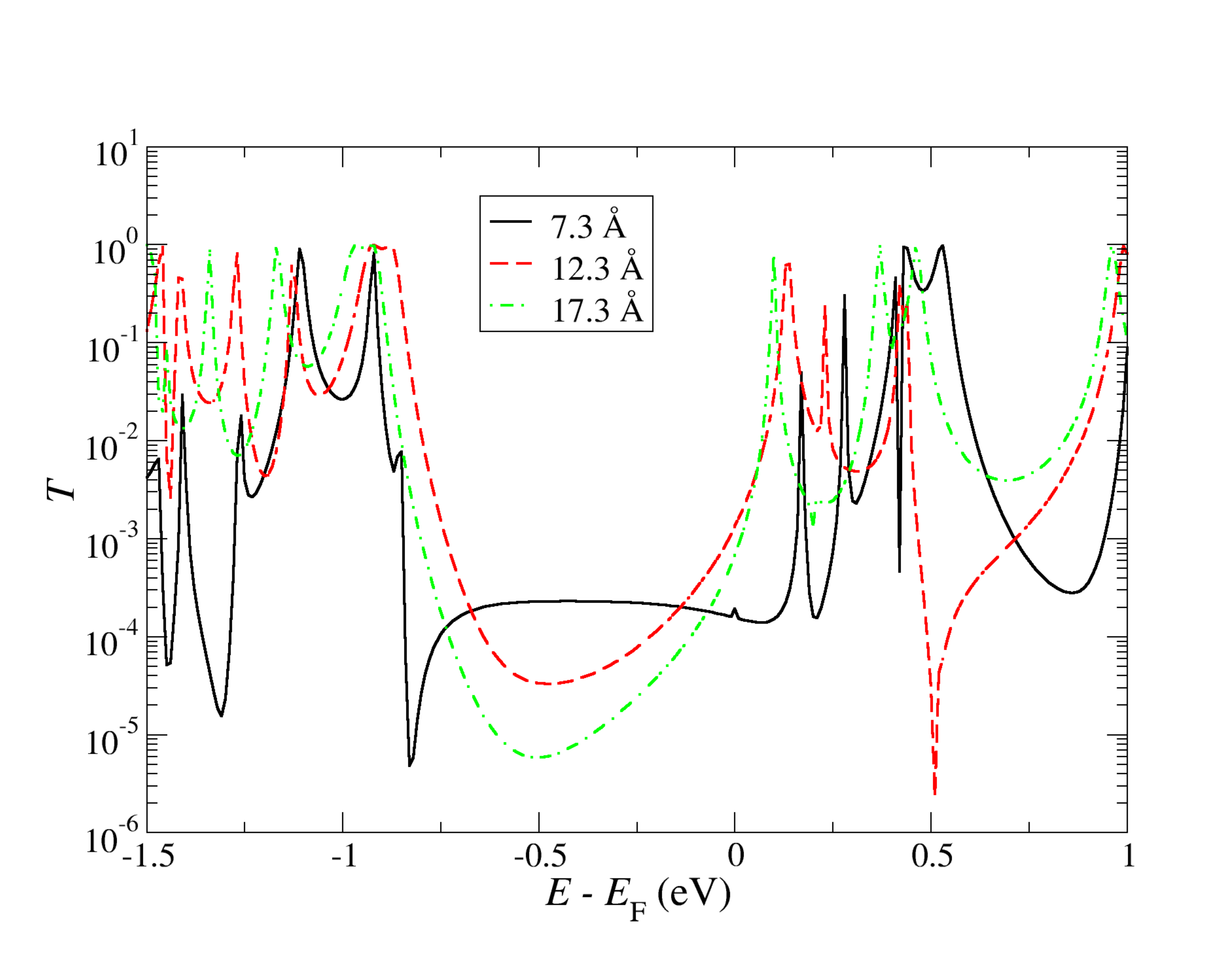}
\caption{(Color online) Transmission curves as a function of energy, referred to the Fermi energy of the Scattering Region.
The different curves correspond to different electrode separations. The gap length is changed by removing or 
adding carbon layers.}
\label{fig:graphene-gap}
\end{figure}

We have studied the impact of the length of the electrode gap on the transport properties of the junction
by attaching additional armchair layers to the edges of both sheets; these layers are made of two carbon rows,
and have a width of 2.502 \AA.
The transmission coefficients for several gap
widths are shown in Fig. (\ref{fig:graphene-gap}). We find several Breit-Wigner resonances associated with molecular
levels of the trimer. Remarkably, these do not shift in energy as the gap width varies\cite{victor-prb-2013}. However, deep dips appear
for several gap widths. These are associated with interference among the different paths whereby electrons can propagate
between the molecule and the electrodes.

\begin{figure}
\includegraphics[trim= 0mm 80mm 0mm 90mm, width=\columnwidth]{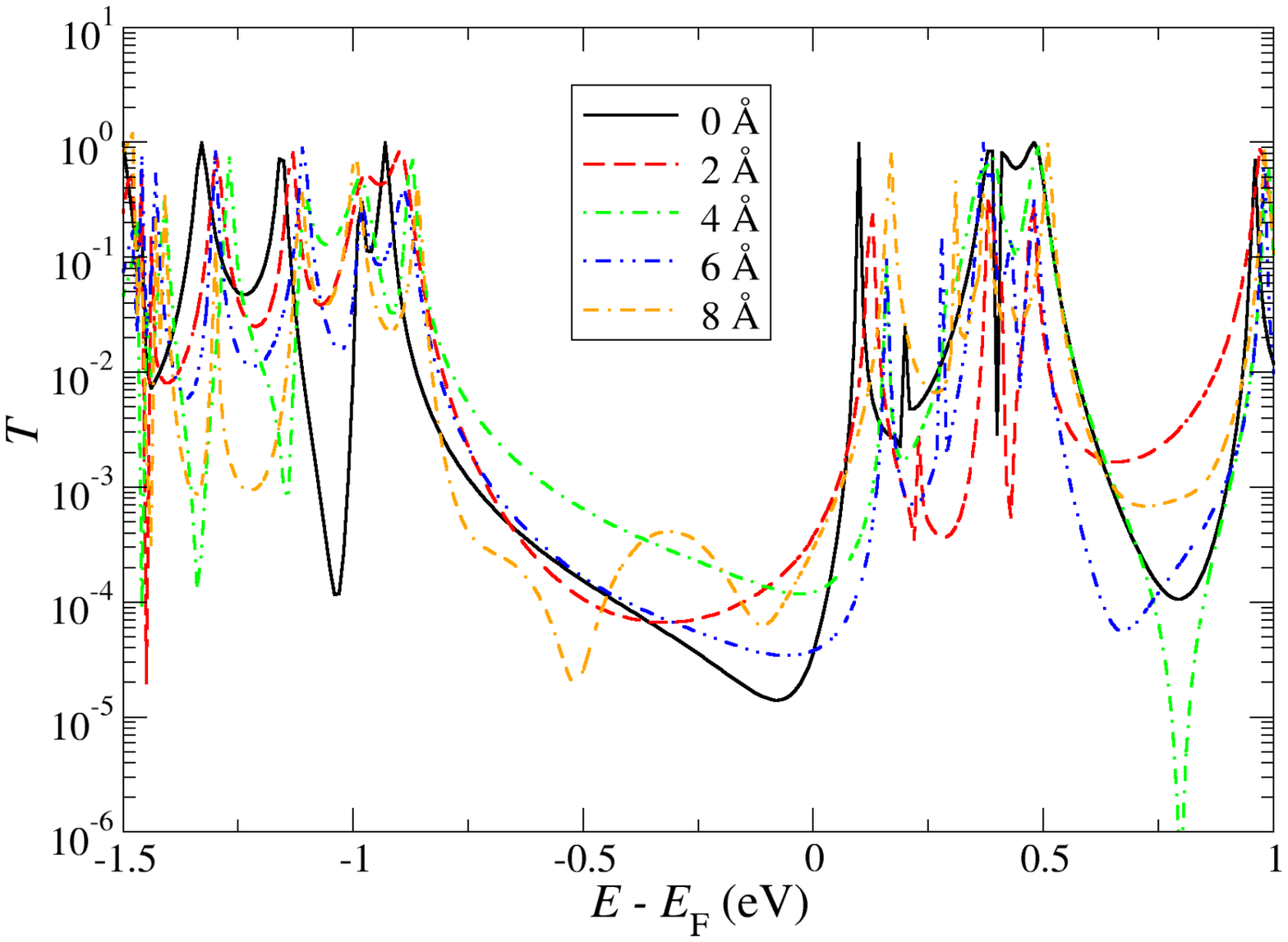}
\caption{(Color online) Transmission coefficient as a function of energy referred to the Fermi energy of the Scattering region.
The different curves correspond to different longitudinal displacements of the molecule referred to the position shown in 
Fig. (\ref{fig:graphene-gap}). The physical gap width is 14.763 \AA. }
\label{fig:graphene-y}
\end{figure}

We have also studied the change in the transmission curves as the molecule is displaced laterally and longitudinally
across the physical gap. We show representative examples of the transmission curves for longitudinal 
displacements in Fig. (\ref{fig:graphene-y}).
The figures show that the energy positions of the molecular Breit-Wigner resonances remain almost constant.
We have found the same behavior for other graphene-based junctions: the energy position of the Breit-Wigner
resonances for a given graphene-based junction does not depend on the molecule position relative to the physical
gap, provided that the bonding mechanism is by physisorption. This universality arises because physisorption
carries no charge transfer between the molecule and the sheets. Furthermore the electrodes are made from the same material and therefore there is no dipole moment associated with
the contacts. Finally, the $\pi-\pi$ hybridization between
molecular orbitals and the electrode states is weaker than for the bonds present in most noble-metal/single-atom contacts,
and does not have a large impact on the nature of the molecular orbitals.

\subsection{LDA+U description of gold porphyrin junctions  }

\begin{figure}
\includegraphics[width=\columnwidth]{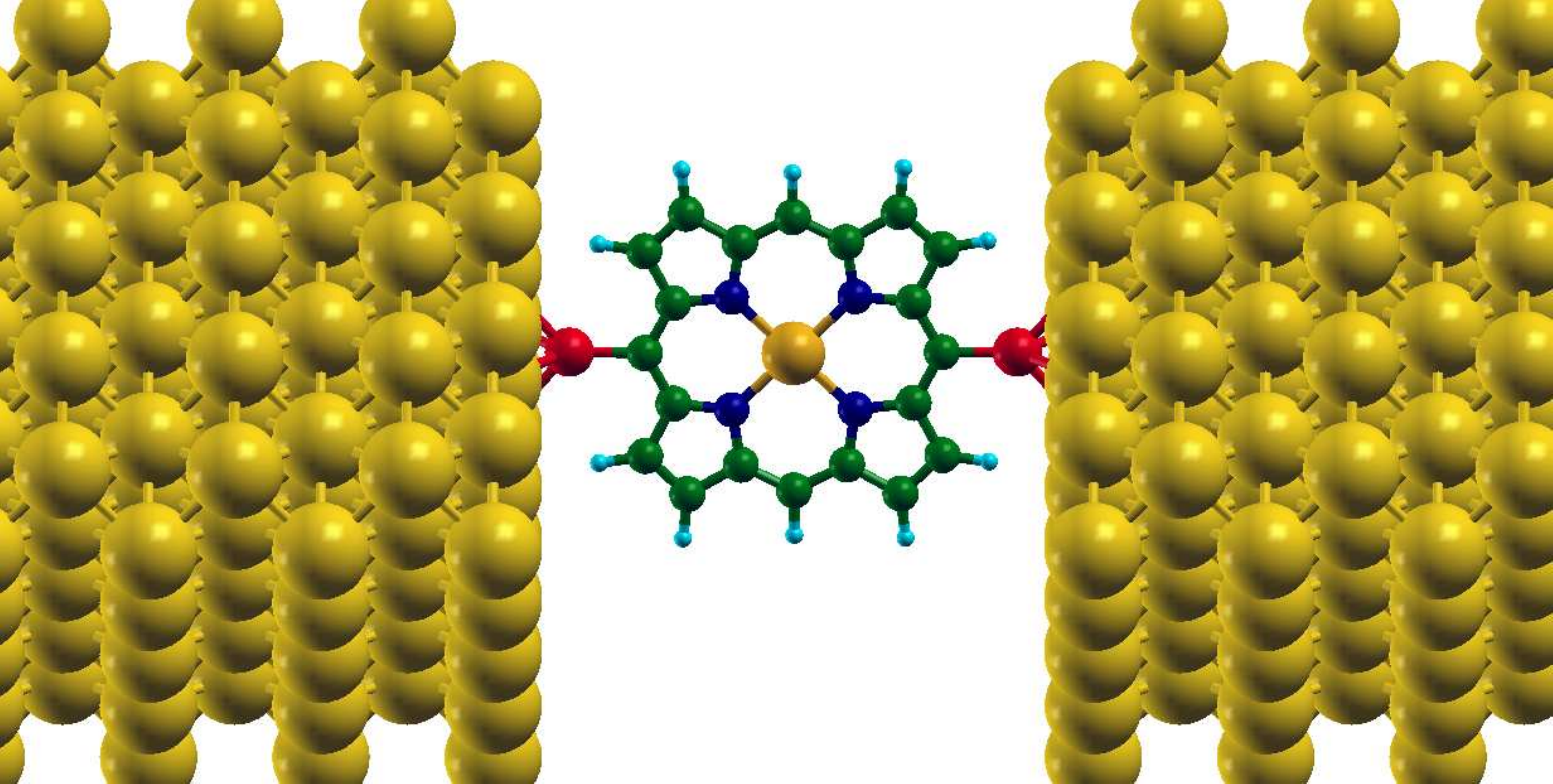}
\caption{(Color online) Schematic view of a metallo-porphyrin molecule sandwiched by gold leads. Yellow, red, cyan, green,
blue and orange represent gold, sulfur, hydrogen, carbon, nitrogen and Co or Cu atoms, respectively.}
\label{fig:ldau-geometry}
\end{figure}

As an example of a GOLLUM calculation using a LDA+U density functional\cite{anisimov91,dudarev98}, we describe in this section 
a case where strong electronic correlations may affect the transport properties of
a nanoscale-scale junction. We discuss the junction shown in Fig. (\ref{fig:ldau-geometry}). Here gold (001) 
electrodes bridge either a porphyrin (P) or a metallo-porphyrin molecule (CuP or CoP). Electron flow 
through any of these three porphyrin molecules is carried by molecular orbitals that hybridize strongly with the 
gold s-orbitals. This gives rise to broad Breit-Wigner resonances
in the transmission coefficients that are identical for the three molecules. However, for the CuP and CoP junctions,
additional electron paths are created whereby electrons hop into and off the localized d-orbitals of the transition
metal atom. The interference between direct and d-orbital-mediated paths creates sharp Fano resonances that can however
be masked by the much wider Breit-Wigner resonances\cite{Ruben-JPCM-2013}. We see below how including strong correlations
in the d-orbitals of the Co and Cu atoms in terms of a LDA+U approach produces strong shifts in the energy dependence of those
resonances.

We have computed the Hamiltonian using the SIESTA code, where we have picked a single zeta basis for the gold atoms at the electrodes,
a double-zeta-polarized basis set for all the atoms in the molecule and a GGA functional. We have  included
a $U$ correction term for the d-orbitals of the Cu and Co atoms in a mean-field fashion, in the spirit of the LDA+U approach\cite{anisimov91,dudarev98}. We present here our results for values of $U$ equal to 0, 2.5 and 4.5 eV.

\begin{figure}
\includegraphics[width=\columnwidth]{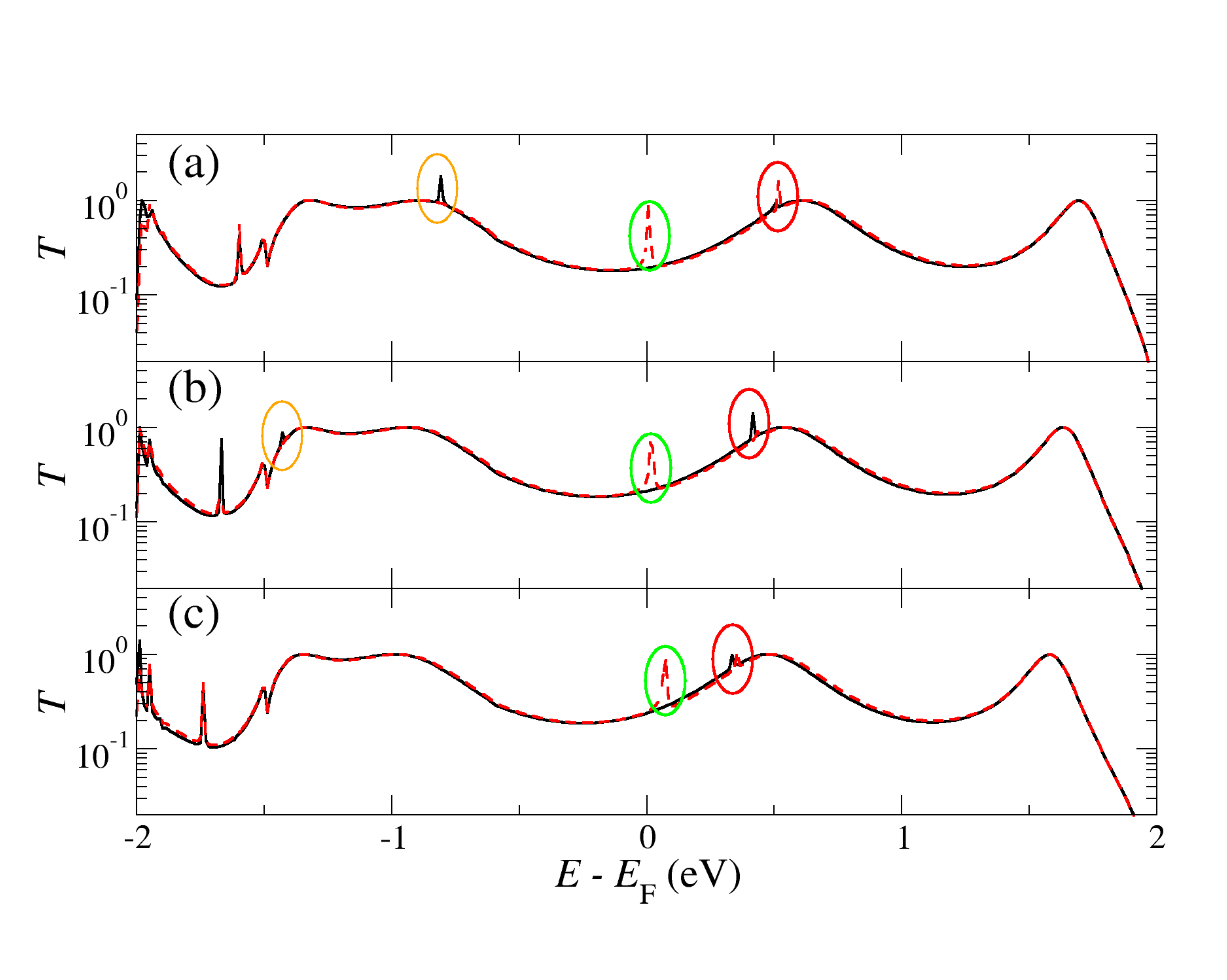}
\caption{(Color online) Transmission coefficient $T(E)$ of the junction shown in Fig. (\ref{fig:ldau-geometry}), 
where the central molecule is CuP. $T$ is plotted as a function of the energy $E$ referred to the Fermi energy $E_F$
of the Scattering Region. The different panels
correspond to the different $U$ corrections added to the DFT Hamiltonian (see text).   The  green  and gold ellipses
encircle masked Fano resonances. They are originated by paths hopping onto the Cu 3d$_{xz}$ or d$_{yz}$ orbitals. 
The red ellipse circles a sharp Breit-Wigner resonance coming from C and N atoms. }
\label{fig:ldau-cup}
\end{figure}

\begin{figure}
\includegraphics[trim= 0mm 20mm 0mm 30mm, width=\columnwidth]{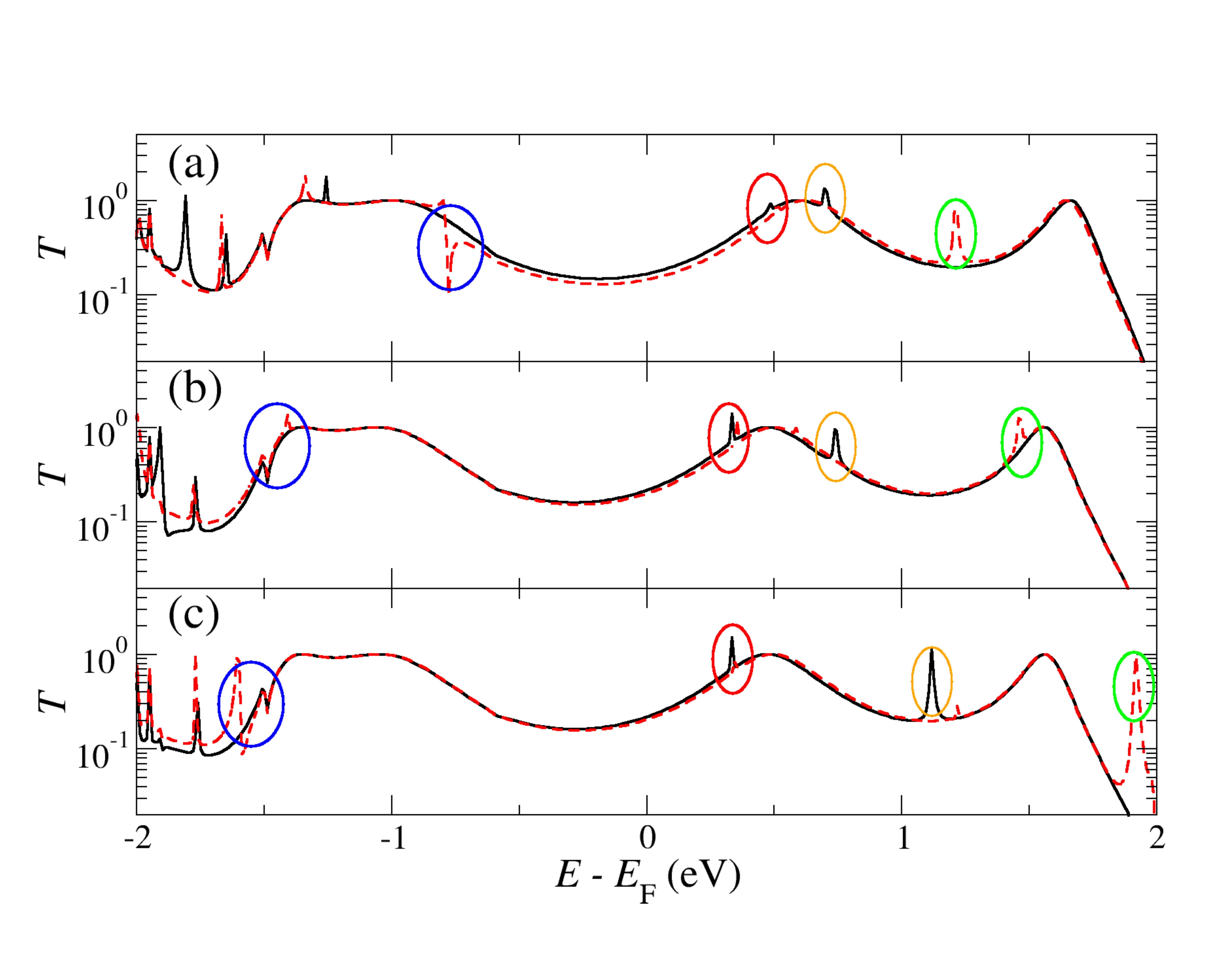}
\caption{(Color online) Transmission coefficient $T(E)$ of the junction shown in Fig. (\ref{fig:ldau-geometry}), 
where the central molecule is CoP. $T$ is plotted as a function of the energy $E$ referred to the Fermi energy $E_F$
of the Scattering Region. The different panels
correspond to the different $U$ corrections added to the DFT Hamiltonian (see text). The  green, gold and
blue ellipses encircle Fano resonances. They are originated by paths whereby electrons hop onto the  Co 3d$_xz$ orbital
(green ellipses) and from a Molecular Orbital composed by the Co 3d$_xy$ and 3d$_yz$ orbitals (blue ellipse).
The red ellipse encircles a sharp Breit-Wigner resonance coming from C and N atoms. }
\label{fig:ldau-cop}
\end{figure}

We find that the transmission coefficients of the three molecules display the same wide Breit-Wigner resonances, that
correspond to molecular orbitals hybridizing strongly with the electrodes. These are shown in Figs. (\ref{fig:ldau-cup})
and  (\ref{fig:ldau-cop}) for CuP and CoP respectively.
In addition, the three molecules show a sharp Breit-Wigner resonance that is marked by a red ellipse in the figures.
This resonance corresponds to a molecular orbital encompassing C and N atoms that is weakly bonded to the electrodes. 
Interestingly, this sharp Breit-Wigner resonance shifts in energy if we change the value of the Coulomb 
interaction $U$ for the CuP and CoP junctions. To understand this phenomenon, we have looked at the
density of states of the junction projected onto each atomic orbital, and the local density of
states integrated in a narrow energy window around the red resonance. We have found that the N- and C-based molecular
orbital hybridizes with the d$_{xy}$ orbital of the copper or cobalt atoms and is therefore affected by the $U$-term.
We have found additional sharp peaks appearing in $T(E)$ for the CuP and CoP junction that do not show up for the
simple porphyrin junction. These are marked by green, blue and gold circles in Figs. (\ref{fig:ldau-cup}) and
(\ref{fig:ldau-cop}). These seem to be sharp Breit-Wigner resonances also.
However, we have demonstrated\cite{Ruben-JPCM-2013} that they are actually Fano resonances, where the Fano dip is masked by
the transmission of the neighboring wide Breit-Wigner resonances. By plotting the density of states of the junction
projected in each orbital, we indeed find that they correspond to the copper d$_{xz}$ or d$_{yz}$ orbitals. Because these Fano resonances
are associated with atomic d-orbitals strongly localized in the transition metal atom, we expect that adding a $U$-term will
have a strong impact on their energy position. Fig. (\ref{fig:ldau-cup}) shows how these resonances  indeed shift
in energy as $U$ is increased. Note that one of the Fano resonances coming from the d$_{xz}$ copper orbital
is strongly pinned to the Fermi energy, while other resonances rapidly move down in energy. 

\subsection{Fixing the transport properties of OPE molecular junctions via the SAINT method}
\begin{figure}
\includegraphics[width=\columnwidth]{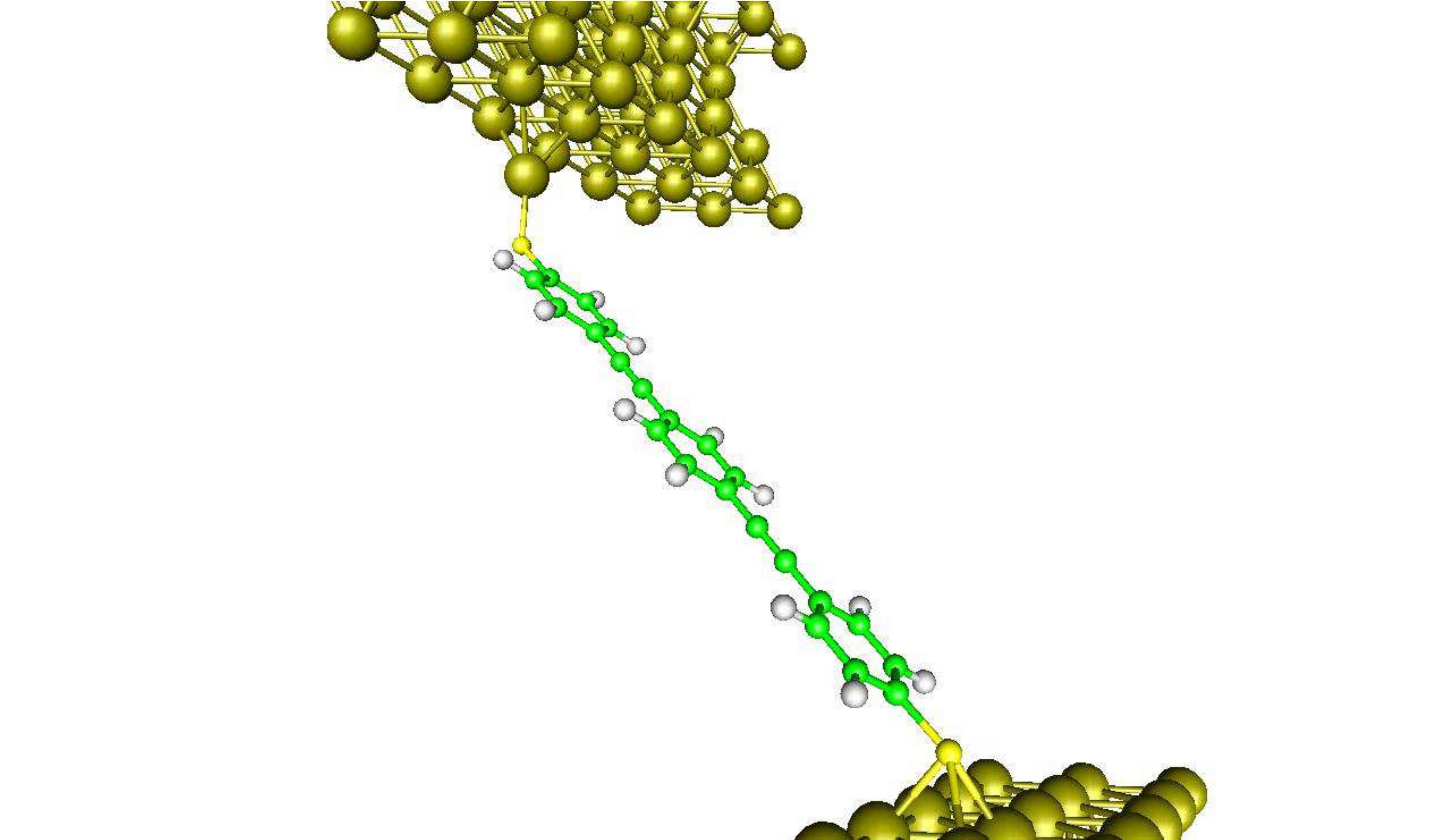}
\caption{\label{Tricene45}(Color online) A (111) gold junction sandwiching a tricene-dithiol molecule.
The molecule is coupled to one ad-atom on one side and to a hollow gold site at the other end, and its 
orientation is titled with respect to the electrodes' normal line.}
\end{figure}

We analyze in this section the transport properties of a series (111) gold junctions that are bridged by 
OPE derivatives. The backbone of these molecules has a varying number of rings ranging from one to three. 
The molecules may be oriented fully perpendicular to the electrodes surfaces, or making a tilting angle, as
we show in Fig. (\ref{Tricene45})
It is well established by now \cite{Ferrer09,Wandlowsky} that gold junctions that contain conjugated
thiol-terminated molecules like OPEs have a larger conductance when the molecule is tilted.
This is due to the increased overlap of the $p_z$ states of the sulfur atoms when the
angle between the molecule and the normal to the surface increases. We show in this section that a 
plain DFT-based calculation predicts that  the largest conductance occurs when the molecule is oriented
perpendicular to the electrodes. This deficiency is remedied by the use of the SAINT method.
This method is an efficient semi-empirical correction that allows us to obtain quantitative agreement between
DFT calculations and experiments\cite{Neaton06,Quek09,Mowbray08,Cehovin08,Wandlowsky}, as we have already stressed
in Section II.

The Hamiltonian of the junction has been obtained with the code SIESTA and a Local Density approximation
(LDA) functional\cite{lda}.  We  have picked a single-zeta basis for the gold
atoms of the electrodes, and a double-zeta-polarized basis for the atoms in the molecule. The PLs of
the electrodes  contain three atomic layers, each having $6\times 3$ atoms. We have chosen
junction geometries where the molecular derivatives are oriented either perpendicularly to the gold surfaces 
or making a 45 degrees angle, as shown in Fig. (\ref{Tricene45}).
Due to this tilting angle, we had to use a non-periodic Scattering Region, as in Fig. (5).
The scattering regions consisted therefore of the molecule, the two surfaces containing two atomic layers 
each and 3 PLs on each branch followed by vacuum. We chose PL2 as the TPL, so all Hamiltonian matrix elements of PL3 
were chopped off.

\begin{figure}
\includegraphics[width=\columnwidth]{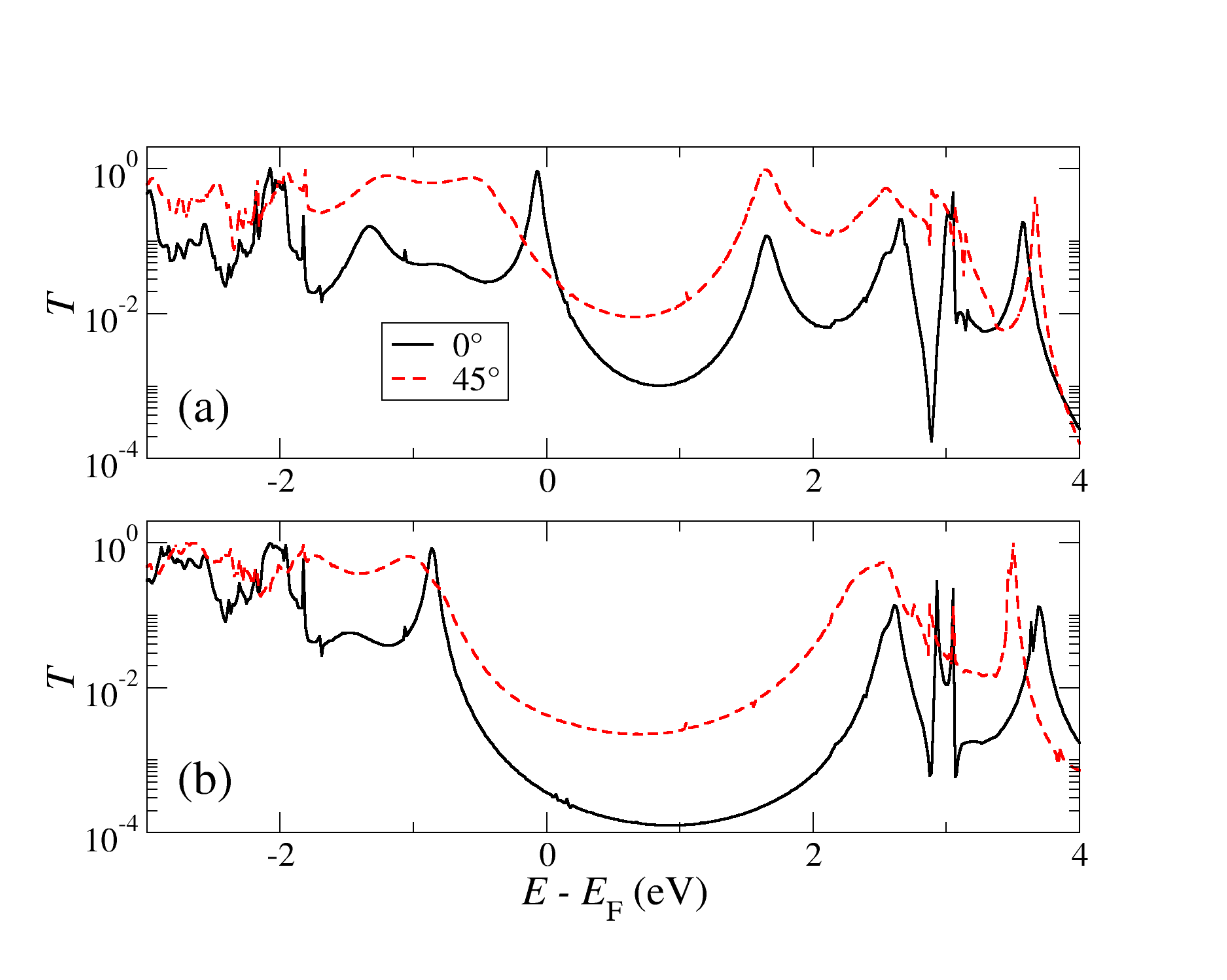}
\caption{\label{T_SAINT_1}(Color online) Transmission of the junction shown in Fig. (\ref{Tricene45}), 
containing an OPE molecule with 3 rings. $T(E)$ has been calculated (a) without and (b) with SAINT corrections. 
Continuous and dashed lines correspond to perpendicular and tilted (45 degrees) molecules, respectively.}
\end{figure}

\begin{figure}
\includegraphics[width=\columnwidth]{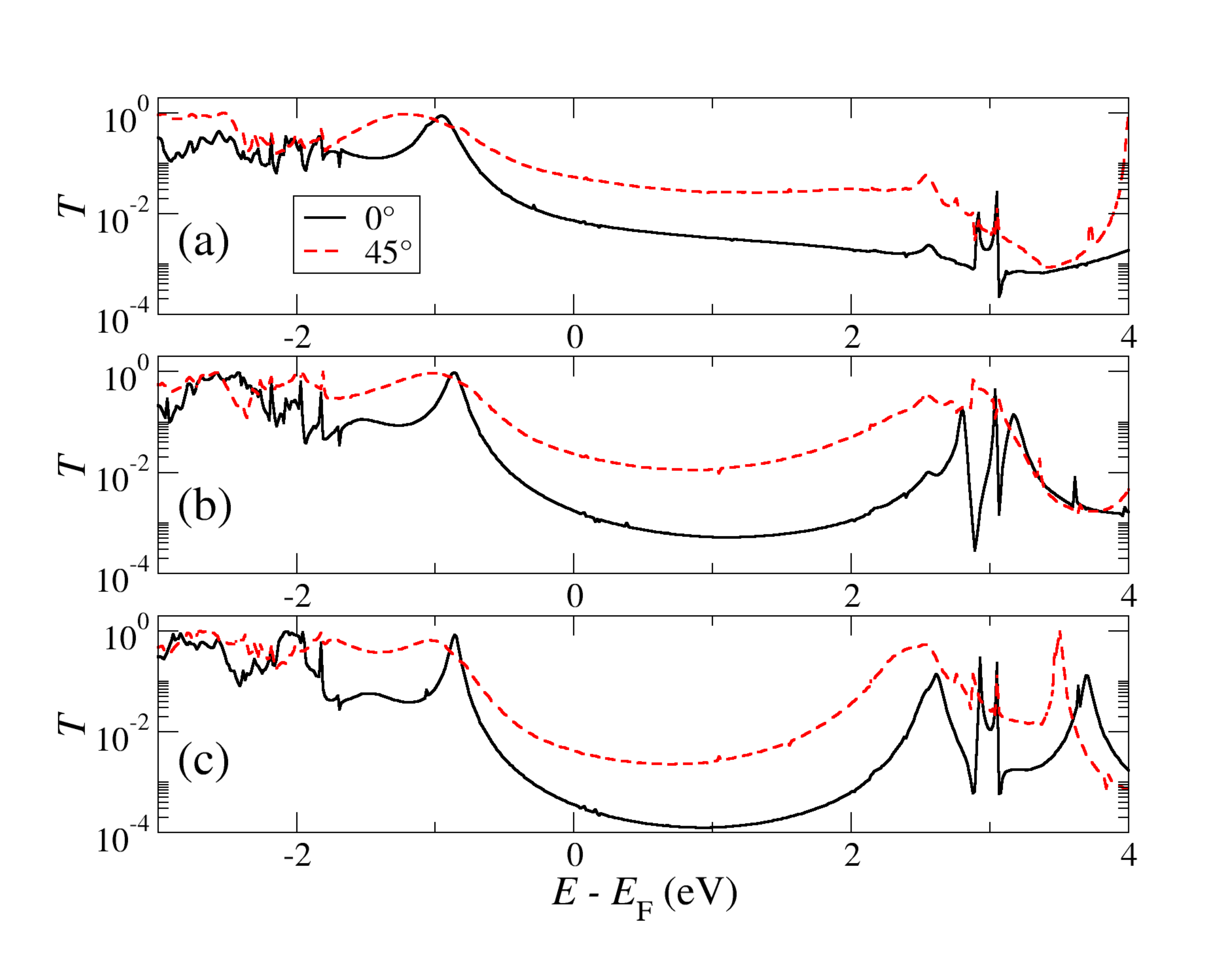}
\caption{\label{T_SAINT_2}(Color online) Transmission of the junction shown in Fig. (\ref{Tricene45}), 
containing an OPE molecule with (a) 1, (b) 2 and (c) 3 rings. $T(E)$ has been calculated with SAINT corrections. 
Continuous and dashed lines correspond to perpendicular and tilted (45 degrees) molecules, respectively.}
\end{figure}

\begin{figure}
\includegraphics[trim = 10mm 10mm 40mm 160mm, width=\columnwidth]{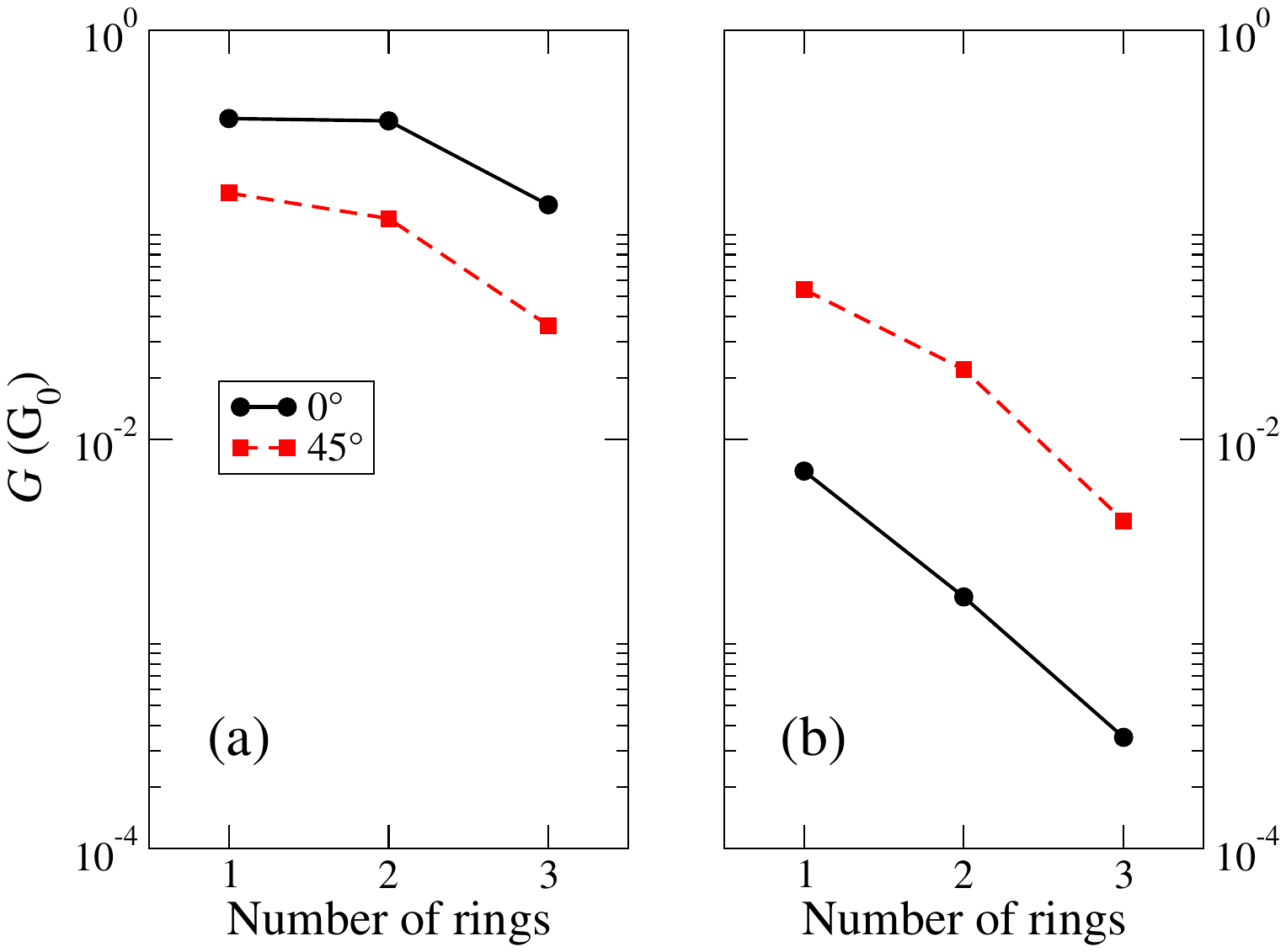}
\caption{\label{G.SAINT}(Color online) Conductance of OPE molecules with a number of rings
between 1 and 3, calculated without (a) and with (b) SAINT corrections. Circles and
squares correspond to perpendicular and tilted (45 degrees) molecules, respectively.}
\end{figure}

We have computed the  transmission curves $T(E)$ for the  referred OPE derivatives using conventional DFT,
and have found for all of them that the conductance (computed from $T(E_F)$) is larger if the molecule is oriented
perpendicular to the electrodes. The upper panel in Fig. (\ref{T_SAINT_1}) demonstrates this behavior for an OPE containing
three rings. The figure shows that the higher conductance is originated by the position of the HOMO level of the molecule,
that is placed only slightly below the Fermi energy of the Scattering Region. In contrast, the HOMO level of the tilted
molecule is shifted farther away from $E_F$. This situation demands for the use of the SAINT correction scheme, that will
reposition the molecular orbital levels at their correct energies. We show in the bottom panel of the figure that this
is indeed the case, and that by the use of the SAINT scheme, the correct experimental trend is recovered, where tilted molecules
show larger conductances. We have verified that the same change happens for OPE molecules containing one and two rings. 
Finally, we plot in Fig. (\ref{T_SAINT_2}) $T(E)$ for the three molecules (containing one, two and three rings) for 
perpendicular and tilted  orientations.
(\ref{G.SAINT}) we
show the transmission of the OPE derivatives with a number of rings between 1 and 3 and two
tilting angles, 0 and 45 degrees. The figures show that all those junctions where the derivative
is oriented perpendicularly to the gold surface (defined here to be 0 degrees) show a larger transmission
at the Fermi level than the tilted cases in contrast with our expectations discussed above \cite{Bra03}.

\begin{table}
\caption{\label{TabSAINT.2} Corrections entering the SAINT scheme for occupied and 
unoccupied levels given in eV.}
\begin{ruledtabular}
\begin{tabular}{ccc}
$\#$ of Rings-tilt angle&$\Delta_\mathrm{o}$&$\Delta_\mathrm{u}$\\
\hline
1-0$^\circ$&-1.7&1.8\\
1-45$^\circ$&-1.5&1.6\\
2-0$^\circ$&-1.3&1.3\\
2-45$^\circ$&-1.2&1.2\\
3-0$^\circ$&-1.2&1.3\\
3-45$^\circ$&-1.1&1.2\\
\end{tabular}
\end{ruledtabular}
\end{table}

The physical mechanism whereby the conductance of the tilted configuration is higher, is due to the higher
hybridization between the molecular orbitals and the electrodes in the tilted configuration. This 
increases the width of the transmission resonances and therefore decreases the effect of opening the gap 
with the SAINT correction. On the other hand, the image charge correction is also larger in the tilted 
configuration, since the molecule is closer to the surfaces, and therefore the reduction in the opening of the
gap is also larger, which means the final gap ends up smaller in the tilted configuration. The
conductance of each case is summarized in Fig. (\ref{G.SAINT}). Notice that the SAINT
correction scheme changes qualitatively the physical picture in this junction.

Our procedure for the SAINT correction scheme is as follows. We first calculate the ionization potential
(IP) and electron affinity (EA) of the molecules in the gas phase. These gas phase corrections 
open the DFT HOMO-LUMO gap. However, these bare shifts need to be corrected because of image charge effects.
The final values for the correction shifts are summarized in table (\ref{TabSAINT.2}). Notice that the 
corrections are very similar in magnitude and have opposite signs\cite{Neaton06}.

\subsection{Kondo and Coulomb blockade effects\label{section_kondo}}

\begin{figure}
\includegraphics[trim = 0mm 80mm 0mm 80mm, width=\columnwidth]{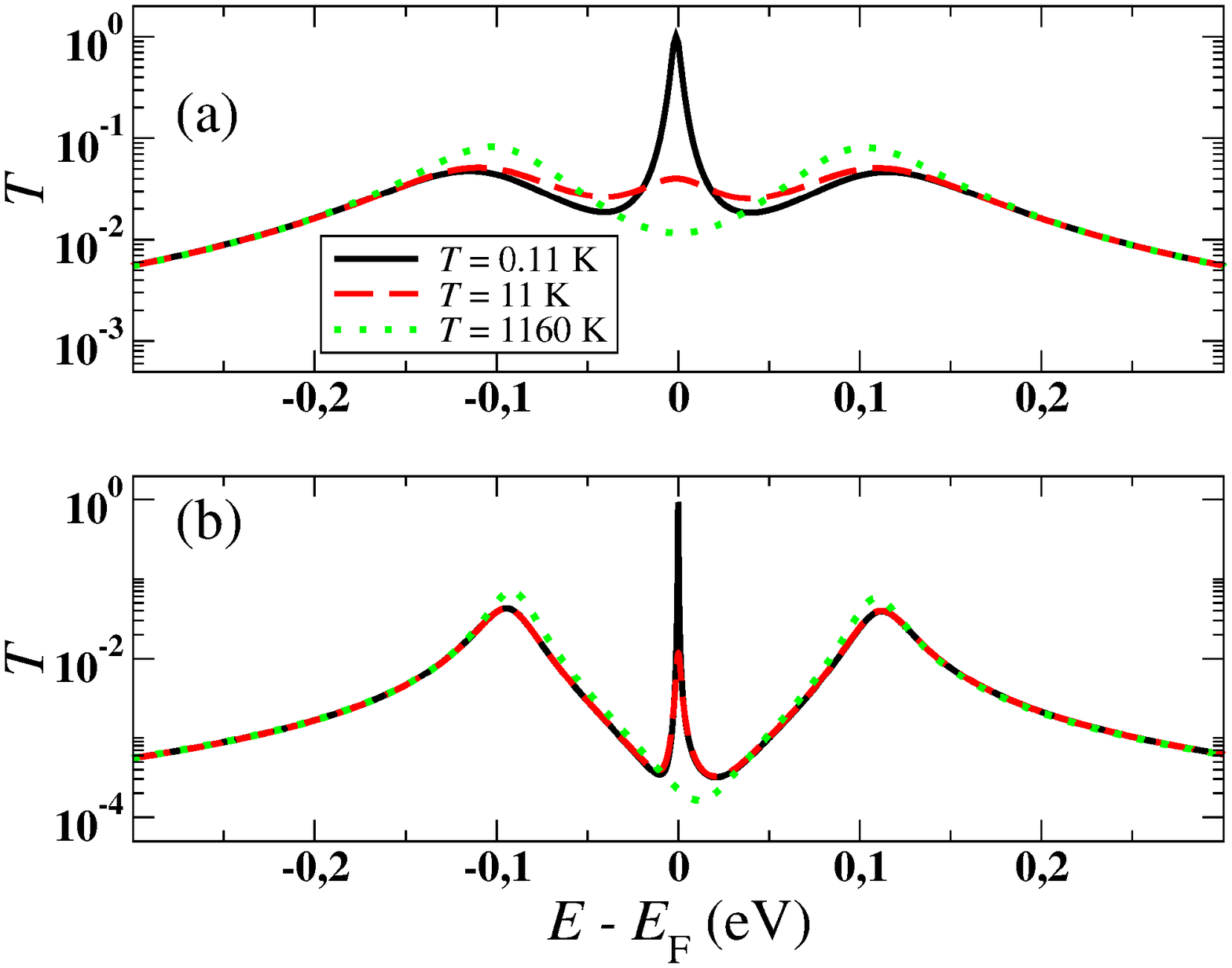}
\caption{(Color online) Zero-voltage transmission coefficients $T(E)$ of (a) a single-level symmetric Anderson model with
input parameters defined in the main text; (b) a hydrogen atom sandwiched by (001) gold electrodes. The panels show
curves computed at several temperatures to reflect the Kondo and the Coulomb blockade regimes.}
\label{fig:anderson-T}
\end{figure}

As noted above, GOLLUM has a simple and flexible input data structure so that model Hamiltonians can be utilized easily.
As a simple example of a simulation exhibiting Kondo and Coulomb blockade behavior, we show here results obtained from 
GOLLUM for a tight-binding single-level Anderson model coupled  to two semi-infinite chains, that corresponds to 
taking $M=2$ in Eq.
(\ref{eq:anderson-hamiltonian}). Due to the coupling to the  two leads, the correlated level  acquires a finite bandwidth
\begin{equation}
\Gamma=2\,\pi \,V^2 \,\rho_L=\frac{ V^2}{t}
 \end{equation}
where $\rho_L$ is the density of states in the leads.  For vanishing $\Gamma$ the model is in the so-called atomic limit
which is characterized by sharp peaks  in the $d$-level density of states $\rho_d$ at $\epsilon_d$ and $\epsilon_d + U$.
This limit corresponds to the Coulomb blockade regime in an actual junction where the conductance is strongly suppressed
except at the charge degeneracy points. However, when the coupling to the leads increases ($\Gamma$ becomes larger
than the temperature $T$, but is still smaller than $U$), virtual processes allow the charge and spin in the molecule
to fluctuate and a resonance close to the Fermi energy appears due to the Kondo effect. This simple model therefore
captures relevant physics of molecular junctions such as the appearance of the Coulomb blockade effect,
and the crossover from the Coulomb blockade to the Kondo regime as the temperature is lowered below the Kondo
temperature
\begin{equation}
 T_K=\sqrt{\frac{U\,\Gamma}{2}}\,\mathrm e^{-\frac{\pi\,|\epsilon_d\,(\epsilon_d+U)|}{2\,U\,\Gamma}}\sim
 \sqrt{\frac{U\,\Gamma}{2}}\,\mathrm e^{-\frac{\pi\,U}{8\,\Gamma}}
 \end{equation}
where the last expression holds in the so-called symmetric limit $\epsilon_d+U/2\sim 0$.

\begin{figure}
\includegraphics[width=\columnwidth]{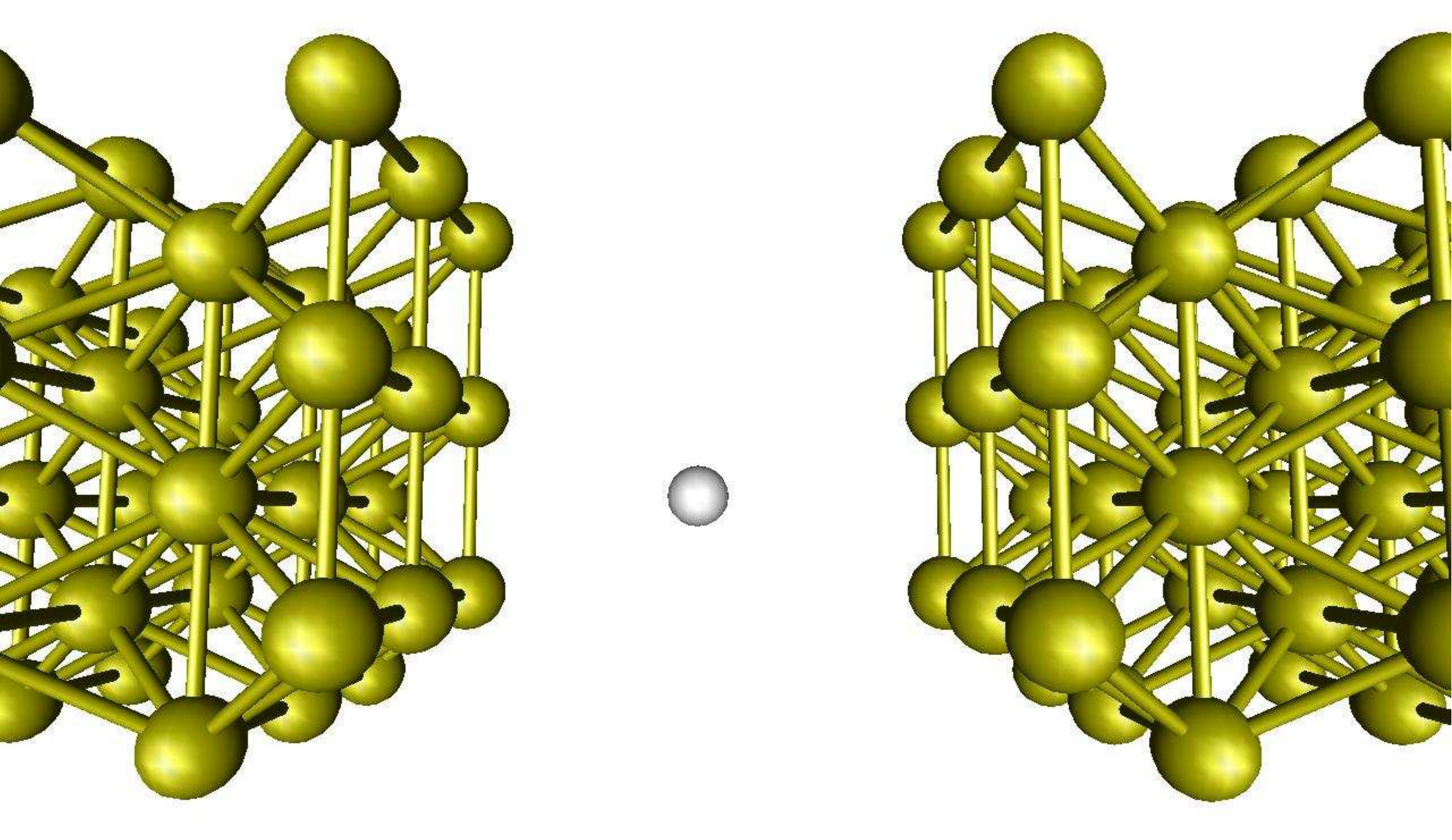}
\caption{A hydrogen atom bridging (001) gold leads. The separation between H and each gold lead is 3.8 \AA.}
\label{fig:anderson-molecule}
\end{figure}

The interpolative impurity solver implemented in GOLLUM provides a good quantitative description of the above phenomena in 
the weak coupling regime and
is also qualitatively correct in the intermediate and strong coupling regimes. It however overestimates the width of the
Kondo resonance. We show in Fig.
\ref{fig:anderson-T} (a) the zero-voltage transmission curve $T(E)$ computed with GOLLUM, and using the single-level Anderson Hamiltonian
(\ref{eq:anderson-hamiltonian}) in the symmetric limit. We take the following parameters: $t=1$ eV, $U=0.2\, t$ and $V=0.1\, t$ so that
$\Gamma= 0.01\, t$ and $\pi U/8\Gamma \sim 8$, placing the junction in the strong correlation regime. These parameters yield a Kondo
temperature $T_K \approx 1.2\times 10^{-5} \,t\approx0.15\, K$.
The figure shows that the interpolative solution provides a transmission curve featuring the lower and upper Hubbard bands placed at their
correct position and having the right width, together with a sharp Kondo resonance at low temperatures which progressively smoothens
and eventually disappears as the temperature is raised. However, the interpolative solution provides a Kondo temperature
$T_K^\mathrm{int}\sim 10 \,K$, e.g.: two orders of magnitude larger than the exact one.

We now show the results obtained from GOLLUM for a similar junction, shown in Fig. (\ref{fig:anderson-molecule}), where a hydrogen atom bridges
two gold (001) electrodes. In this case,  the input Hamiltonian is generated by the DFT
code SIESTA and the leads are repeated periodically in the plane perpendicular to the transport direction, using
PL unit cells in AB stacking and $3\times 3$ atoms in each atomic layer. We have used a single-zeta basis set and a generalized
gradient approximation functional. We have adjusted the distance $d$ between the hydrogen atom and the leads to reproduce a coupling
similar to that set for the above Anderson model which is achieved with $d=2.8$ \AA. The generated transmission curve
is shown in Fig. \ref{fig:anderson-T} (b) for the same three temperatures used for the Anderson Hamiltonian. We find that the shape
of the transmission curves remains qualitatively the same. However, both the lower and upper Hubbard bands and the Kondo resonance
are now sharper at their tips.

\begin{figure}
\includegraphics[trim = 0mm 70mm 0mm 70mm, width=\columnwidth]{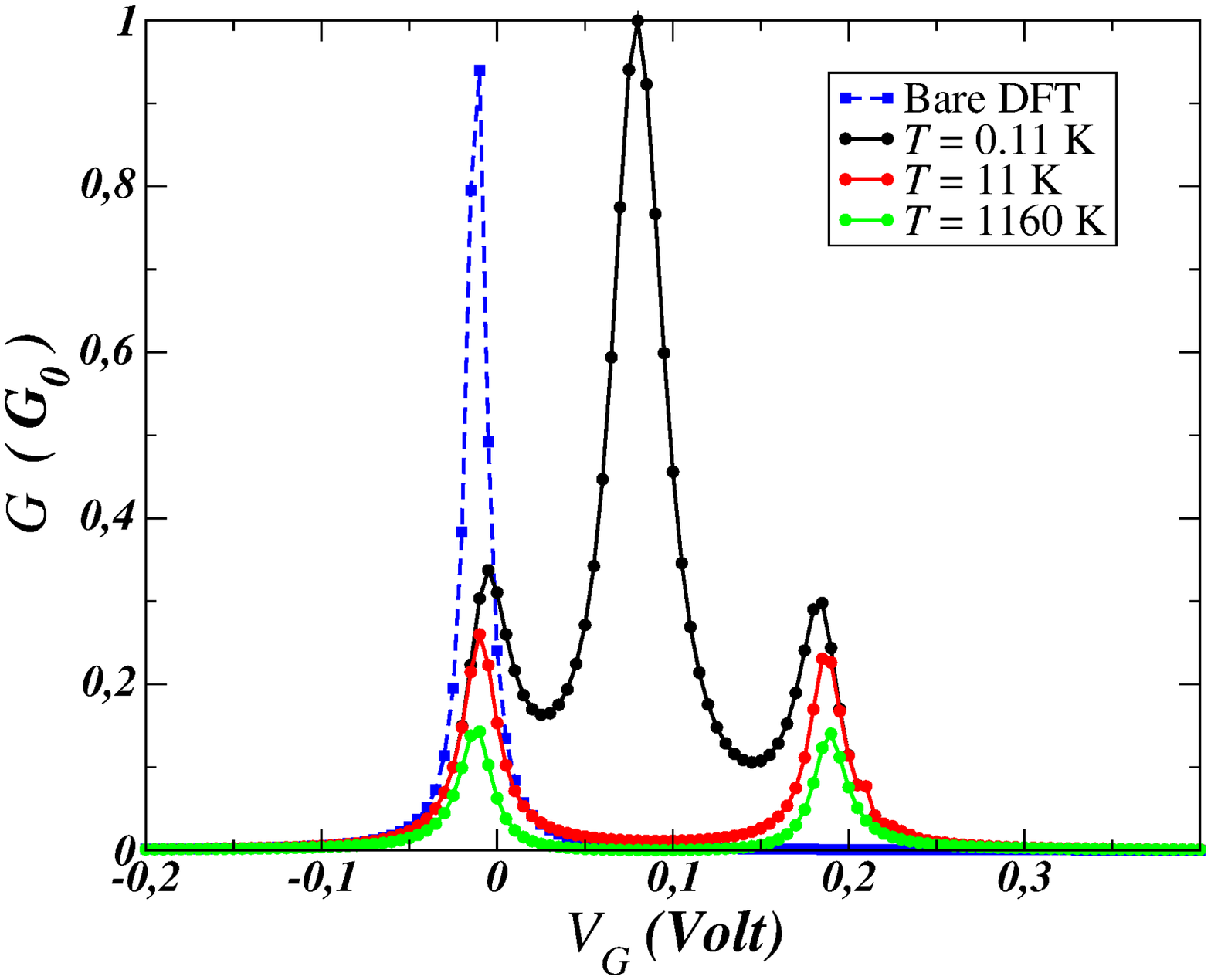}
\caption{Conductance as a function of gate voltage and temperature for a (001) gold junction bridged by a single 
hydrogen atom.}
\label{fig:anderson-gate}
\end{figure}

Using GOLLUM, we subsequently apply a gate voltage $V_g$ to the gold-hydrogen-gold junction and compute the low-voltage conductance 
$G$ as a function of $V_g$ to compare the results using plain DFT versus DFT in combination with the interpolative method. 
The results are shown in Fig. (\ref{fig:anderson-gate}).
We have not included the double-counting term to make the comparison between both approaches more explicit. The figure nicely shows how
the interpolative method splits the single
DFT peak into two Coulomb blockade peaks and also a Kondo peak. The figure also shows how the Kondo peak disappears at temperatures
above $T_K^\mathrm{int}$ leaving only the Coulomb blockage features.

\begin{figure*}
\includegraphics[width=1.\textwidth]{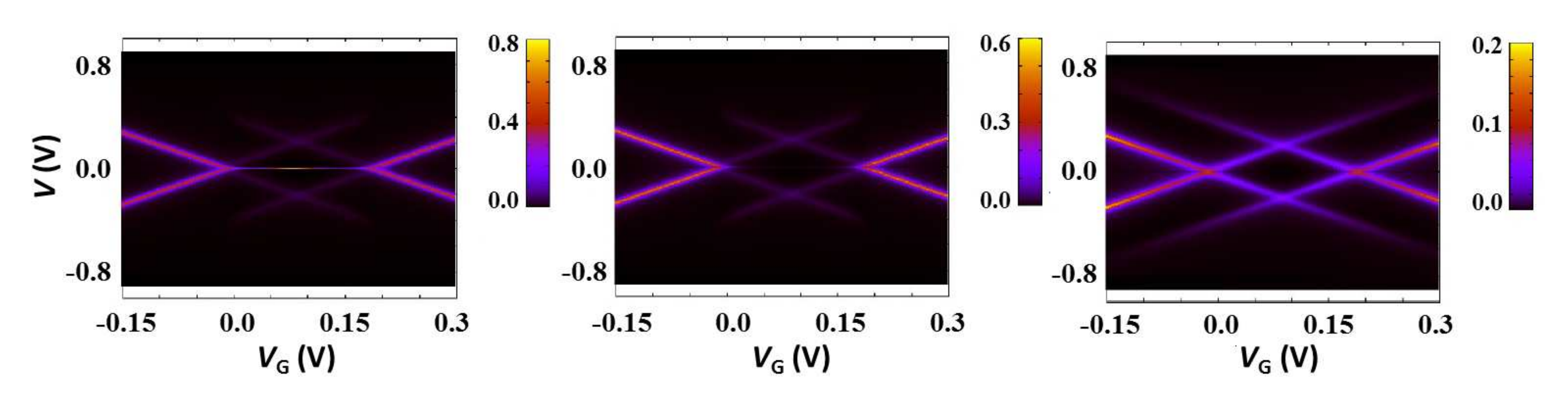}
\caption{Density-contour plots of the low-voltage conductance $G$ in units of $G_0$ 
of the gold-hydrogen-gold junction shown in Fig. (\ref{fig:anderson-molecule}). $G$ is plotted as a function of the bias $V$
in the vertical axis and the gate voltage $V_g$ in the horizontal axis. The conductance is plotted at a temperature 
(left panel) $T=0.11$ K; (middle panel) $T=11$ K; and (right panel) $T=1160$ K.}
\label{fig:anderson-diamond}
\end{figure*}

Finally, we subject the gold-hydrogen-gold junction to the combined effect of finite bias $V$ and gate $V_g$ voltages.
Fig. (\ref{fig:anderson-diamond}) shows density-contour plots of the low-voltage conductance as a function of $V$ and $V_g$.
This figure demonstrates that GOLLUM can nicely reproduce Coulomb blockade diamonds, as well as the Kondo line, that 
disappears as the temperature is raised above $T_K^\mathrm{int}$.

\subsection{A junction displaying NDR behavior: a comparison between GOLLUM and SMEAGOL.}
\begin{figure}
\includegraphics[width=\columnwidth]{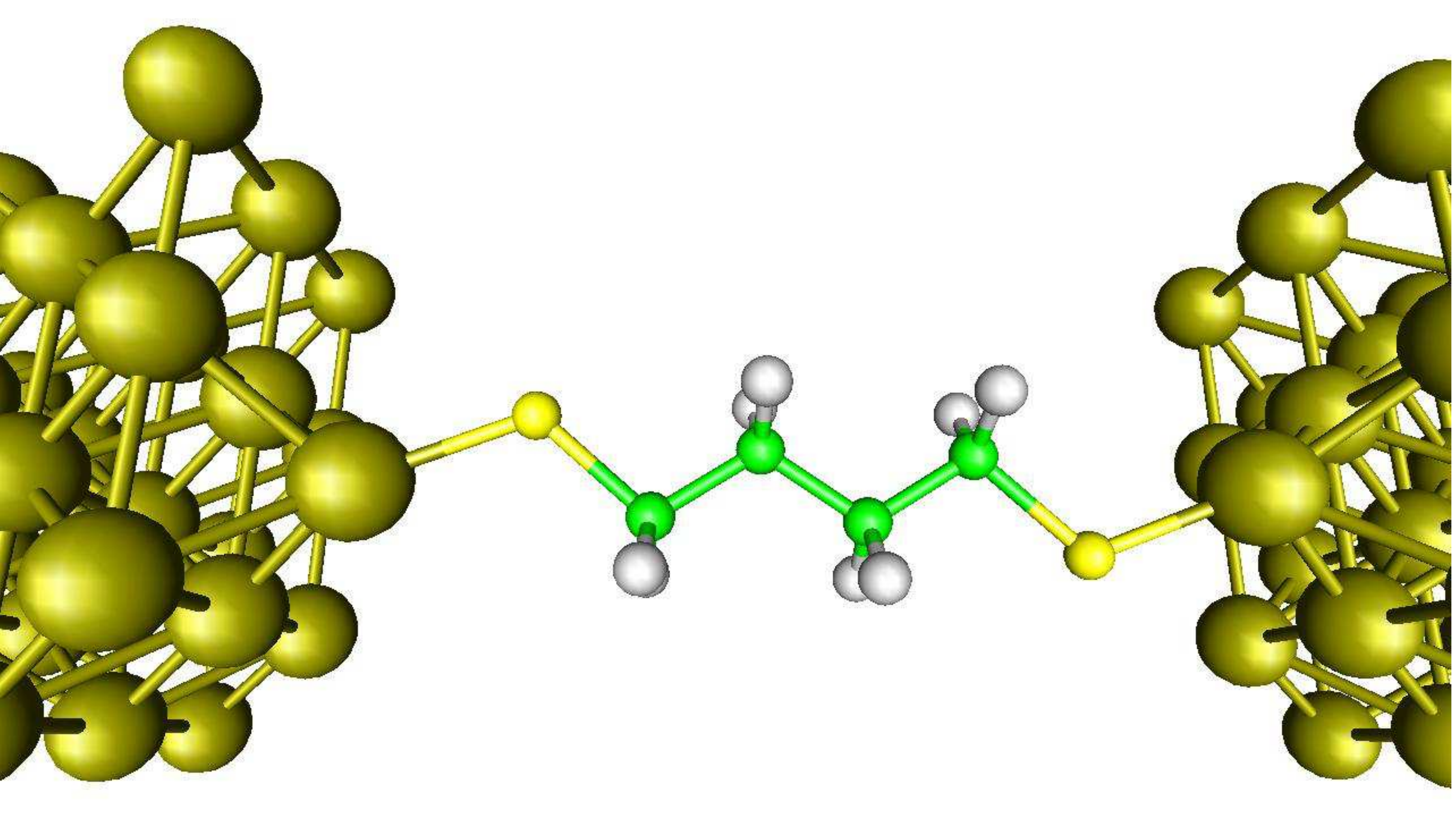}
\caption{\label{Alkane4}(Color online) A junction where a Butane-dithiol molecule is sandwiched by (001) gold electrodes
and subjected to a finite bias potential.}
\end{figure}

We demonstrate with two examples how GOLLUM incorporates finite-voltage effects. In the first, we have computed the 
current-voltage characteristics of a gold (001) junction sandwiching an alkane molecule, that we show in Fig. (\ref{Alkane4}).
We have computed the zero-bias Hamiltonian of the junction using the SIESTA code, with a double-zeta-polarized basis for
all the atoms, and a GGA functional. The PLs contain two atomic layers, with $3\times 3$ atoms each. We have applied 
periodic boundary conditions across the plane perpendicular to the transport direction and have computed $H(k_\perp)$
at the $\Gamma$ point. 

\begin{figure}
 \includegraphics[trim = 20mm 20mm 40mm 160mm, width=\columnwidth]{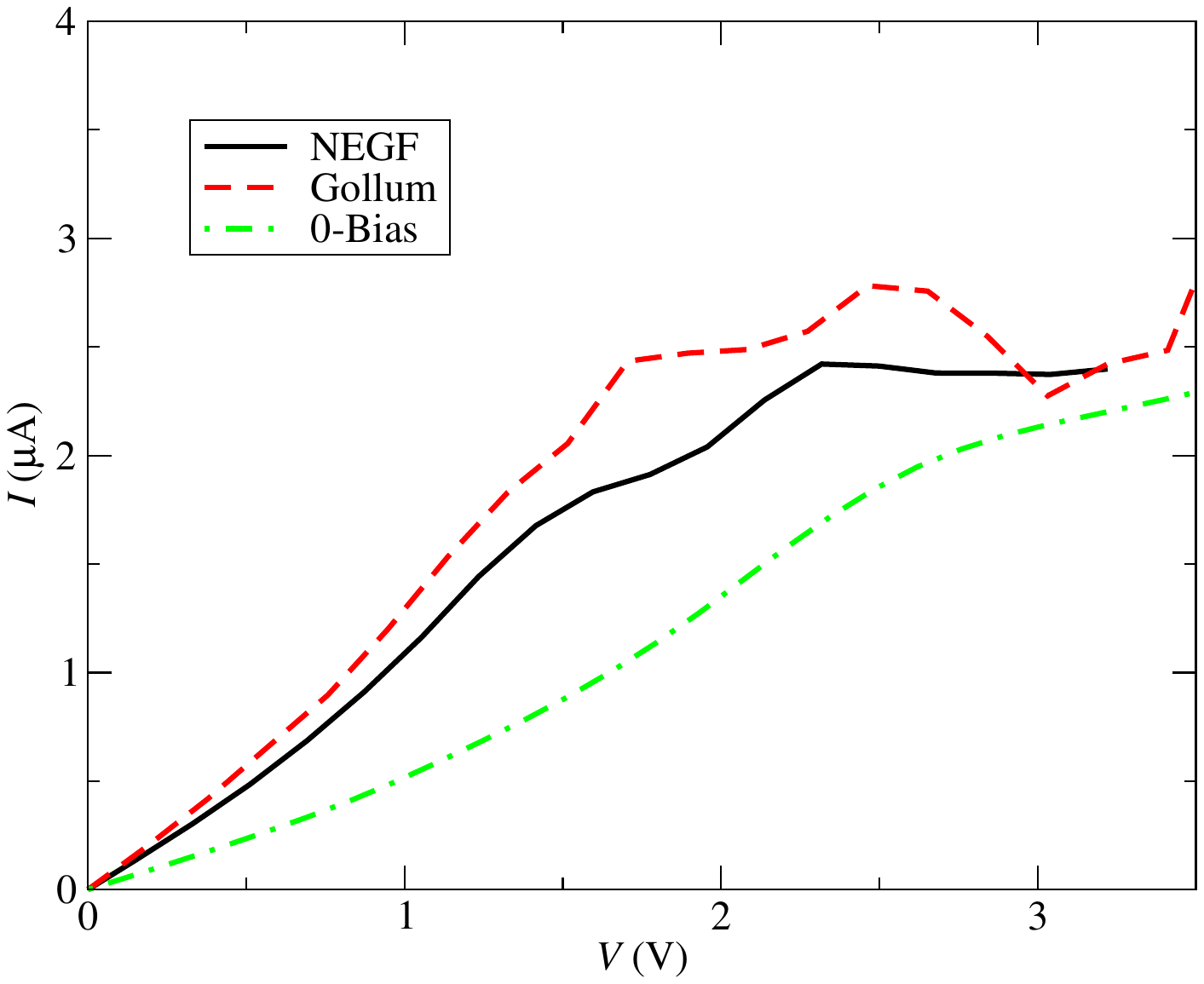}
 \caption{\label{alkane5} Current-voltage curves of the junction shown in Fig. (\ref{Alkane4}). The solid black line
 represents the result obtained from a full NEGF calculation using the code SMEAGOL. The red-dashed line shows the 
 the curve obtained from GOLLUM using the method discussed in section II.B. The green dot-dashed line corresponds to
 integrating the zero-voltage transmission coefficient.}
\end{figure}

For every given voltage $V$,  we modify the EM Hamiltonian as described in Eq. (\ref{voltage}). All the orbitals $n$ at 
the left branch in the EM region in Fig. (\ref{Alkane4}) are shifted 
by $V_n=+V/2$, starting at the TPL and stopping at the linking sulfur atom. Similarly,  the orbitals at 
the right branch in the EM region are shifted by $V_n=-V/2$ all, starting at the linking sulfur atom and including those 
at the TPL. The current $I(V)$ is then computed from the modified Hamiltonian ${\cal K}^{EM}(V)$. The resulting 
$I-V$ curve is shown as a red dashed line 
in Fig. (\ref{alkane5}). For comparison, the black line in the same figure shows the $I-V$ curves 
obtained with a full NEGF simulation using the
code SMEAGOL. Finally, the dot-dashed green line shows the current-voltage curve obtained from GOLLUM
by integrating the zero-voltage transmissions obtained from ${\cal K}^{EM}(0)$. The figure demonstrates that our proposed method
reproduces rather accurately the features found in the full NEGF calculation, including the NDR feature at $V\approx 2$ volt,
while the plain equilibrium calculation fails to reproduce the gross features of the current-voltage characteristics. It 
underestimates the low-voltage conductance by a factor of two.

\begin{figure}
 \includegraphics[trim = 20mm 20mm 40mm 160mm, width=\columnwidth]{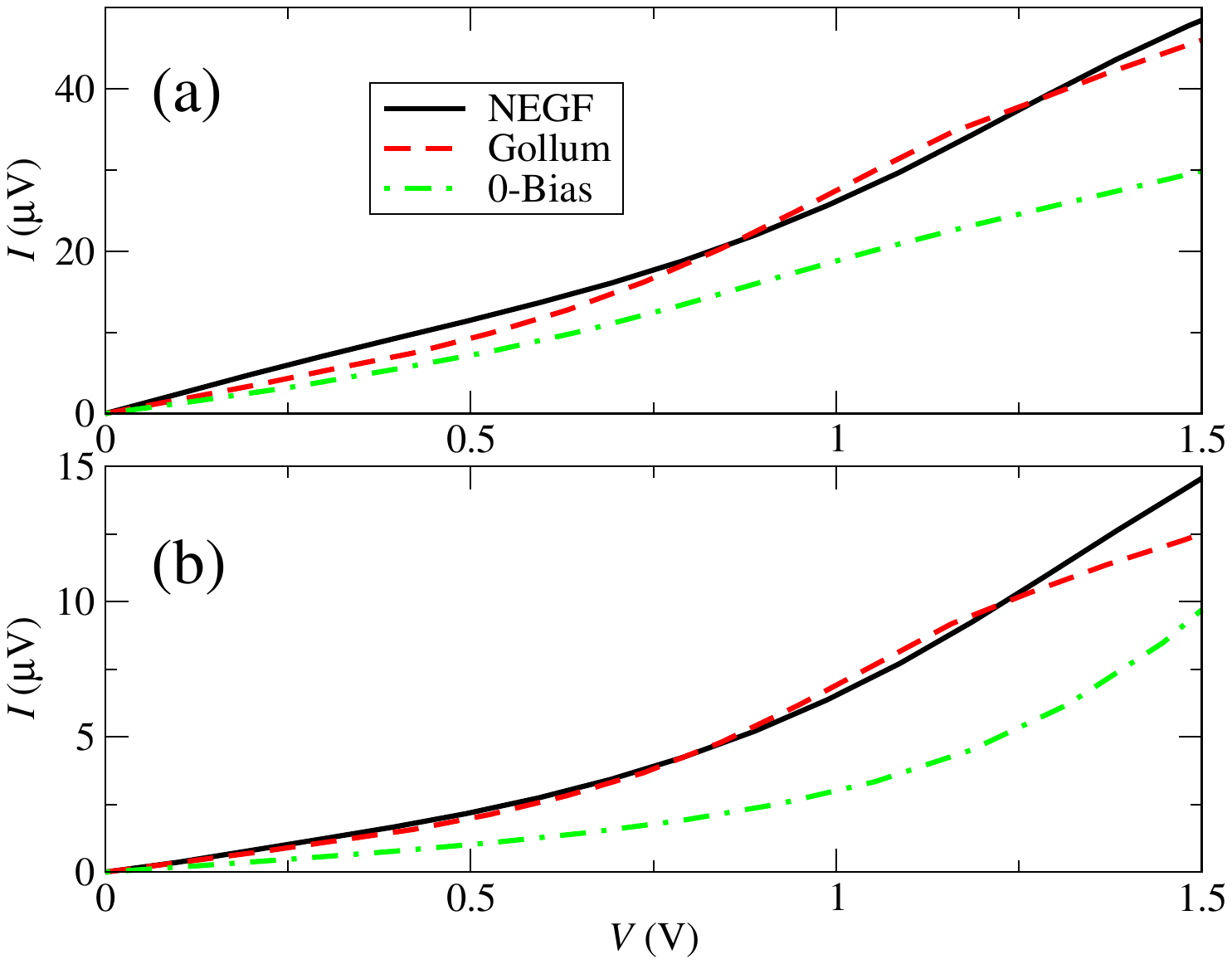}
 \caption{\label{porphirin} Current-voltage curves of a junction similar to that shown in Fig. (\ref{fig:ldau-geometry}),
 where gold (111) electrodes sandwich a porphyrin molecule, whose sulfur end-atom attaches to the electrodes at a 
 hollow position. (a) corresponds to the equilibrium distance between the electrodes and the molecule; 
 (b) corresponds to a junction where each electrode and sulfur atom is pulled 0.3 angstrom away from the molecule backbone. 
 The solid black line represents the result obtained from a full NEGF calculation using the code SMEAGOL. The red-dashed line 
 shows the the curve obtained from GOLLUM using the method discussed in section II.B. The green dot-dashed line corresponds to
 integrating the zero-voltage transmission coefficient.}
\end{figure}

As a second example, we have calculated the current-voltage characteristics of a (111) gold junction sandwiching a porphyrin
molecule. The junction geometry is similar to that shown in Fig. (\ref{fig:ldau-geometry}). The sulfur atoms attach to the electrodes
at a hollow site. The porphyrin molecule does not have in the present case a metallic atom at the center, but has two saturating
hydrogen atoms instead. We have performed two series of calculations. In the first, the geometry and physical gap distance has been relaxed 
and the calculations have been done at the most stable configuration. In the second, we have pulled the electrodes and sulfur atoms away.
To do so, we have increased the sulfur-molecule distance by 0.3 \AA.
The current-voltage curves are shown in Fig. (\ref{porphirin}). These characteristics do not show non-trivial features. We note again that
the plain calculation using zero-voltage transmissions fails to reproduce the NEGF curve, underestimating the conductance by a factor close to
2. In contrast, our prescription provides characteristics that reproduce accurately the NEGF results.

To understand the difference between the dot-dashed green lines and the finite-voltage results of Figs. (\ref{alkane5}) and 
(\ref{porphirin}), we have analyzed the evolution of the transmission coefficients $T(E,V)$ as 
the voltage bias $V$ is ramped.
We have found running NEGF simulations that the main non-equilibrium effect that affects the current  for the two junctions
above is an energy shift of the molecular HOMO resonance, that moves up
as the voltage is increased. Our prescription not only captures the effect, but also follows accurately the evolution
of the resonance shifts dictated by the NEGF calculation, at least at low voltages. By shifting the HOMO resonance upwards 
in energy, 
a larger weight of the resonance enters into the energy integration window used to compute the current integral, hence increasing the current.  In contrast, this effect can not be 
captured at all if one uses the plain $T(E,V=0)$ transmission coefficients.

\subsection{Temperature dependence of the thermoelectric properties of a C$_{60}$ molecular junction.}
In this section, we show how GOLLUM can compute the thermoelectric properties of complex junctions formed by trapping a C$_{60}$
molecule between gold electrodes.
In a previous paper \cite{Evangeli13} we have demonstrated both experimentally and theoretically that $C_{60}$-based nanojunctions show promisingly-high values for the thermopower and figure of merit. However the temperature
dependence of these values and the fluctuations caused by the exact geometrical details have not been 
thoroughly investigated,
partly due to the computationally-expensive nature of the
calculations. Here we show that the fast and efficient implementation of GOLLUM allows us to undertake a more complete exploration
of the transport properties of these C$_{60}$-based junctions.

\begin{figure}
 \includegraphics[width=\columnwidth]{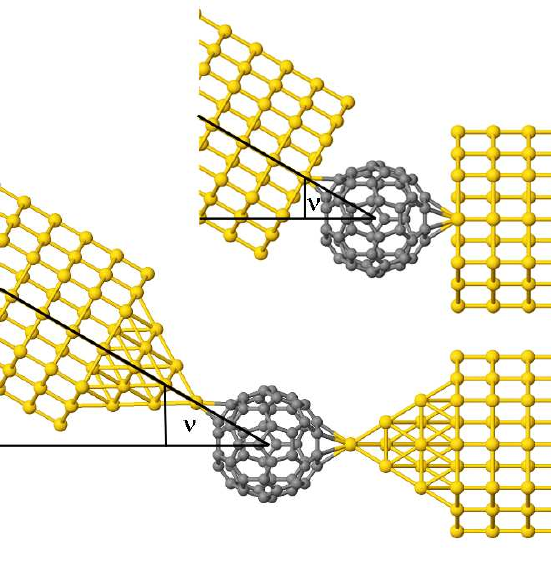}
 \caption{Geometry of a junction having gold (111) electrodes, which sandwich a $C_{60}$ molecule. The electrodes are
  terminated as either a flat surface (upper panel) or a pyramid (lower panel), and can be tilted an angle $\nu$ relative
  to each other.}
 \label{fig:c60}
\end{figure}

The systems of interest consist of two (111) gold leads that can be tilted an angle $\nu$ relative to each other. Each lead is
terminated using either a flat surface or a pyramid, as shown in Fig. (\ref{fig:c60}). We have performed DFT calculations
using the code SIESTA, with a double-zeta-polarized basis set, and the LDA functional\cite{lda}. We have relaxed the molecular
geometries using a force tolerance of 0.02 eV/\AA\, and have found the equilibrium
distance between the leads and the C$_{60}$ molecule to be of about 0.22 nm, depending on the exact orientation of the molecule. This result is in
good agreement with other previously reported  distances \cite{Wang04}. We have kept this distance fixed in all subsequent
transport calculations. However, we have taken for completeness five possible orientations
of the $C_{60}$ molecule relative to the electrodes. These are: (a) a C-C bond between a hexagon and a pentagon facing the Au surface;
(b) a hexagon facing the Au surface; (c) a pentagon facing the Au surface; (d) a bond between two hexagons facing the Au surface; and
(e) a single atom facing the Au surface. We have also tilted one of the electrodes in steps of 15 degrees between 0 and 60 degrees to
see the interference effects caused by the exact position of the tip on the surface of the fullerene,
recalculating the thermoelectric properties at each step. The starting position (for $\nu=0$) for the C$_{60}$ against the electrodes is such that one of its pentagons is facing the Au surfaces.

\begin{figure}
 \includegraphics[width=\columnwidth]{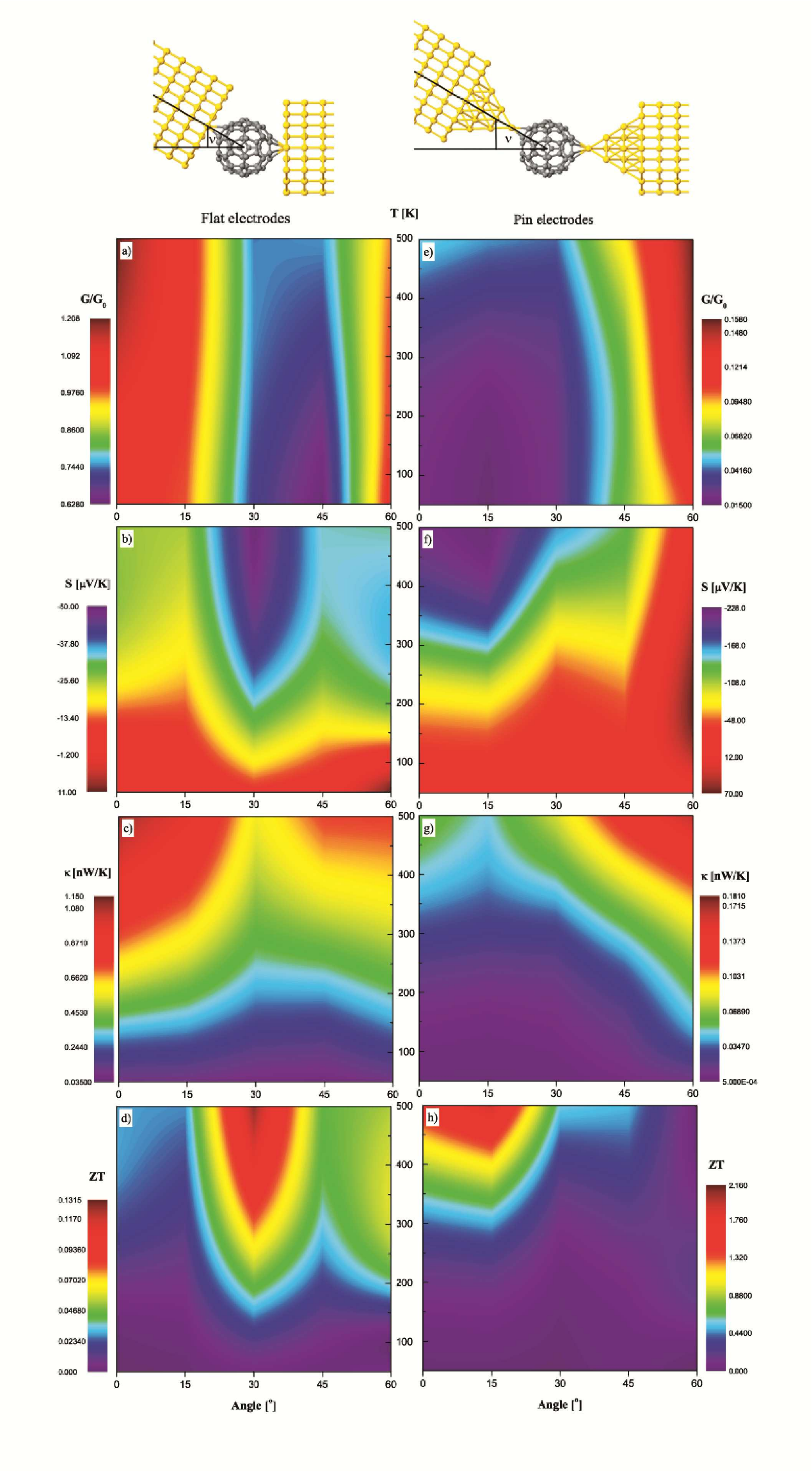}
 \caption{(Color online) The panels show our thermoelectric results for the junction in Fig. (\ref{fig:c60}) 
 having flat or pyramid-terminated electrodes; $\nu$ is the rotation angle. Figs. (a)-(b)-(c)-(d) show
 two-dimensional contour plots of the conductance $G$ (measured in units of $G_0$), the thermopower $S$ (measured in
 $\mu V/K$), the thermal conductance $\kappa$ (measured in nW / K), and the figure of merit $ZT$ (dimensionless).  
 The vertical and horizontal axes are the temperature T measured in Kelvin and the tilting angle $\nu$ 
 for flat-electrode junctions.  Figs. (e)-(f)-(g)-(h) show the same magnitudes for pyramid-terminated electrodes. 
 Note that the color code is different for each figure.}\label{fig:c60-1}
\end{figure}

The results of our calculations are shown in Figure (\ref{fig:c60-1}). By taking horizontal cuts through these surfaces, 
we can see clear evidence of quantum interference as the angle changes and the tip  is repositioned
around the fullerene surface. For the conductance $G$, these oscillations are almost
temperature independent, whereas in the case of the  thermopower $S$, the thermal conductance $\kappa$ and the electronic figure of merit ZT, these
oscillations are almost negligible at low temperatures and then grow with $T$.
Comparing the conductance obtained with flat electrodes against that with pyramid-terminated electrodes, we can clearly see that  
$G$ decreases substantially
when using the  pyramid-like electrodes, in agreement with Ref. (\onlinecite{Zheng09}). Furthermore in the case of flat
electrodes the conductance fluctuates with the angle $\nu$ by about half an order of magnitude, whereas for the pyramid-terminated
electrodes we find a larger change of almost one order of magnitude. This again demonstrates that the pyramid-terminated electrode scans
the molecular surface like an STM tip,  with improved detail, while part of these features are blurred when using
a flat electrode. The flat-electrode junctions possess conductances values of about 0.6-1.2 $G0$, while junctions with pyramid
tips have conductances of order 0.015-0.15 $G0$. 
We overestimate the experimental values for $G$ \cite{Evangeli13,Bohler07,Kiguchi07,Neel07},
by about one order of magnitude, a known problem associated with  the underestimation of  the HOMO-LUMO gap
inherent to the plain DFT approach. As expected, the thermopower $S$ is more sensitive to the angle $\nu$ when using 
pyramid-terminated junctions compared with the case of flat-surfaced junctions. Interestingly, $S$ is quite high especially 
at the higher temperatures, achieving values of of about $100$ to $200 \mu V/K$.

\subsection{Multi-terminal calculations}
\begin{figure}
 \includegraphics[width=\columnwidth]{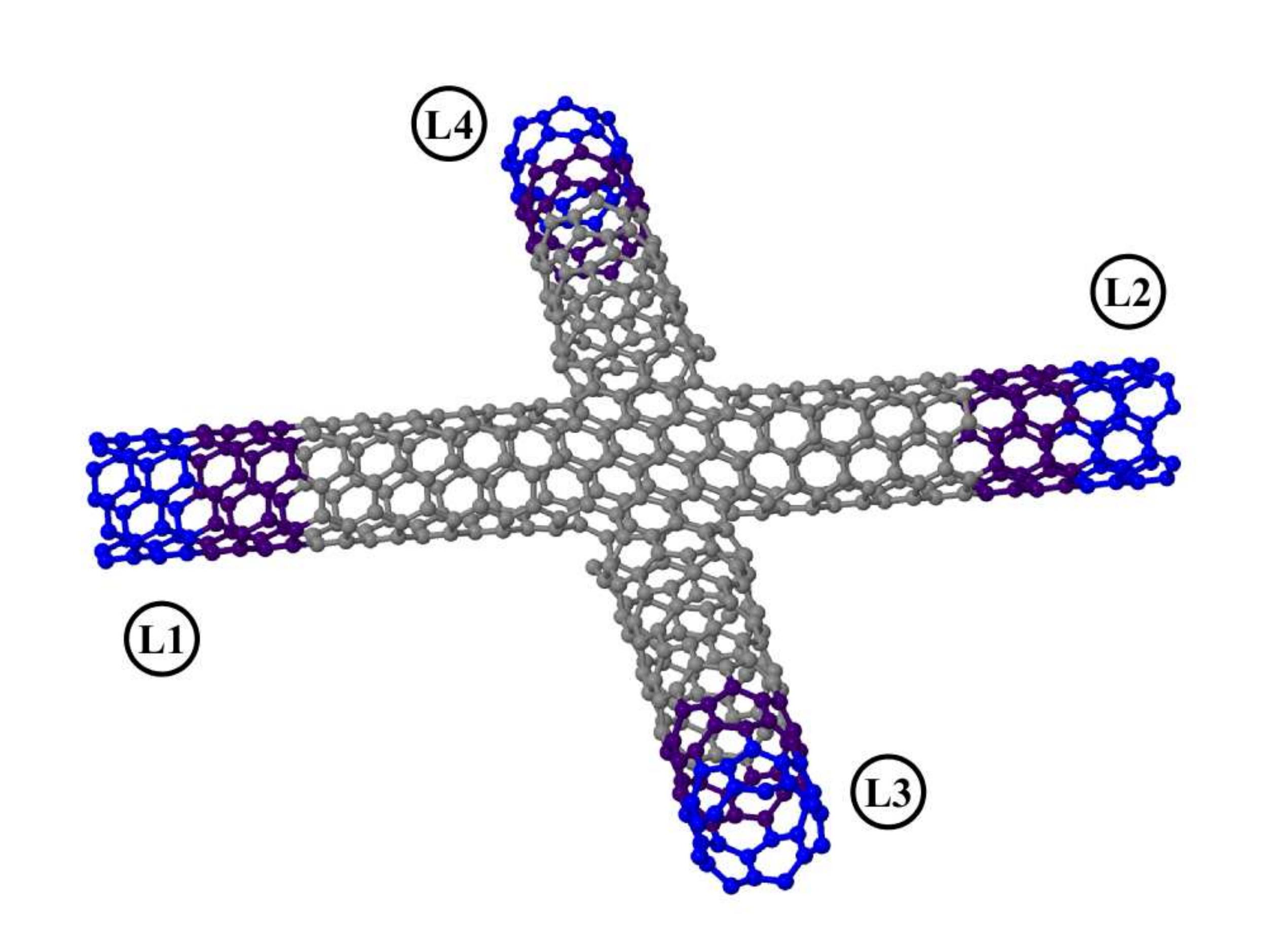}
 \caption{(Color online) Four-probe cross bar carbon nanotube device. Here, L1 and L2 are armchair CNTs, 
 while L3 and L4 are zigzag CNTs, both with diameters of 5.65 \AA. The distance between the edges 
 of L1 and L2 and of L3 and L4 are 50.51 and 49.87 \AA, respectively.}
  \label{fig:four-probe-cnt}
\end{figure}

Ab-initio force-relaxation simulations show that it is possible to sculpt complex three-dimensional
structures of nanoscale-scale dimensions by cutting shapes into a graphene bilayer \cite{laith12}. We find that the edges of
the two graphene sheets coalesce in order to saturate dangling bonds and to maximize the degree of $sp^2$ hybridization. For example,
by cutting a cross shape in a bilayer graphene sheet, the resulting sculpturene is the three-dimensional crossbar carbon nanotube (CNT) shown
in Fig. (\ref{fig:four-probe-cnt}), which is an example of a  four-terminal electronic device. This four-terminal device is composed
of two armchair and two zigzag CNT electrodes. These become perfectly periodic CNT leads (shown with blue) far enough from
the junction. To perform a GOLLUM-based four-terminal calculation, we obtain the mean-field Hamiltonian of this structure using
the SIESTA code, with a double-zeta-polarized basis set and a GGA functional\cite{gga}. We start with the referred 
cross-shaped bilayer graphene sheet and after relaxing the inter-atomic forces to tolerances below 0.02 eV/\AA, we find 
the crossbar shaped device shown in Figure (\ref{fig:four-probe-cnt}).

\begin{figure}
 \includegraphics[width=0.9\columnwidth]{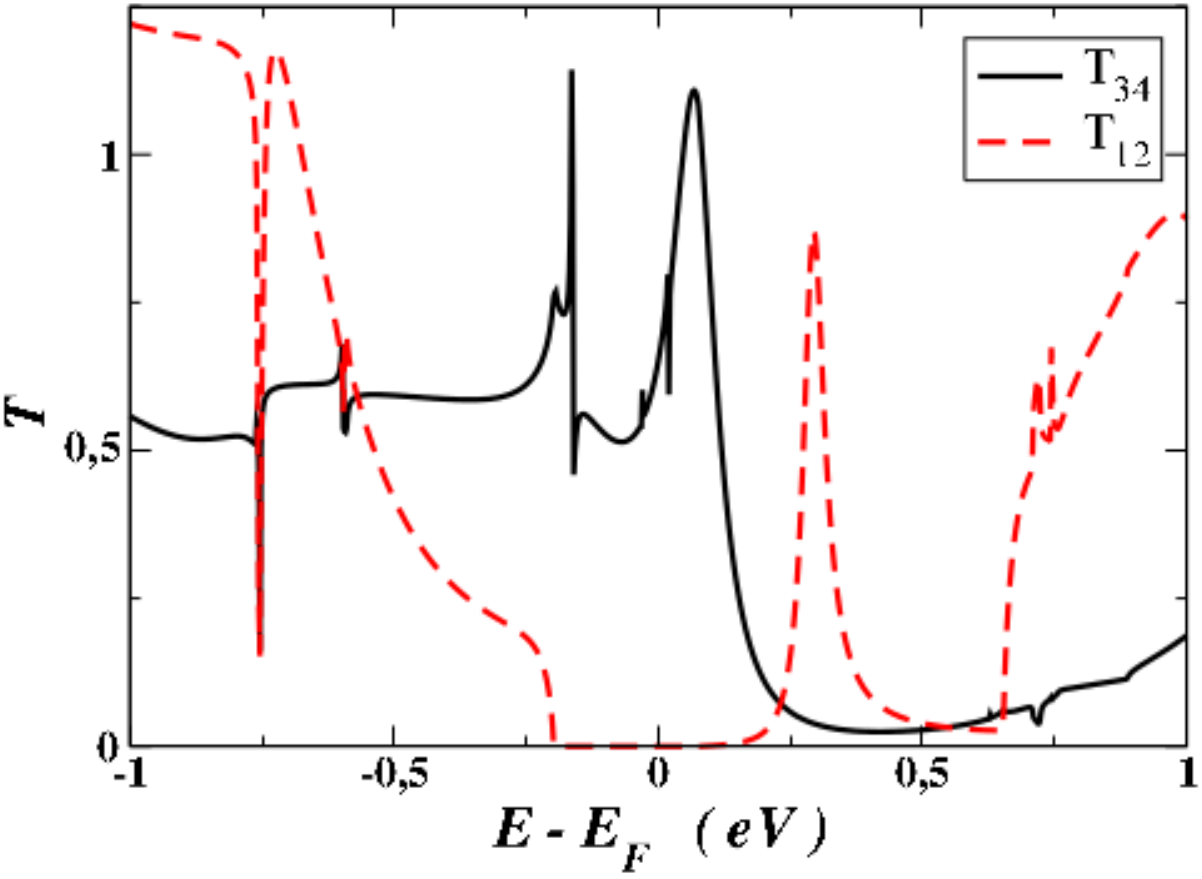}
 \caption{(Color online) Transmission coefficients between leads 1 and 2 $T_{12}$ and leads 3 and 4 $T_{34}$
 (solid black and dashed red lines respectively). Energies are measured in eV and referred to the Fermi energy of the
 Scattering Region.}
  \label{fig:T-12-34}
\end{figure}

We feed the resulting ab-initio Hamiltonian into GOLLUM, and compute the transmission coefficients $T_{ij}$ between every 
possible
combination of pairs of leads. Notice that the armchair CNT leads are semiconducting, while the zigzag CNTs are metallic. We therefore
expect different qualitative behaviors for the transmission properties among the different arms. This is shown in Fig. (\ref{fig:T-12-34}),
where the transmission coefficients between the two armchair arms $T_{12}$ are much smaller than those connecting the zigzag arms $T_{34}$.
In addition, the figure indicates that the central cross area, where the two arms join together is not transparent, but introduces strong
scattering. Similarly, the transmission from lead 1 to lead 3 also shows a reduced transmission at low energies due to the
semiconducting behavior of the armchair lead 1, see fig (\ref{fig:T-12-34}). This figure shows that $T_{13}=T_{42}$ due to
the junction symmetry.

\begin{figure}
  \includegraphics[width=0.9\columnwidth]{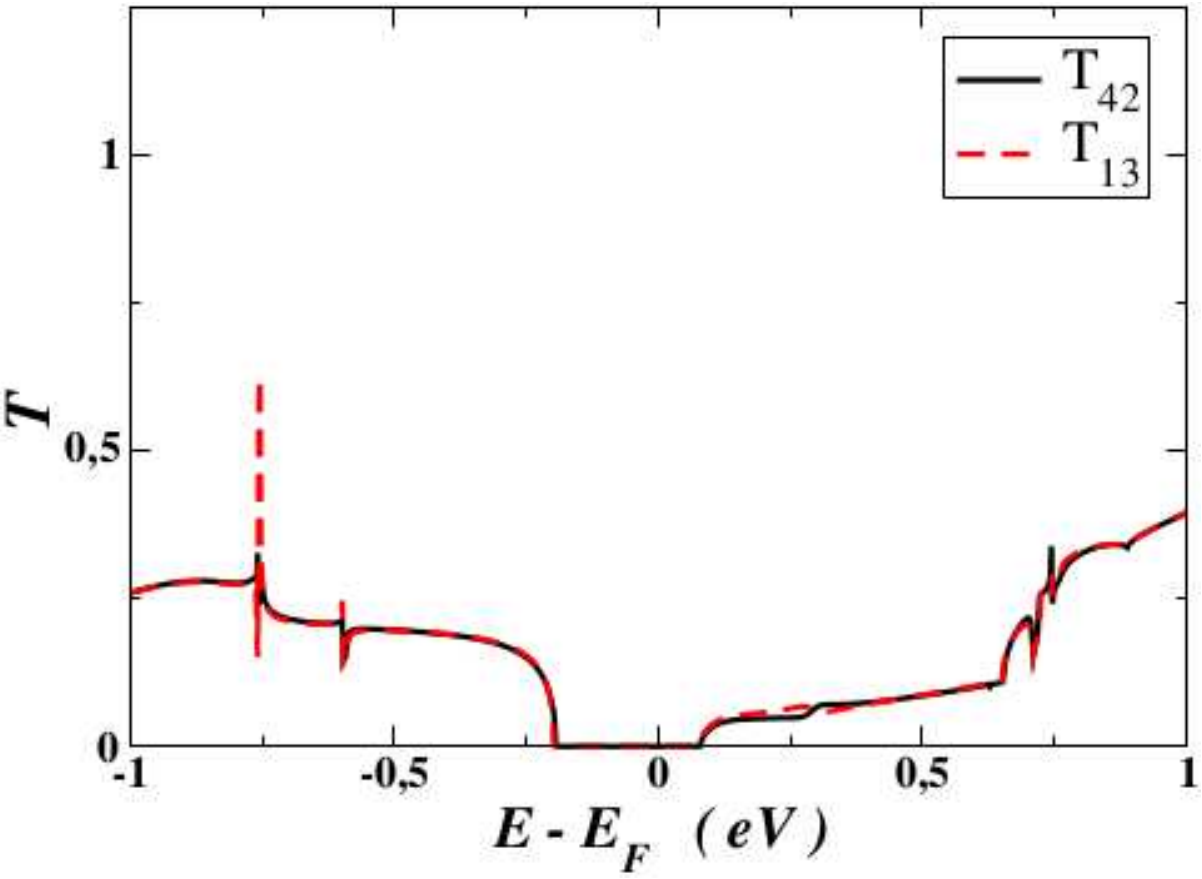}
 \caption{(Color online) Transmission coefficients between lead 4 and lead 2 ($T_{42}$, solid black line) and between lead 1
 and lead 3 ($T_{13}$, dashed red line). Energies are measured in eV and referred to the Fermi energy of the Scattering Region.}
  \label{fig:T-1-234}
\end{figure}

\subsection{Environmental effects on quantum transport}
In the literature, most theoretical  analyses of phase-coherent transport properties  assume that the junction is immersed in vacuum and
therefore ignore the effects of the surrounding environment. In contrast, many experiments are carried out under ambient conditions,
which can have a marked effect on transport properties\cite{water}. If surrounding environmental molecules possess a dipole moment, then the scattering region will be
subject to a fluctuating electrostatic field. Previous work to investigate the impact of environmental water
on the transport properties of a a single-molecule junction\cite{water} also took into account the effect of a solvation shell
of water molecules surrounding the junction. GOLLUM describes these effects systematically, by noting that for nanostructures such as single-molecule
junctions, the timescale for such fluctuations is typically longer than the time taken for an electron to pass through the device and
therefore one can adopt the Born-Oppenheimer approximation and freeze the environment during each electron transit.
However, successive electrons experience different environmental snapshots and therefore are subjected to different instantaneous mean field Hamiltonians
leading to different instantaneous conductances. The measured (time-averaged) electrical conductance will hence be an ensemble average over these snapshots.
To obtain a series of environmental snapshots, GOLLUM uses classical molecular dynamics to describe the environmental molecules
and for each snapshot, feeds the resulting geometries into a DFT code to compute the corresponding self-consistent Hamiltonian. The resulting mean field Hamiltonian is then used to compute
the scattering matrix and related transport properties.

\begin{figure}
 \includegraphics[width=0.9\columnwidth,height=5cm]{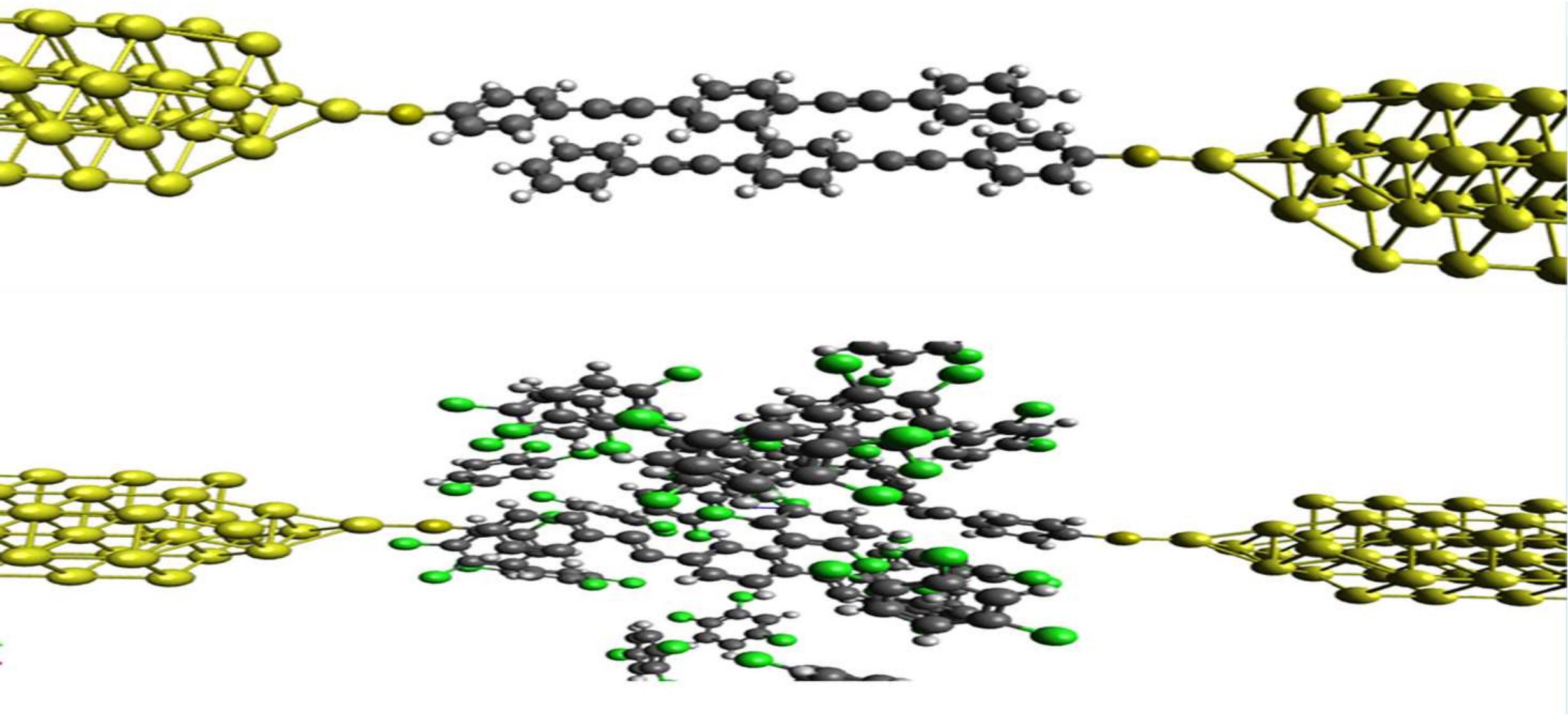}
 \caption{(Colour online) (top) Geometry of a pi-stacked molecule connected to gold electrodes. (bottom) 
 Single snapshot of a
MD calculation where the molecule is surrounded by TCB solvent molecules.}
\label{fig:solvent-molecule}
 \end{figure}

\begin{figure}
 \includegraphics[width=\columnwidth]{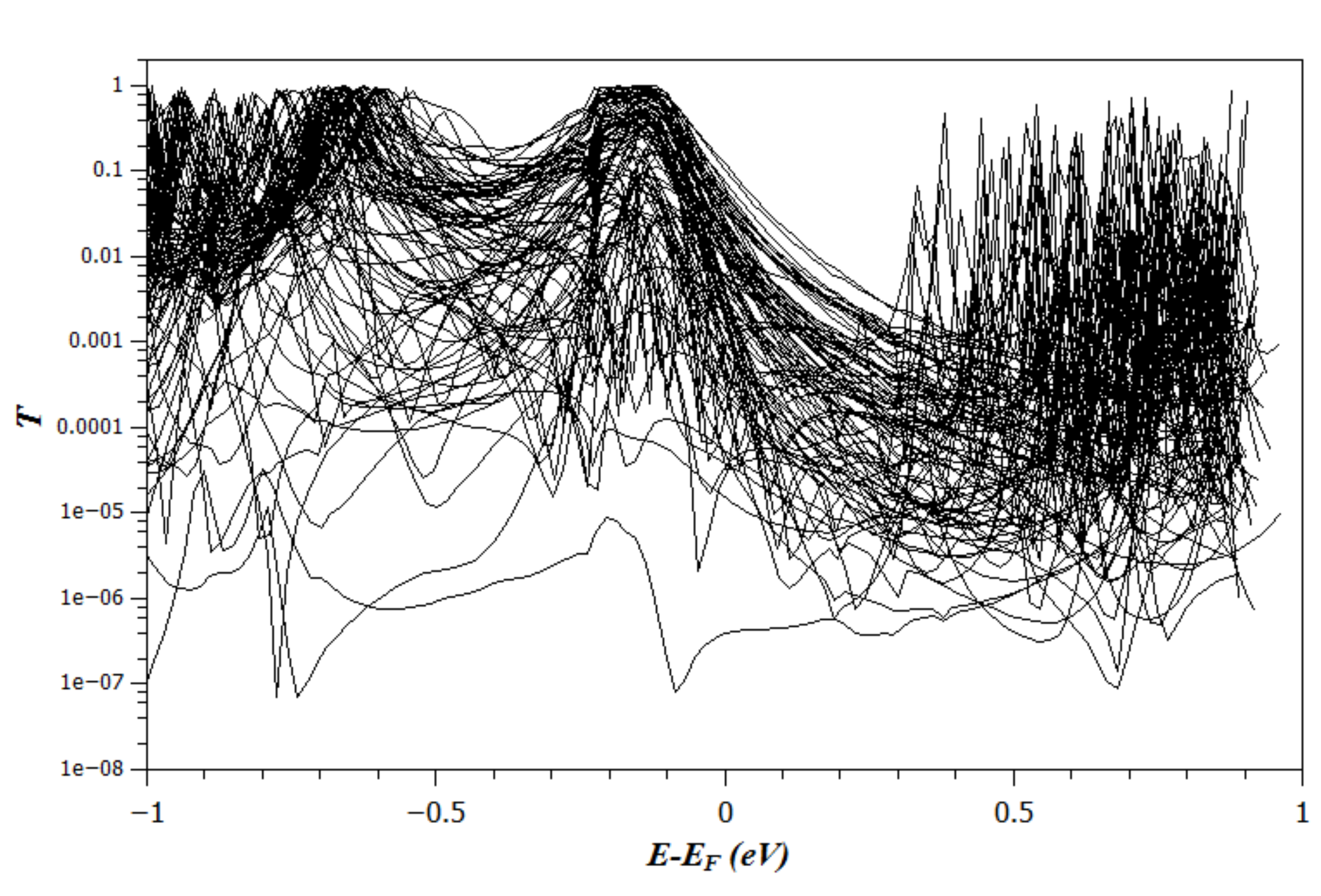}
 \caption{(Colour online) Transmission curves of the junction shown in Fig. (\ref{fig:solvent-molecule}) for 100 snapshots.}
\label{fig:solvent-trace}
 \end{figure}

\begin{figure}
 \includegraphics[width=\columnwidth]{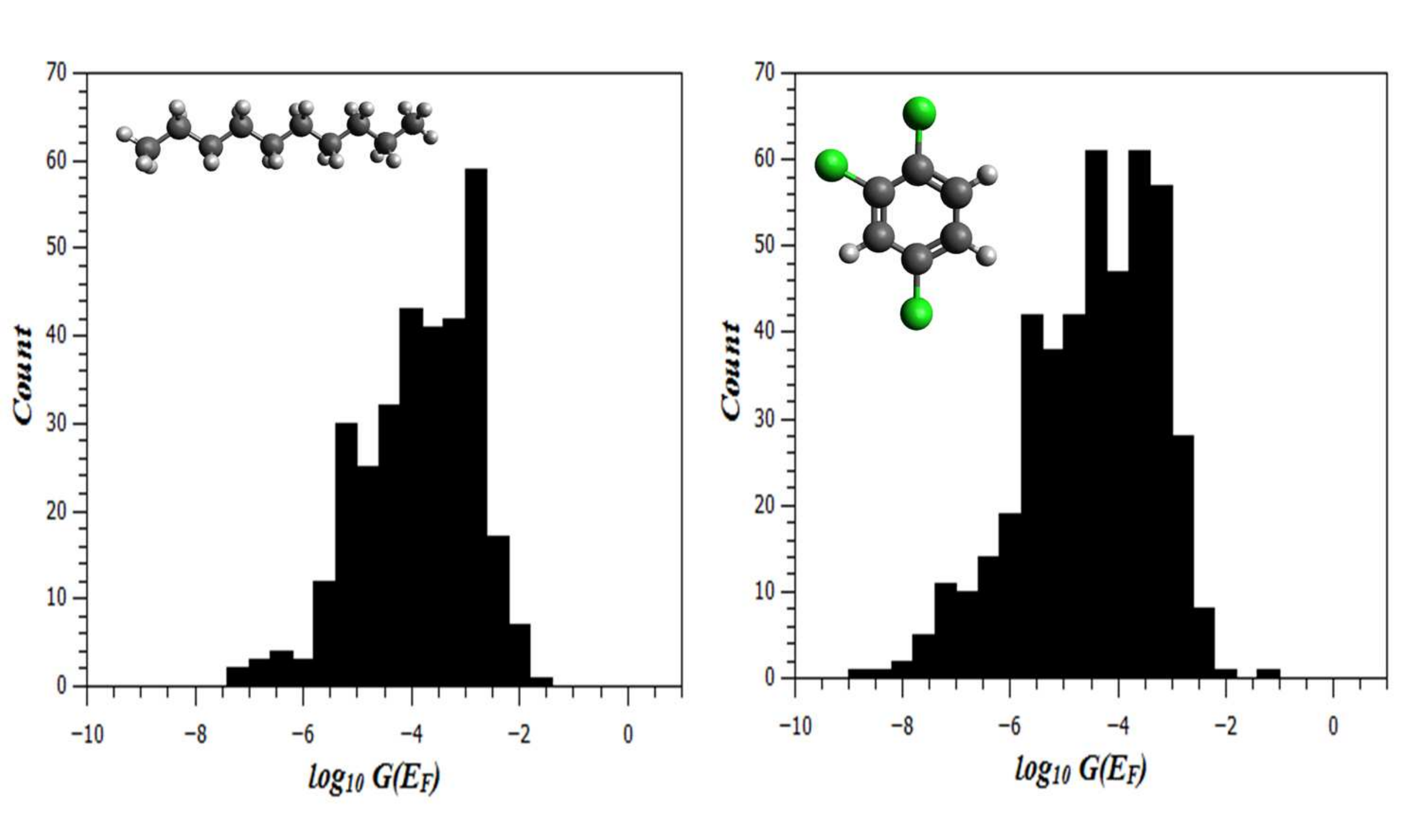}
 \caption{(Colour online) Conductance histograms of the junction shown in Fig. (\ref{fig:solvent-molecule}) 
 for two different solvents: decane (left) and TCB (right).}
\label{fig:solvent-histogram}
 \end{figure}

To illustrate this approach, we compute here the ensemble averaged conductance of the junction shown in Fig.
(\ref{fig:solvent-molecule}), where two pi-stacked monothiol terminated oligophenyleneethynylenes form
a bridge between two gold (001) electrodes. The bridging molecule is surrounded by two different solvents: decane and 
1,4-dioxane,1,2,4-trichlorobenzene (TCB). An example of a junction surrounded by a shell of
TCB molecules is shown in the lower panel of Fig. (\ref{fig:solvent-molecule}). 
We have tested here the classical molecular dynamics packages LAMMPS\cite{lammps}
and DLPOLY\cite{dlpoly}, but GOLLUM is flexible enough to accept coordinates from other classical Molecular Dynamics packages.
In what follows, we show results obtained with LAMMPS, where we have used  the Dreiding force field to describe the intra- and
inter-molecular interactions and have employed the REAXFF forcefield to obtain the initial charges.
To create the environment, we place two hundred solvent molecules surrounding the backbone molecule. We perform the simulations
using a constant temperature and volume (NVT) ensemble and subsequently a constant temperature and pressure (NPT) thermostat.
We equilibrate the junction for 150 ps with 0.1 fs time steps continuously raising the temperature to 290 K. We do not include the gold
electrodes in the molecular dynamics simulation, so to simulate the binding of the anchor groups, we hold the positions of the two
terminating atoms which connect to the electrodes fixed.  We record between 350 and 500 snapshots of the junction, that have been
taken  every 2ps. For each snapshot, we feed the atomic coordinates into the DFT code SIESTA and generate the DFT Hamiltonian. We then
feed the Hamiltonians into the transport code to compute the electrical conductance. Some example transmission curves for 100 snapshots
of the junction with a TCB solvent are shown in the right panel in Fig. (\ref{fig:solvent-molecule}). We note that the the room-temperature 
dynamics of the
atoms at the junction lead to a large spread in the transmission curves, and therefore to  many different values of
the low-voltage conductance. We therefore assemble conductance histograms to help identify  the most probable conductance values.
The resulting histograms in the presence of decane and TCB solvents are shown in Fig. (\ref{fig:solvent-histogram}). The fact that
the most probable conductance values are different  shows that ambient-conditions or liquid-immersed molecular electronics experiments
are affected by the surrounding solvent. Notice that in these simulations we kept fixed the molecule-electrode geometry, so the 
spread in $G$ is due entirely to environmental effects.

\subsection{Nanopore-based DNA nucleobase sensing}
The fact that the transport properties of nanoscale junctions depends on the surrounding environment leads to a wide range of
possible sensing applications.
In this section, we demonstrate the versatility of GOLLUM by showing how it can be used to predict the change in conductance of a
nanopore,  when a single DNA strand is trans-located through it.
Deoxyribonucleic acid (DNA) is a molecule that encodes the genetic instructions used in the development and functioning of all known
living organisms and many viruses. DNA molecules are double-stranded helices, consisting of two long biopolymers composed of simpler
units called nucleotides. Each nucleotide is composed of one of the four nucleobases guanine (G), adenine (A), thymine (T)
and cytosine (C), which are attached to a backbone made of alternating sugar and phosphate groups. The two polymer strands are bound together
by non-covalent bonds that link base pairs and are easily separated to form two single-stranded DNA molecules (ssDNA) molecules.
DNA sequencing aims at identifying the sequence of the DNA bases in a sample of ssDNA.

Many researchers are actively seeking new methods to sequence DNA with improved reliability and scalability and that are economically viable.
Biological nanopores made from protein such as a-hemolysin have been shown experimentally to sense the presence of DNA \cite{Kasianowicz96,Mathe05},
but are also very sensitive to temperature and pH, and can only be used within a limited voltage bias window\cite{Schneider10}.
As an alternative, solid state devices which can be integrated into existing semiconducting circuitry technology and that are robust
to the chemical environment have been proposed as sensors \cite{Schneider10,Merchant10,Garaj10,Sathe11}.

\begin{figure}
 \includegraphics[width=\columnwidth]{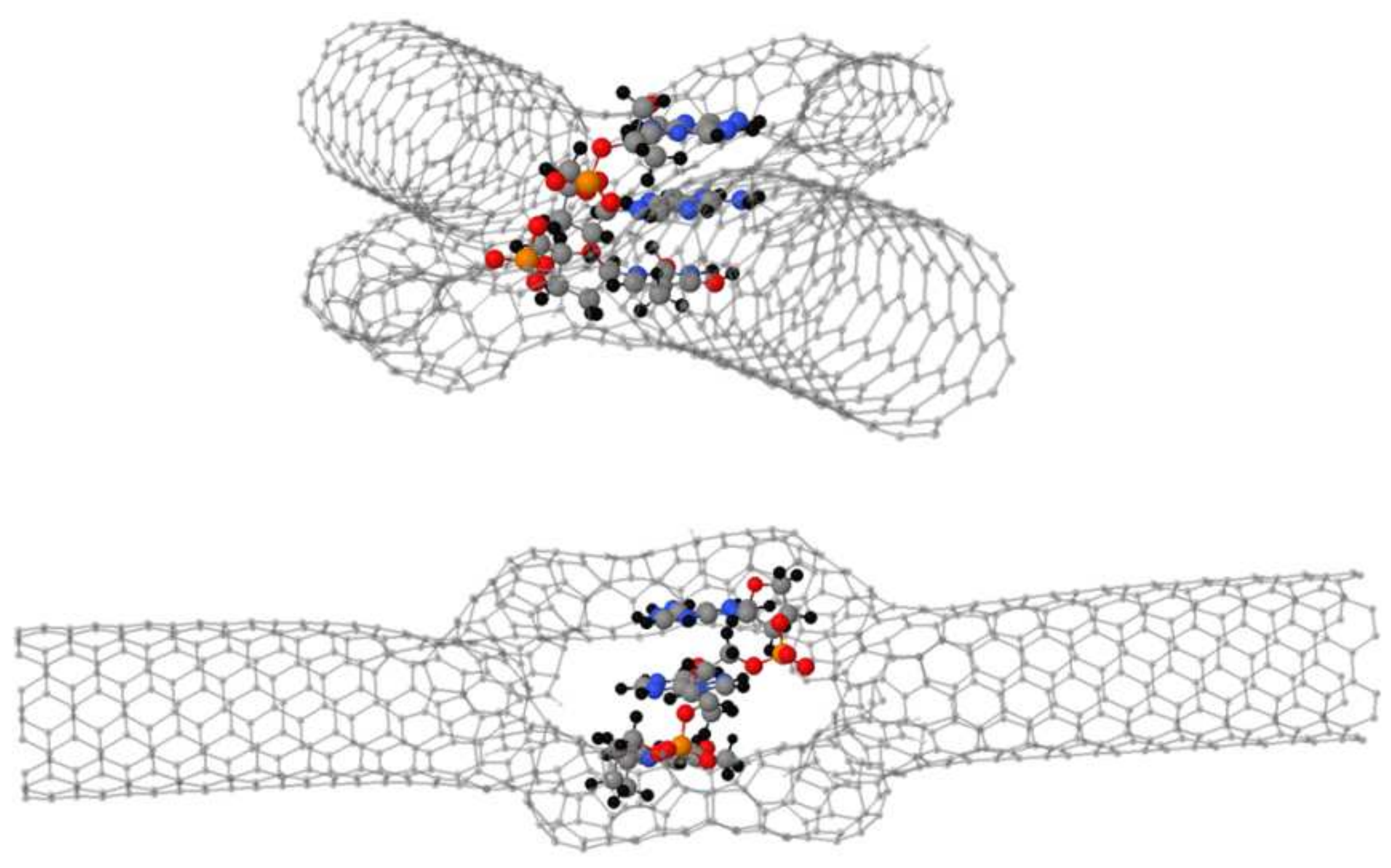}
 \caption{(Colour online) sketch of a possible CNT based DNA sensor: a piece of a ssDNA is being
 translocated through a torus-shaped sculpturene.}
\label{fig:ttht}
 \end{figure}

\begin{figure}
 \includegraphics[width=\columnwidth]{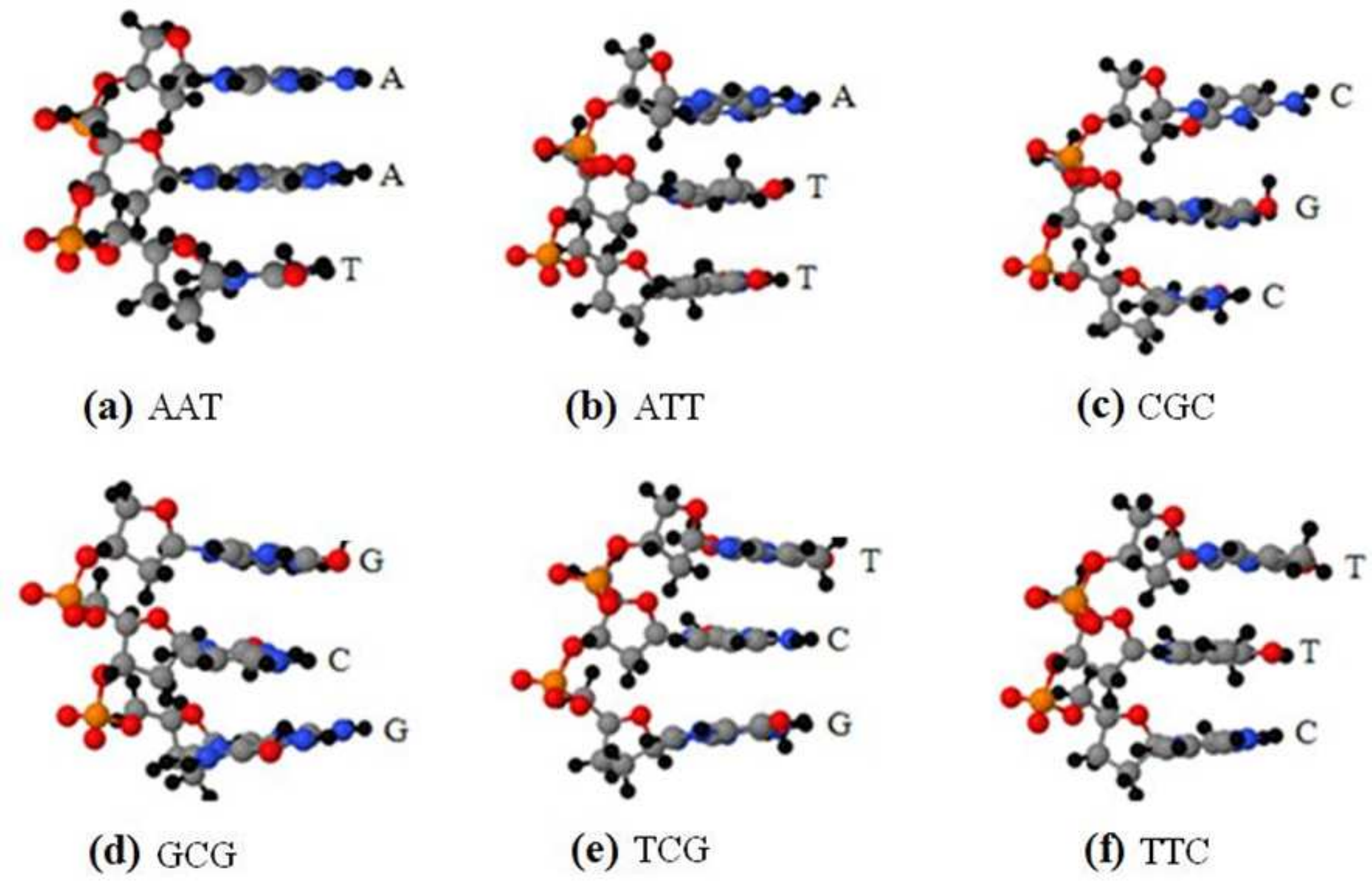}
 \caption{(Color online) The six nucleotide sequences of three base pairs joined by a DNA backbone, discussed in the text:
 (a) AAT; (b) ATT; (c) CGC; (d) GCG; (e) TCG and (f) TTC.}
\label{fig:base-sequences}
 \end{figure}

\begin{figure}
 \includegraphics[width=\columnwidth]{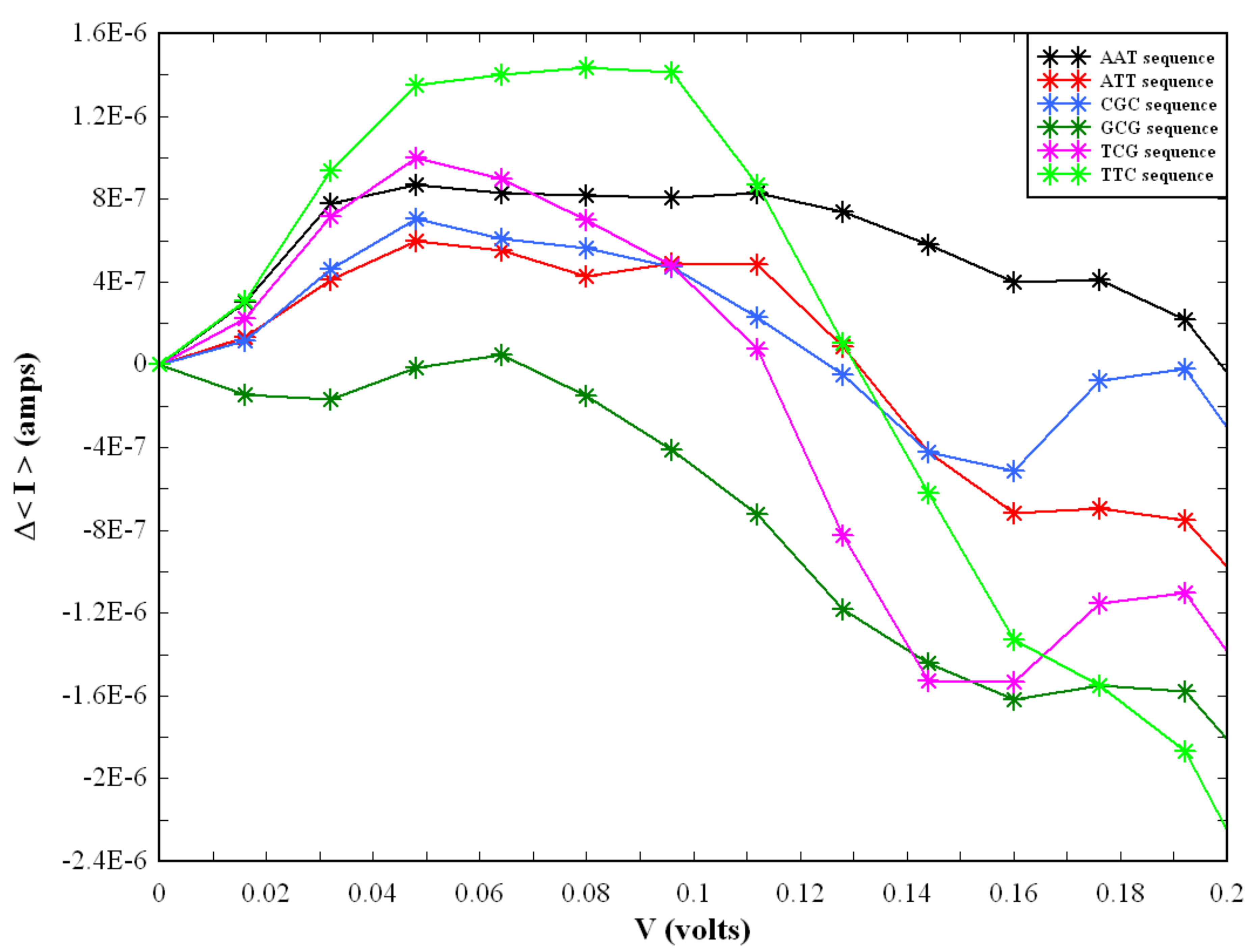}
 \caption{(Color online) $I-V$ curves for each of the six sequences shown in Fig. (\ref{fig:base-sequences}), where the current has
 been averaged over different pore-nucleotide relative angles and the current of the empty pore has been subtracted.}
\label{fig:ivcurves-sequences}
 \end{figure}

To demonstrate the versatility of GOLLUM, we examine here the potential for DNA nucleobase sensing of the sculpturene device shown in
Fig. (\ref{fig:ttht}), which comprises a torus-like nanopore connected to two CNT electrodes\cite{laith12}.
The torus in the figure has an inner pore with a diameter of 1.6 nm, whereas the leads are two (6,6) armchair nanotubes having
a diameter of about 5 \AA.
We have selected for our study six short strands of ssDNA containing three bases, that are shown in Fig. (\ref{fig:base-sequences}).
We have first  relaxed the coordinates of the nucleotides that are threading the pore,
using the DFT code SIESTA with a double-zeta basis set and a LDA functional\cite{lda}. Since the pore
diameter is slightly larger than the strand width, the strand and its nucleotides can adopt different conformations and orientations
inside the pore. We accumulate snapshots of these different conformations and orientations for each of the six ssDNA strands as they
trans-locate the pore. For each snapshot, we compute the current-voltage curve and subtract the current for the empty
pore $\Delta I=I(V)-I_0(V)$. The current averaged over snapshots $\Delta \langle I\rangle$ for each ssDNA strand is plotted
in Fig. (\ref{fig:ivcurves-sequences}). The sizable height of the curves  demonstrate that the conductance of the pore is
sensitive to the gating effect produced by the presence of ssDNA strands inside the pore. Furthermore, the different behavior
of the curves means that, armed with a proper statistical analysis, the sculpturene device can distinguish different nucleotide
sequences, so that this kind of device could be utilized potentially as a discriminating DNA sensor.


\subsection{Theoretical simulation of the pulling curves and histograms of break-junction experiments.}
A large body of experiments in single-molecule electronics is performed using  the mechanically-controlled break junction (MCBJ) technique,
in which a metallic strip is pulled slowly until it breaks into two separate pieces. This process enables the formation of electrodes with  molecular-scale
gaps, which can be bridged by a single molecule. Experimentally, these two electrodes are repeatedly pulled away
or pushed towards each other. By applying a small bias voltage and recording the current passing through the junction, the
low-voltage conductance can be measured as a function of the distance between the electrodes. When the distance is small enough, a single
molecule can bridge the gap between the electrodes and its conductance can be measured. In the literature, most theoretical studies are
confined to small numbers of ideal geometries and binding configurations. In this section, our aim is to demonstrate that the versatility of
GOLLUM allows us to compute whole 'pulling curves' of conductance versus electrode separation.

\begin{figure}
 \includegraphics[width=\columnwidth]{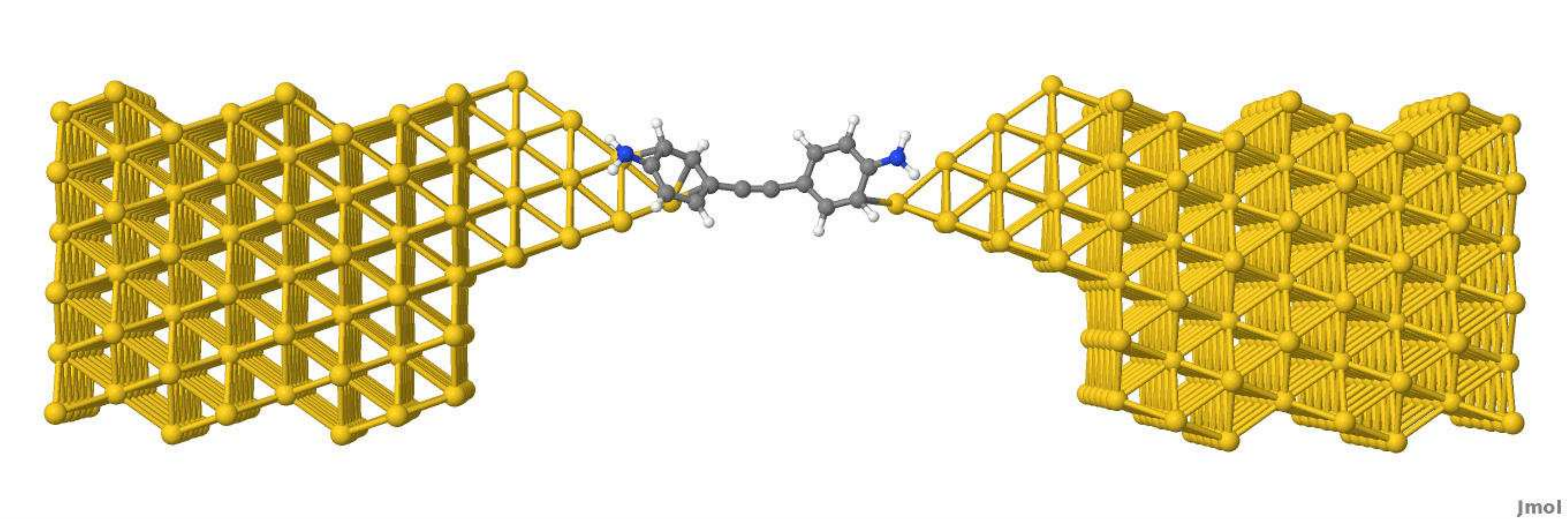}
 \caption{(Colour online) A snapshot of an opening cycle in a gold-bipyridine MCBJ simulation.}
\label{fig:pulling-example}
 \end{figure}

\begin{figure}
 \includegraphics[width=\columnwidth]{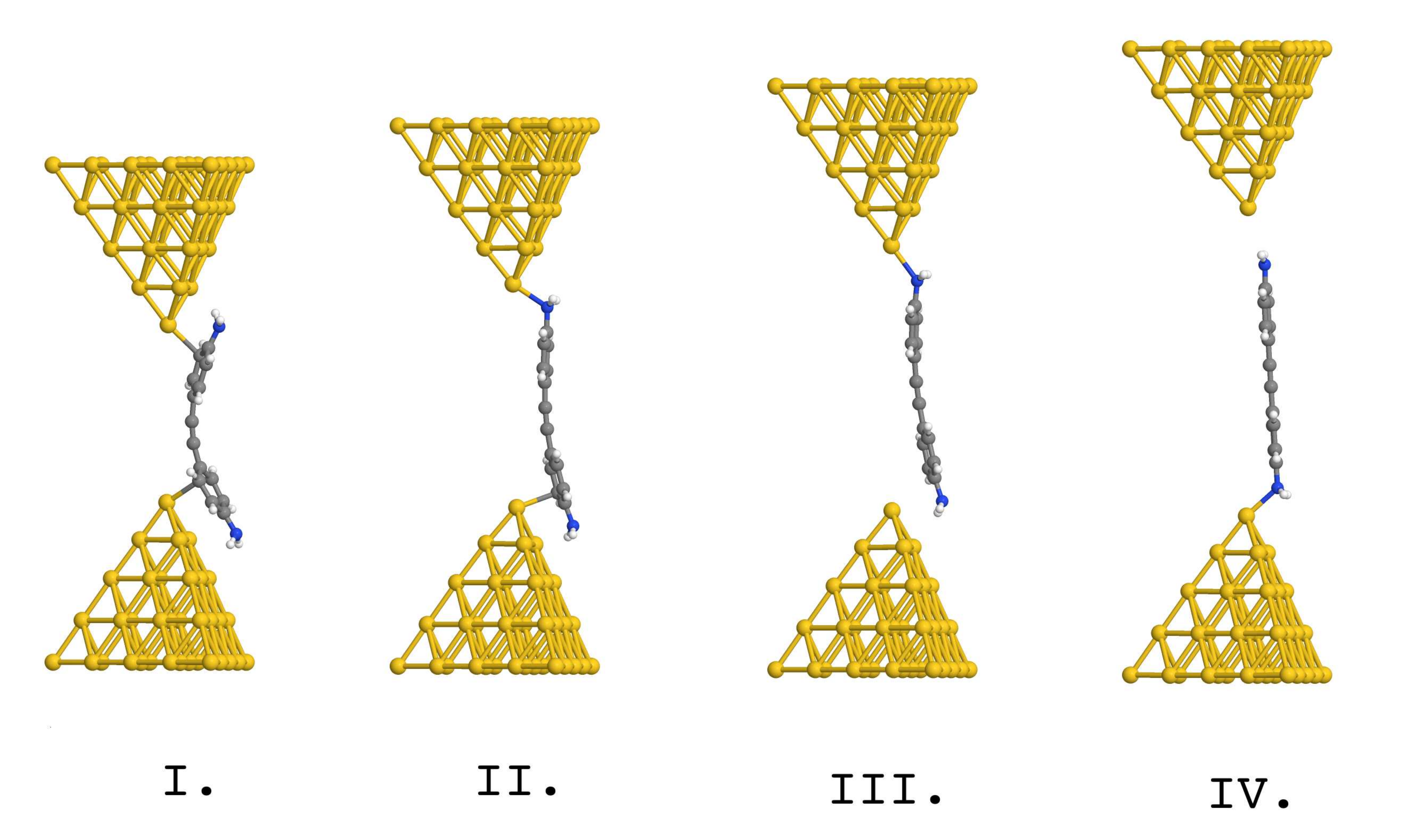}
 \caption{(Color online) Relaxed configurations of the junction shown in Fig. (\ref{fig:pulling-example}) for 
 four different distances $d$.}
\label{fig:pulling-geometries}
 \end{figure}

The single-molecule junction that we discuss here is shown in  Fig. (\ref{fig:pulling-example}) and consists of gold (111) electrodes.
The electrodes in the simulation are terminated by pyramids and bridged by a bipyridine molecule.
To simulate a stretching process we have created
one hundred geometries of the junction, each with a different distance $d$ between the center of the end atoms of the
two leads. To optimize the atomic arrangement of each of the hundred geometries, we start from an idealized setup consisting of the
two pyramids surrounded by vacuum (e.g.: not attached to the gold leads). We place the molecule slightly shifted to one side to break
the symmetry and keep an Au-N bond-length of about 2 \AA. We then relax the inter-atomic forces with  the SIESTA code using a GGA
functional\cite{gga} and a double-zeta-polarized basis set until each individual force is smaller than 0.02 eV/\AA. We keep
fixed the atomic positions of the bottom two layers of the pyramids during this geometry optimization.
Fig. (\ref{fig:pulling-geometries}) shows four of the hundred relaxed configurations achieved.

We then reattach the crystalline  gold leads, and impose periodic boundary conditions along the plane perpendicular to the transport direction.
We use a SZ basis for the gold atoms of the leads, together with a simplified pseudo-potential, where only the 6s
channel is included to speed up the simulations. However, we use a double-zeta-polarized basis set for the atoms at the gold pyramids and
in the molecule. SIESTA then creates the Hamiltonian of each of the hundred junctions that we feed into GOLLUM.

\begin{figure}
 \includegraphics[width=\columnwidth]{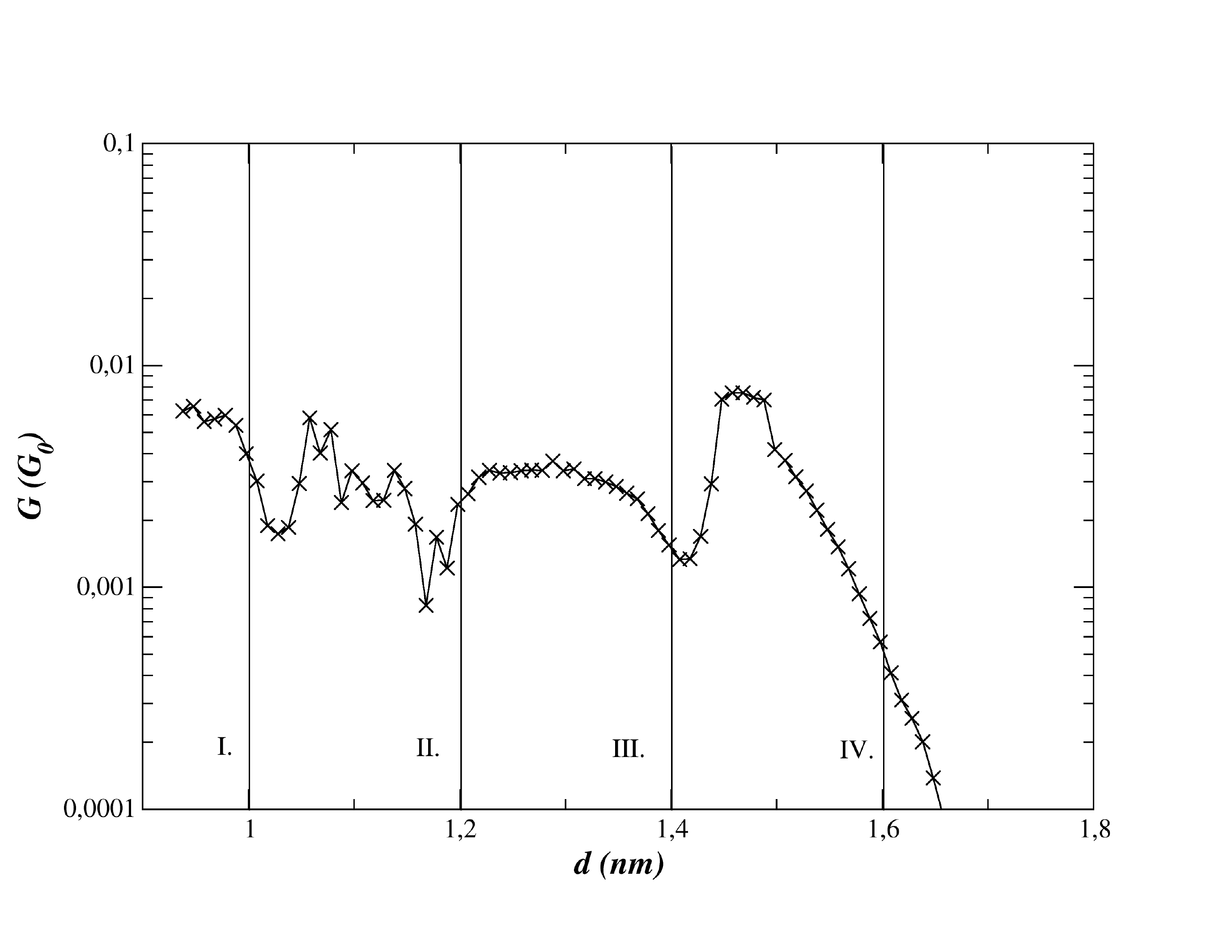}
 \caption{(Color online) Conductance of the junction shown in Fig. (\ref{fig:pulling-example}). 
 $G$ is measured in units of $G_0$ and plotted as a function of the distance $d$. Each of the conductance points displayed
 correspond to one of the hundred relaxed configurations of the MCBJ simulation. The figure also shows the four different distances
 $d$ corresponding to the relaxed geometries displayed in Fig. (\ref{fig:pulling-geometries}).}
\label{fig:pulling-curve}
 \end{figure}

Figure (\ref{fig:pulling-curve}) shows the low-voltage conductance $G$ versus the electrode separation $d$. This 'pulling curve'
shows that during the pulling process the conductance possesses a plateau, in agreement with many experiments using MCBJs.
This simulation also  reveals  that the aromatic rings contact directly the gold surface, therefore increasing the molecule-gold
coupling and the molecular conductance.

\subsection{Quantum Pumping in Carbon Nanotube Archimedes Screws}
So far, all calculated quantities have been obtained from the modulus squared of the scattering matrix elements. To demonstrate
that GOLLUM  also provides information about transport properties associated with the phases of the scattering matrix, we
now examine an  example of a quantum pump. Quantum pumps are time-dependent electron scatterers, which are able
to transport electrons between two external reservoirs subjected to the same chemical potential. The pump process is adiabatic if
the frequency of the pump cycle is smaller than the inverse of the characteristic timescale of the scatterer, the Wigner delay
time\cite{WignerDelay_1955}. Experimental \cite{Pothier1992-SEQP,SwitkesM_1999_1} and theoretical
\cite{ZhouF_1999_1,AvronJE_2000_1,AvronJE_2001_1} studies of adiabatic quantum pumps have examined the conditions for optimal
pumping and the effects of noise and dissipation.

Adiabatic pumping can be understood in terms of the parametric derivative of the
full scattering matrix $S$ at fixed chemical potential \cite{ButtikerM_1994_1,BrouwerPW_1998_1}.
An adiabatically-slowly time-varying scatterer connected by ideal channels to external reservoirs, produces a current
\begin{equation}
\partial_{t}Q_{j}(t)=\frac{e}{h}\,\mathcal{E}_{jj}(t,E_F)
\end{equation}
pumped into the $j$th channel, where $\mathcal{E}_{jj}$ is the energy shift matrix defined by
\begin{equation}
\mathcal{E}\left(t,E_{F}\right)=
i\hbar\,\,\frac{\partial {\cal S}\left(t,E_{F}\right)}{\partial t}\,\,{\cal S}^{\dagger}\left(t,E_{F}\right),
\end{equation}
with $S$ being the full scattering matrix and $E_{F}$ the Fermi
energy. Pumping can occur if the $s$-matrix depends on time through
a parameter $\varphi(t)$. Hence currents can be expressed
in terms of parametric derivatives
\begin{equation}
\partial_{t}Q_{j}=\partial_{\varphi}Q_{j}(\varphi)\,\,\partial_{t}\varphi(t)
\end{equation}
where  $\partial_{\varphi}Q_{j}$ is the parametric current
entering channel $j$.

\begin{figure}
\includegraphics[width=0.9\columnwidth]{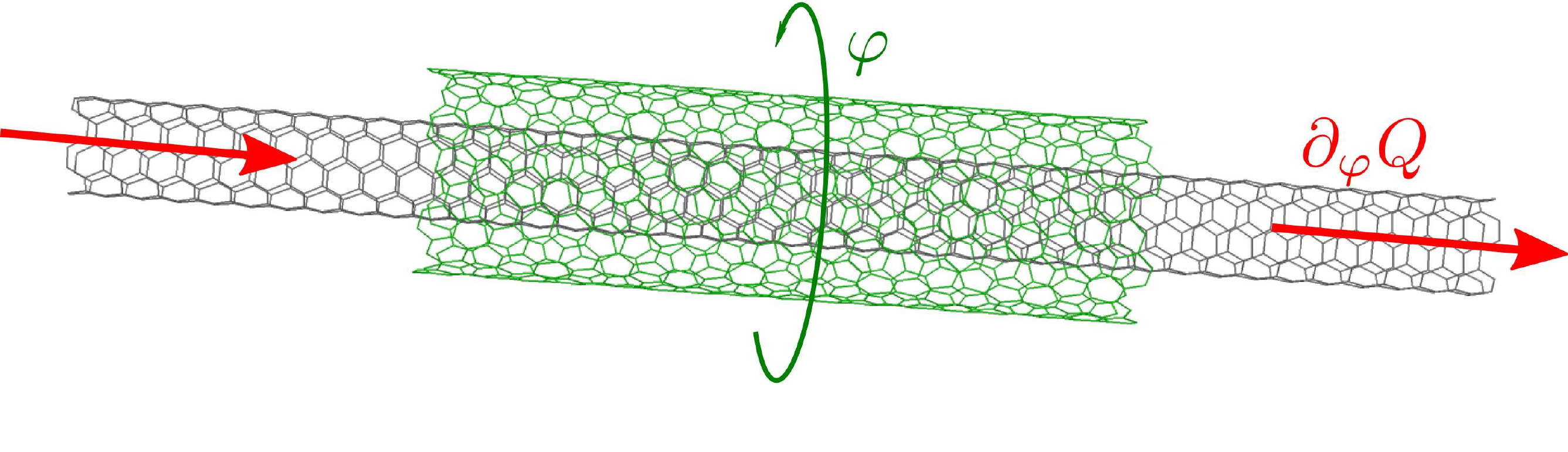}
\caption{(Color online)  The device geometry of a double-wall nano-electromechanical
quantum pump. An outer carbon nanotube of length $L\approx50$~\AA{} surrounds
concentrically an inner tube, with an inter-layer spacing $W\approx 3.4$~\AA{}
corresponding to the van der Waals distance. The inner wall remains
fixed, while the outer tube is rotated about the tube axis by the angle
$\varphi$. A slow variation of $\varphi$ results in a parametric current
$\partial_{\varphi}Q$. Finite charge can be pumped in one rotational cycle
depending on the chiralities of the constituent tubes.}
\label{fig:shuttle}
\end{figure}

Since GOLLUM gives us access to the full scattering matrix $\cal{S}$, it offers
the possibility of investigating adiabatic pumping in nanostructures.
We demonstrate this capability by calculating the charge pumped in a double-walled
carbon nanotube nano-electromechanical device shown in Fig. (\ref{fig:shuttle})
\cite{acsnano_cntpump}, that mimics the experimental setup of Ref. (\onlinecite{FennimoreAM_2003_1}).
 Since an electron current travelling along the inner tube can cause a chiral outer tube to rotate \cite{motor}, the quantum pump shown in Fig. (\ref{fig:shuttle}) represents the inverse effect, in which rotation of the outer tube causes a current to flow along the inner tube. The position and orientation of the inner tube is kept fixed,
while the shorter outer tube rotates slowly. The angle $\varphi$
describes the real space rotation angle of the outer tube and also
plays the role of the pumping parameter in this system.

\begin{figure}
\includegraphics[width=0.9\columnwidth]{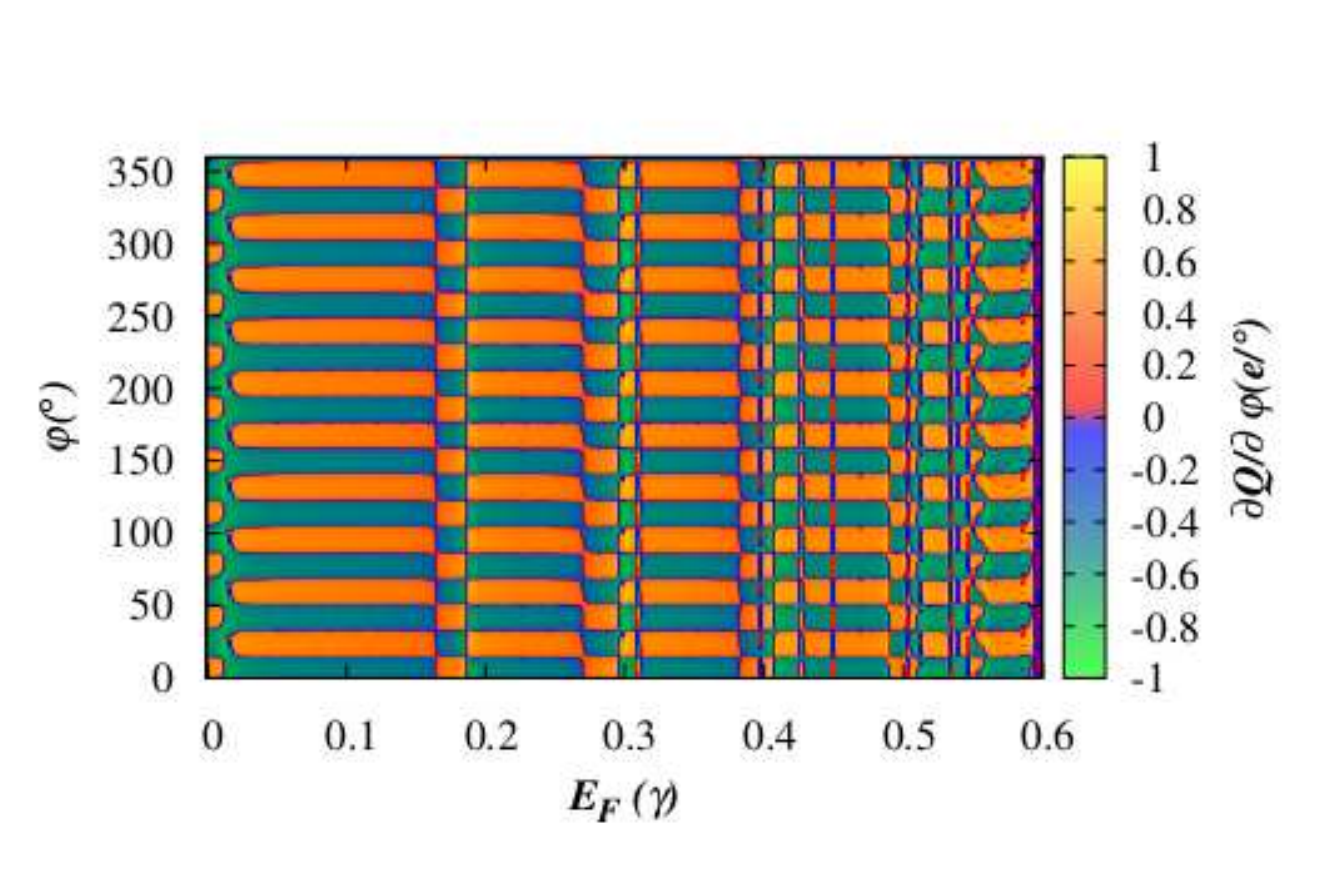}
\caption{(Color online) Contour plots providing the parametric current $\partial_{\varphi}Q$
for a device consisting of a (5,5) carbon nanotube surrounded by a (14,6) outer nanotube.
The outer tube rotates slowly around the inner one. The contour plot shows the current 
as a function of the rotational angle $\varphi$ measured in degrees and the Fermi energy 
$E_{F}$ measured in units of the hopping integral $\gamma$ between carbon atoms. 
The orange color indicates charge being pumped from left to right and vice versa.}
\label{fig:paremplot}
\end{figure}

To reveal the rich behavior of this family of quantum pumps, Fig. (\ref{fig:paremplot})
shows the parametric current, as a function of the rotational angle
$\varphi$ for a typical device. Depending on the particular angle,
charge may be pumped either from left to right or vice versa. The
integral of this parametric emissivity within a full parametric cycle
of $360^{\circ}$ is the number of electrons pumped per cycle.

\begin{figure}
\includegraphics[width=0.99\columnwidth]{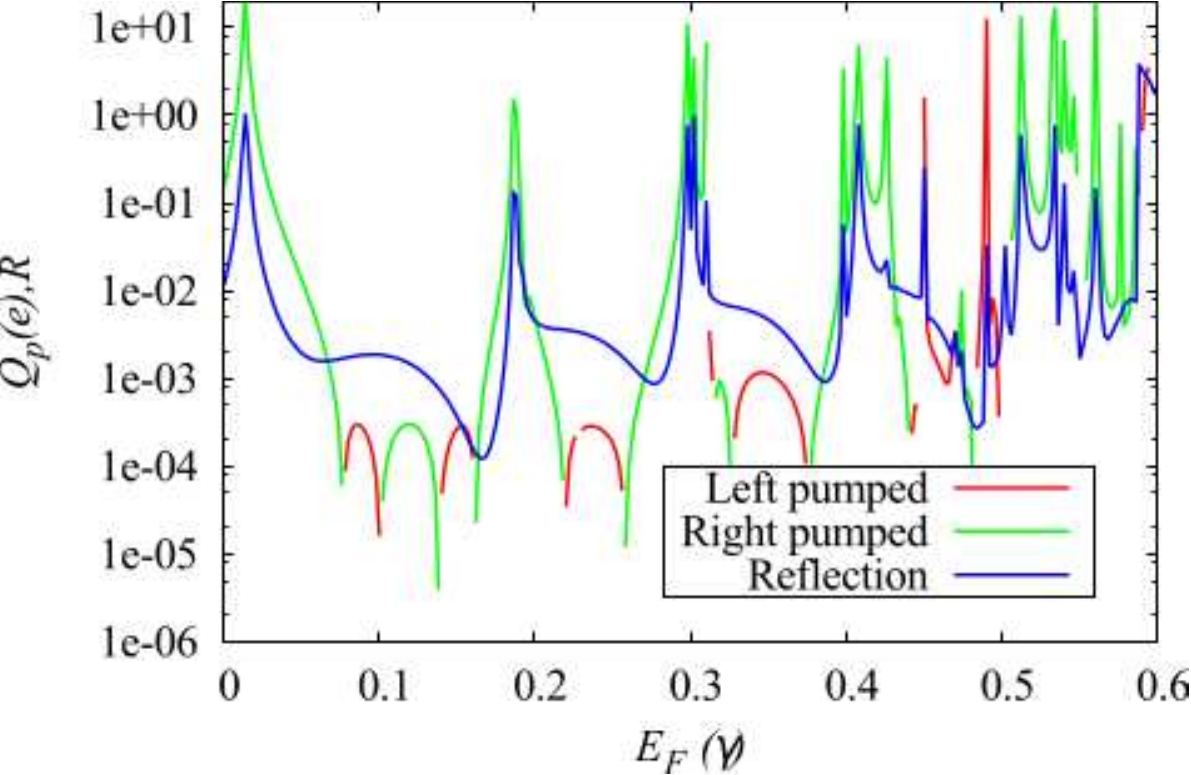}
\caption{(Color online) The calculated charge pumped per cycle $Q_p(e)$ through the device discussed
in Fig. (\ref{fig:paremplot}) as a function of the Fermi energy measured in units of the hopping integral
$\gamma$ between carbon atoms at the CNTs. The green solid
line shows the charge pumped towards the left; the red solid line shown the charge pumped towards the right. 
At certain energies, the pumped charge is very high. These peaks correlate with the Fabry-Perot resonances
in the reflection coefficient $R$. $R$ is shown as a solid blue line.}
\label{fig:14-6plot}
\end{figure}

In Fig. (\ref{fig:14-6plot}) we show the charge pumped in a (5,5) carbon
nanotube with a (14,6) outer nanotube rotating slowly about it. The
average pumped charge clearly drops by several orders of magnitude
as the Fermi energy is increased from zero. Therefore for a most efficient
pumping, the Fermi energy should be close to the Dirac point. Note
however, that the pumped charge could again increase if the Fermi
level is large enough to open another channel. Beyond this average
behavior, there exist numerous sharp peaks in the pumped charge. The
location of these peaks correlates with Fabry-Perot resonances
in the reflection coefficient. This suggests that the largest
pumping occurs at those resonances. In other words, when the transmission
is high, pumping is low, and vice versa.

\subsection{Transport in disordered systems: ballistic, diffusive and localized behavior}
\begin{figure}
\includegraphics[trim=0cm 5cm 0cm 5cm, clip=true, width=1.0\columnwidth]{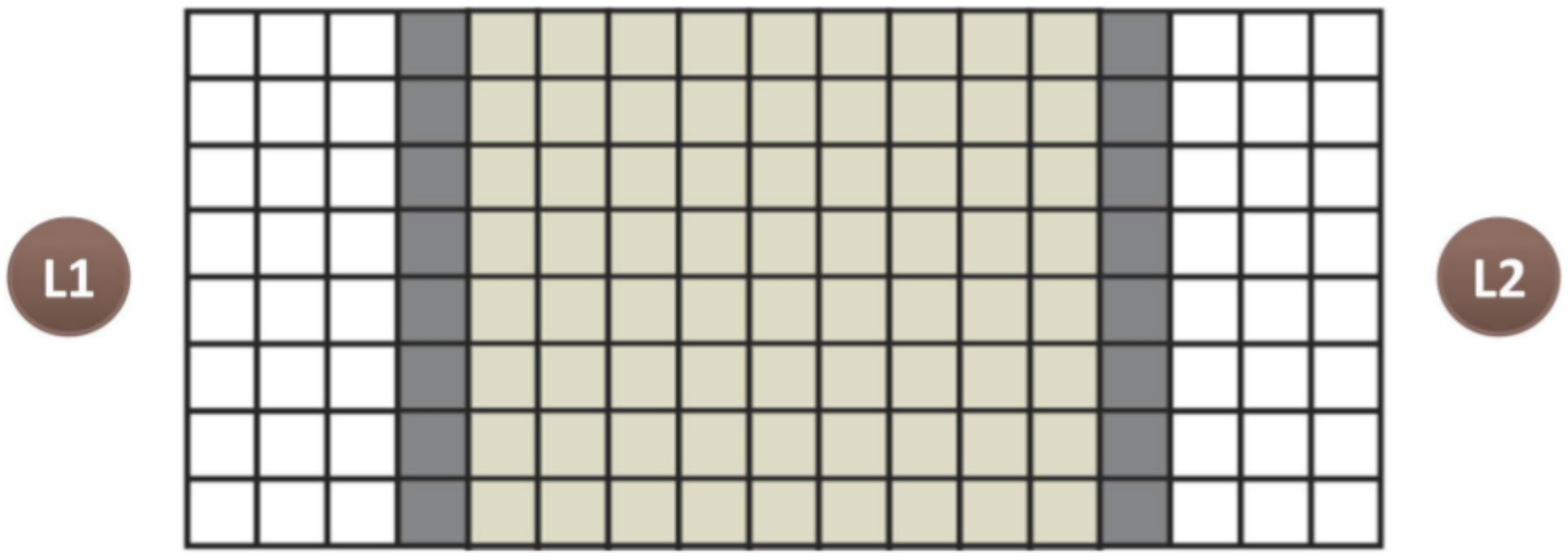}
\caption{A two-terminal tight-binding system defined on a square lattice, comprising two leads connected to a disordered scattering region.}
\label{fig:BD1}
\end{figure}

Finally, to demonstrate that GOLLUM can handle the disordered systems, we calculate the ensemble-averaged conductivity $\sigma$ of a 
two-terminal system on a square lattice, with leads attached to a disordered  scattering region as shown in Fig. (\ref{fig:BD1}). The 
width of the system is $W=11$ unit cells and the length is varied between $L=1$ to $L=500$ unit cells. 
The conductivity is defined as $\sigma=T(E_{F})W/L$, where $T(E_{F})$ is the transmission from the left lead to right lead evaluated at the Fermi energy, $E_{F}=0.5$ eV. The tight-binding  Hamiltonian of the system has a single orbital per site, with nearest neighbour couplings.  $\gamma=-1$ eV. The site energies within the leads are $\varepsilon_{0}=0$ eV, while the random site energies within the scattering region are uniformly distributed over the interval $[-1.6, 1.6]$ eV.

\begin{figure}
\includegraphics[width=1.0\columnwidth]{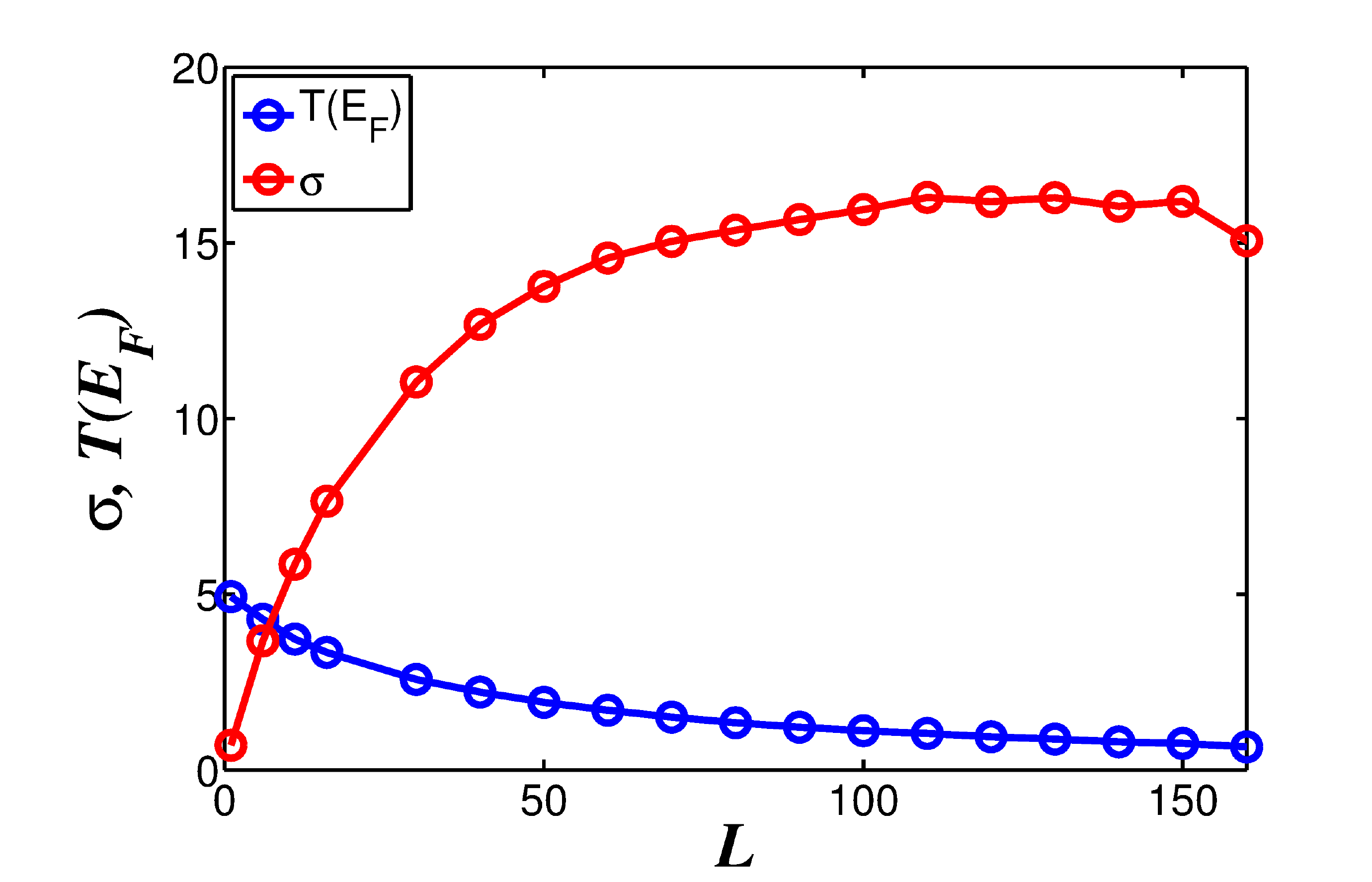}
\caption{Conductivity $\sigma$ and transmission coefficient $T(E_F)$ as a function of the number $L$ of unit cells in the transport
direction, for the two probe square lattice shown in the previous 
figure.}
\label{fig:BD2}
\end{figure}

Figure (\ref{fig:BD2}) shows the the ensemble-averaged conductivity ($\sigma$) and transmission coefficient ($T(E_F)$) for the system shown in Fig. (\ref{fig:BD1}). It contains three regions. Within the ballistic regime between $L=0$ and approximately $L=20$, the conductivity increases linearly with  length. In the diffusive region ($L=40 - 80$), the conductivity exhibits ohmic behavior and is almost independent of length. Finally for $L$ greater than 100, there is a cross over to the Anderson localized regime.

\subsection{Impact of the spin-orbit interaction in the transport properties of nickel chains}
We end this article by showing how the spin-orbit interaction induces gaps at certain band crossings in the 
one-dimensional electronic structure of infinite nickel chains. These gaps may appear or not depending on the orientation
of the atomic spins relative to the axis of the chain. They lead to dips in the transmission coefficient
$T(E)$ of the chain at the gap energies.

We have simulated linear nickel chains using the DFT program SIESTA. The chains have a single atom per unit cell and are oriented along the z-axis. We have used a standard set of pseudopotential parameters,  a double-zeta basis set and simple LDA for the exchange-correlation potential.

\begin{figure}
\includegraphics[trim = 0mm 0mm 80mm 200mm, clip=true, width=1.0\columnwidth]{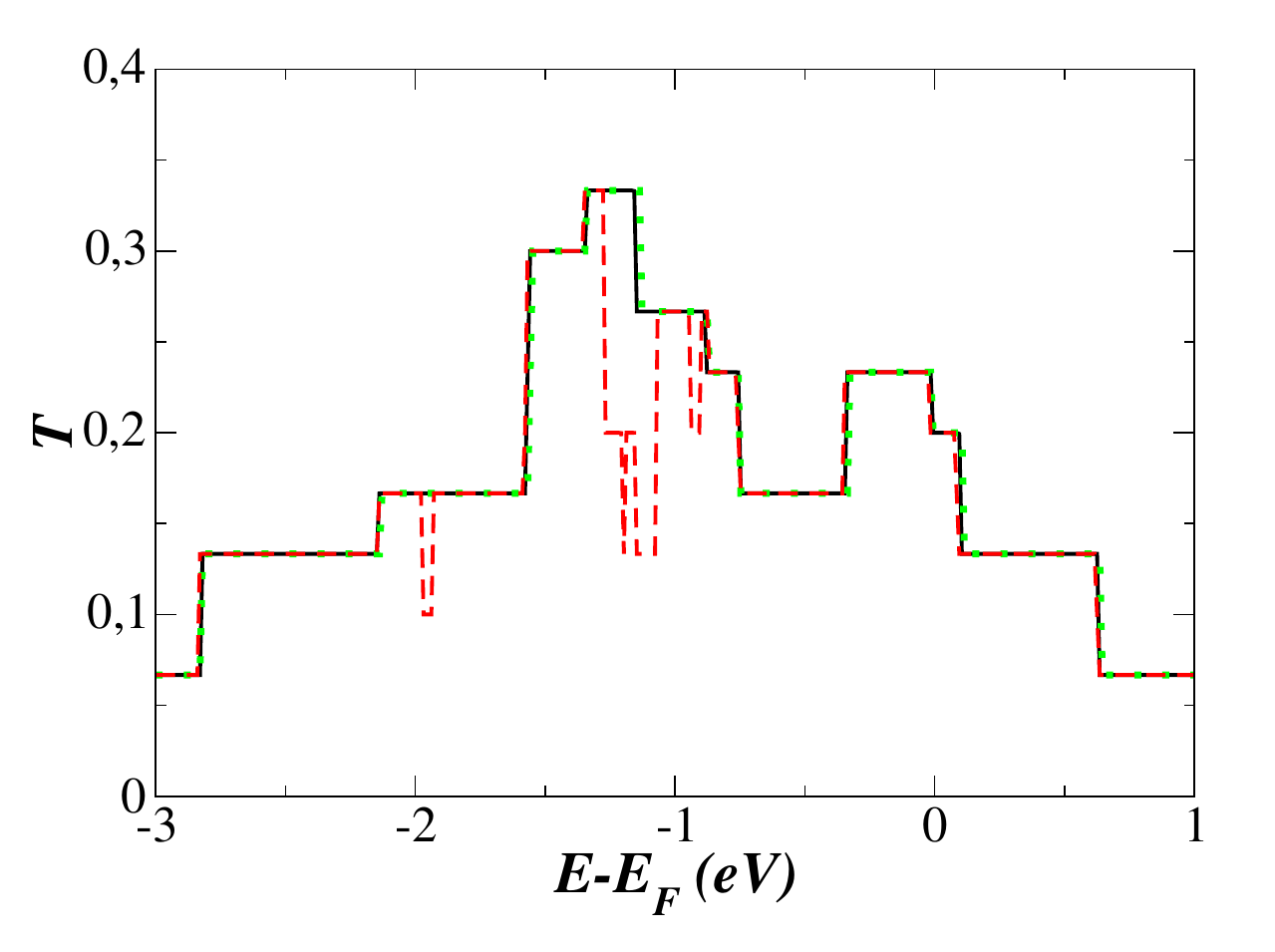}
\caption{Transmission coefficient of a linear nickel chain oriented along the z-axis. The solid black line shows $T(E)$ when the 
atomic spins are oriented along the chains axis. The dashed red line shows $T(E)$ when the atomic spins are oriented
in the plane perpendicular to the chain axis. A blue dotted line showing $T(E)$ when the spin-orbit interaction is switched off
falls on top of the black line. }
\label{fig:ni-chain-so}
\end{figure}

We have checked that the electronic structure and $T(E)$ are the same for any spin orientation if the spin-orbit
interaction is set to zero, as is should due rotational invariance. However, if the spin-orbit interaction is switched on, then
a finite yet small magnetic anisotropy barrier appears. We have found that if we choose the atomic spins to lie along the chain axis,
then there are no spin-orbit gaps close to the Fermi energy. As a consequence, the transmission coefficients with and without
the spin-orbit interaction are indistinguishable from each other (as shown in Fig. (\ref{fig:ni-chain-so})).
In contrast, when the atomic spins are oriented in a plane perpendicular to the chain axis, then several small gaps open 
around the Fermi energy. These gaps are seen as dips in Fig. (\ref{fig:ni-chain-so}).

\section{Conclusion}
We have developed a new quantum transport code, which is fast, easy to use and versatile. This flexibility has been demonstrated by presenting a wide range of example calculations, encompassing charge, spin and thermal transport, corrections to density functional theory such as LDA+U and spectral adjustments, transport in
the presence of non-collinear magnetism, the quantum-Hall effect,  Kondo and Coulomb blockade effects, finite-voltage transport, multi-terminal transport, quantum pumps, superconducting nanostructures, environmental effects and
pulling curves and conductance histograms for mechanically-controlled-break-junction experiments. Further
developments are in the pipeline, including the incorporation of  phonon transport. GOLLUM will soon be freely
available from the following web site {\it http://www.physics.lancs.ac.uk/gollum} and the authors of this article are available to help potential users access the code.

The research presented here was funded by the Spanish Ministerio de Econom\'{\i}a y Competitividad through the grant FIS2012-34858, by the UK EPSRC and by the
European Commission  FP7 ITN "MOLESCO" Project No. 606728
VMGS thanks the Spanish Ministerio de Econom\'{\i}a y Competitividad for a Ram\'on y Cajal fellowship (RYC-2010-06053).
LO has been supported by the Hungarian Scientiﬁc Research Fund No. K108676. 
JF wishes to thank useful discussions with N. Lorente and A. Kormanyos.

\appendix
\section{Procedures used to regularize $K_1$}
We have described in section II.A.2 the method employed by GOLLUM to find the surface Green's function $G_S$ of each lead.
However, the solution of Eq. (\ref{GEP}) gives with some frequency numerical inaccuracies which render the method useless
as it stands. These inaccuracies are caused by the highly non-singular behavior of the Hamiltonian matrix $K_1$ connecting
adjacent PLs. We discuss here the adaptation of the method described in Refs. (\onlinecite{smeagol,Rungger08})
that GOLLUM uses to regularize $K_1$. Mathematically, we perform an SVD decomposition of this $N\times N$ matrix,
\begin{equation}
 K_1=U\,S\,V^\dagger
 \label{SVD}
\end{equation}
where $U$ and $V$ are unitary matrix and $S$ is a diagonal matrix containing the eigenvalues $\lambda$ of $K_1$. Numerical
algorithms usually arrange them in descending order. The condition number of $K_1$ is defined as the ratio
$\kappa=\lambda^\mathrm{max}/\lambda^\mathrm{min}$ between the maximum and minimum eigenvalues of $K_1$. $\kappa$ determines
how singular is $K_1^{-1}$ and therefore the propensity to suffer inaccuracies when handling $K_1$. Small eigenvalues
$\lambda$ appear whenever $K_1$ is very sparse. Physically, this is originated for example if the PL are very long so that
a large fraction of hopping integrals (matrix elements of $K_1$) is zero. Our first procedure to regularize $K_1$
consists in adding a real or complex random matrix to $K_1$. We have found that this is frequently enough to render a
regular $K_1$ matrix. If this first procedure fails this is because the orbitals involved do not play a
role in the transport properties of the lead and should be decimated out, so that the dimensions of the $K_!$ matrix are reduced.

Notice that reducing the dimensions of $K_1$ has the advantage that the computation of the surface Green's function $G_S$ is
much lighter. However, the procedure must be performed with some care as finally $G_S$ must connect with the corresponding TPL of the
EM branch, whose matrices have dimensions $N\times N$. Explicitly, we write the Schroedinger equation of the infinite chain
of the corresponding Lead:
\begin{equation}
K_0\,C_n + K_1\,C_{n+1} + K_{-1}\,C_{n-1}=0
\label{lead-equation}
\end{equation}
where $C_n=e^{i k n a } \,C(k)$, $n$ labels the PL and runs from $-\infty$ to $+\infty$ and we assume that the chain will be
chopped off at the $n_0$ PL and then connected to the TPL of the EM. We first perform the SVD decomposition of $K_1$ described
in Eq. (\ref{SVD}). We then set to zero all eigenvalues smaller than a given tolerance $tol$. We have checked that setting
$tol=10^{-8} - 10^{-9}$ provides unproblematic $K_1$ and $K_{-1}=K_1^\dagger$. Let us assume that we set to zero $D$ eigenvalues
of $K_1$, so that $M=N-D$ remain non-zero. We now construct the matrices
\begin{eqnarray}
 K_1&=&U\,\left(\begin{array}{cc}K_{1,M}&0\\0&0\end{array}\right)\,V^\dagger\nonumber\\
 K_1'&=&(K_{-1}')^\dagger=V^\dagger\,K_1\,V=V^\dagger \,U\,S=\left(\begin{array}{cc}K_{1,M}&0\\P&0\end{array}\right)\nonumber\\
 K_0'&=&V^\dagger\,K_0\,V=\left(\begin{array}{cc}K_{0,M}&W\\W^\dagger&K_{0,D}\end{array}\right)\nonumber\\
 \tilde{K}_1&=&\tilde{K}_{-1}^\dagger=V^\dagger\,K_1\nonumber=\left(\begin{array}{cc}V_1&V_2\\V_3&V_4\end{array}\right)=
          \left(\begin{array}{c}A\\B\end{array}\right)\nonumber\\
           C_n'&=&V^\dagger\,C_n=\left(\begin{array}{c}C_{M,n}'\\C_{D,n}'\end{array}\right)
\end{eqnarray}
We transform Eq. (\ref{lead-equation}) for all sites up to site $n_0-1$ as follows:
\begin{widetext}
\begin{equation}
 V^\dagger\,\left(K_0\,C_n + K_1\,C_{n+1} + K_{-1}\right)\,V\,V^\dagger\,C_{n-1}=
 K_0'\,C_n' + K_1'\,C_{n+1}' + K_{-1}'\,C_{n-1}'=0
\end{equation}
\end{widetext}
while the equation for sites $n_0$ and $n_0+1$ will be
\begin{eqnarray}
 K_0'\,C_{n0}'+K_{-1}'\,C_{n0-1}'+\tilde{K}_1\,C_{n0+1}=0\nonumber\\
 K_0\,C_{n0+1}+\tilde{K}_{-1}\,C_{no}'+K_1\,C_{n+2}=0
\end{eqnarray}
We now decimate out $C_D'$ up to site $n_0$, arriving to the new set of equations
\begin{widetext}
\begin{eqnarray}
&& \underbrace{\left(K_{0,M}-W^\dagger\,K_{0,D}^{-1}\,W-P^\dagger\,K_{0,D}^{-1}\,P\right)}_{K_0^\mathrm{new}}\,C_{M,n}'+
 \underbrace{\left(K_{1,M}-W^\dagger\,K_{0,D}^{-1}\,P\right)}_{K_1^\mathrm{new}}\,C_{M,n+1}'+
 \underbrace{\left(K_{1,M}^\dagger-P^\dagger\,K_{0,D}^{-1}\,W\right)}_{K_{-1}^\mathrm{new}}\,C_{M,n-1}'=0\,\,\,\,n\,<\,n_0\nonumber\\
&& \left(K_{0,M}-W^\dagger\,K_{0,D}^{-1}\,W-P^\dagger\,K_{0,D}^{-1}\,P\right)\,C_{M,n0}'+
 \left(A-W^\dagger\,K_{0,D}^{-1}\,B\right)\,C_{n0+1}'+
  \left(K_{1,M}^\dagger-P^\dagger\,K_{0,D}^{-1}\,W\right)\,C_{M,n0-1}'=0\nonumber\\
&& \left(K_0-B^\dagger\,K_{0,D}\,B\right)\,C_{n0+1}'+\left(A^\dagger-B^\dagger\,K_{0,D}\,W\right)+K_1\,C_{n0+2}'=0
 \end{eqnarray}
This set of equations generates the new Hamiltonians of each lead, and connects it with the extended molecule.
Overlap matrices must also be decimated as they enter into the expression for the group velocity,
Eq. (\ref{group-velocity}). The denominator in this equation looks like
\begin{equation}
C(k)^\dagger\left(S_0+S_1\,e^{i k a}+S_{-1}\,e^{-i k a}\right)\,C(k)=
 C(k)'^\dagger\left(S_0'+S_1'\,e^{i k a}+S_{-1}'\,e^{-i k a}\right)\,C(k)'
\end{equation}
where we have applied the SVD transformation to the overlap matrices
\begin{eqnarray}
 S_0'&=&V^\dagger\,S_0\,V=\left(\begin{array}{cc}A&B^\dagger\\B&C\end{array}\right)\nonumber\\
 S_1'&=&V^\dagger\,S_1\,V=\left(\begin{array}{cc}D&0\\F&0\end{array}\right)
\end{eqnarray}

By decimating out the unwanted degrees of freedom $C_D$, we arrive at the following expressions
\begin{eqnarray}
 S_0^\mathrm{new}&=&A-W^\dagger\,K_{0,D}^{-1}\,B+B^\dagger\,K_{0,D}^{-1}\,W+W^\dagger\,K_{0,D}^{-1}|,C\,K_{0,D}^{-1}\,W+
 P^\dagger\,K_{0,D}^{-1}\,C\,K_{0,D}^{-1}\,P-P^\dagger\,K_{0,D}^{-1}\,E-E^\dagger\,K_{0,D}^{-1}\,P\nonumber\\
 S_1^\mathrm{new}&=&D-B^\dagger\,K_{0,D}^{-1}\,P-W^\dagger\,K_{0,D}^{-1}\,E+W^\dagger\,K_{0,D}^{-1}C\,K_{0,D}^{-1}\,P
\end{eqnarray}
\end{widetext}

\end{document}